\documentclass[longauth]{aa} 

\usepackage{graphicx}
\usepackage{txfonts}
\graphicspath{{figures/}}
\usepackage{amsmath}
\usepackage{fourier}
\usepackage{url}
\usepackage{longtable}
\usepackage{tikz}
\usepackage{chemformula}
\usepackage{placeins}
\usepackage{ulem}

\usepackage[colorlinks=true,linkcolor=blue, citecolor=blue, filecolor=blue,urlcolor=blue]{hyperref}

\usepackage{acronym}
\acrodef{PC}[PC]{phase curve}
\acrodef{LC}[LC]{light curve}
\acrodef{LD}[LD]{limb darkening}
\acrodef{GP}[GP]{Gaussian Process}
\acrodef{MCMC}[MCMC]{Markov Chain Monte Carlo}
\acrodef{HJ}[HJ]{Hot Jupiter}
\acrodef{UHJ}[UHJ]{ultra-hot Jupiter}
\acrodef{FFI}[FFI]{Full Frame Images}
\acrodef{TPF}[TPF]{Target Pixel File}
\acrodef{CBV}[CBV]{Co-trending Basis Vectors}
\acrodef{DRP}[DRP]{Data Reduction Pipeline}
\acrodef{IRFM}[IRFM]{Infra-Red Flux Method}
\acrodef{SED}[SED]{Spectral Energy Distribution}
\acrodef{FAP}[FAP]{False Alarm Probability}
\acrodef{MAP}[MAP]{Maximum A Posteriori}
\acrodef{GLS}[GLS]{Generalized Lomb-Scargle}
\acrodef{PSF}[PSF]{Point Spread Function}
\acrodef{AIC}[AIC]{Akaike Information Criterion}
\acrodef{PPD}[PPD]{posterior probability distribution}
\acrodef{POET}[POET]{Photometry for Orbits, Eclipses, and Transits}
\acrodef{BCD}[BCD]{basic calibrated data}
\acrodef{GCM}[GCM]{global circulation model}

\DeclareSymbolFont{UPM}{U}{eur}{m}{n}
\DeclareMathSymbol{\umu}{0}{UPM}{"16}
\let\oldumu=\umu
\renewcommand\umu{\ifmmode\oldumu\else$\oldumu$\fi}

\newcommand\micron{\umu\rm m}
\newcommand\microns{\micron\xspace}
\newcommand\Spitzer{Spitzer\xspace}

\usepackage[flushleft]{threeparttable}

\begin{document}

\nolinenumbers

\newcommand{\teff}{T$_{\rm eff}$\xspace}
\newcommand{\logg}{$\log${(g)}\xspace}
\newcommand{\wtb}{WASP-3~b\xspace}
\newcommand{\wtA}{WASP-3\xspace}
\newcommand{\cheops}{CHEOPS\xspace}
\newcommand{\tess}{TESS\xspace}
\newcommand{\aas}{AAS\xspace}
\newcommand{\pycheops}{{\tt pycheops}}
\newcommand{\scalpels}{SCALPELS\xspace}
\newcommand{\vsini}{\ensuremath{v \sin i_\star}\xspace}
\newcommand{\ageo}{$A_{\rm g}$\xspace}
\newcommand{\asph}{$A_{\rm S}$\xspace}
\newcommand{\abond}{$A_{\rm B}$\xspace}
\newcommand\decl{$\delta_{\rm ecl}$\xspace}
\newcommand\tday{$\rm T_{\rm d}$\xspace}


%
   \title{The CHEOPS view on the climate of WASP-3~b.\thanks{The CHEOPS program IDs are CH\_PR100016 and CH\_PR100052.}}


\author{
G. Scandariato\inst{1} $^{\href{https://orcid.org/0000-0003-2029-0626}{\includegraphics[scale=0.5]{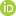}}}$, 
L. Carone\inst{2}, 
P. E. Cubillos\inst{2,3}, 
P. F. L. Maxted\inst{4} $^{\href{https://orcid.org/0000-0003-3794-1317}{\includegraphics[scale=0.5]{figures/orcid.jpg}}}$, 
T. Zingales\inst{5,6} $^{\href{https://orcid.org/0000-0001-6880-5356}{\includegraphics[scale=0.5]{figures/orcid.jpg}}}$, 
M. N. Günther\inst{7} $^{\href{https://orcid.org/0000-0002-3164-9086}{\includegraphics[scale=0.5]{figures/orcid.jpg}}}$, 
A. Heitzmann\inst{8} $^{\href{https://orcid.org/0000-0002-8091-7526}{\includegraphics[scale=0.5]{figures/orcid.jpg}}}$, 
M. Lendl\inst{8} $^{\href{https://orcid.org/0000-0001-9699-1459}{\includegraphics[scale=0.5]{figures/orcid.jpg}}}$, 
T. G. Wilson\inst{9} $^{\href{https://orcid.org/0000-0001-8749-1962}{\includegraphics[scale=0.5]{figures/orcid.jpg}}}$, 
A. Bonfanti\inst{2} $^{\href{https://orcid.org/0000-0002-1916-5935}{\includegraphics[scale=0.5]{figures/orcid.jpg}}}$, 
G. Bruno\inst{1} $^{\href{https://orcid.org/0000-0002-3288-0802}{\includegraphics[scale=0.5]{figures/orcid.jpg}}}$, 
A. Krenn\inst{2} $^{\href{https://orcid.org/0000-0003-3615-4725}{\includegraphics[scale=0.5]{figures/orcid.jpg}}}$, 
E. Meier Valdes\inst{10}, 
V. Singh\inst{1} $^{\href{https://orcid.org/0000-0002-7485-6309}{\includegraphics[scale=0.5]{figures/orcid.jpg}}}$, 
M. I. Swayne\inst{11,4} $^{\href{https://orcid.org/0000-0002-2609-3159}{\includegraphics[scale=0.5]{figures/orcid.jpg}}}$, 
Y. Alibert\inst{10,12} $^{\href{https://orcid.org/0000-0002-4644-8818}{\includegraphics[scale=0.5]{figures/orcid.jpg}}}$, 
R. Alonso\inst{13,14} $^{\href{https://orcid.org/0000-0001-8462-8126}{\includegraphics[scale=0.5]{figures/orcid.jpg}}}$, 
T. Bárczy\inst{15} $^{\href{https://orcid.org/0000-0002-7822-4413}{\includegraphics[scale=0.5]{figures/orcid.jpg}}}$, 
D. Barrado Navascues\inst{16} $^{\href{https://orcid.org/0000-0002-5971-9242}{\includegraphics[scale=0.5]{figures/orcid.jpg}}}$, 
S. C. C. Barros\inst{17,18} $^{\href{https://orcid.org/0000-0003-2434-3625}{\includegraphics[scale=0.5]{figures/orcid.jpg}}}$, 
W. Baumjohann\inst{2} $^{\href{https://orcid.org/0000-0001-6271-0110}{\includegraphics[scale=0.5]{figures/orcid.jpg}}}$, 
W. Benz\inst{12,10} $^{\href{https://orcid.org/0000-0001-7896-6479}{\includegraphics[scale=0.5]{figures/orcid.jpg}}}$, 
N. Billot\inst{8} $^{\href{https://orcid.org/0000-0003-3429-3836}{\includegraphics[scale=0.5]{figures/orcid.jpg}}}$, 
L. Borsato\inst{6} $^{\href{https://orcid.org/0000-0003-0066-9268}{\includegraphics[scale=0.5]{figures/orcid.jpg}}}$, 
A. Brandeker\inst{19} $^{\href{https://orcid.org/0000-0002-7201-7536}{\includegraphics[scale=0.5]{figures/orcid.jpg}}}$, 
C. Broeg\inst{12,10} $^{\href{https://orcid.org/0000-0001-5132-2614}{\includegraphics[scale=0.5]{figures/orcid.jpg}}}$, 
M. Buder\inst{20}, 
M.-D. Busch\inst{21}, 
A. Collier Cameron\inst{22} $^{\href{https://orcid.org/0000-0002-8863-7828}{\includegraphics[scale=0.5]{figures/orcid.jpg}}}$, 
A. C. M. Correia\inst{23} $^{\href{https://orcid.org/0000-0002-8946-8579}{\includegraphics[scale=0.5]{figures/orcid.jpg}}}$, 
Sz. Csizmadia\inst{24} $^{\href{https://orcid.org/0000-0001-6803-9698}{\includegraphics[scale=0.5]{figures/orcid.jpg}}}$, 
M. B. Davies\inst{25} $^{\href{https://orcid.org/0000-0001-6080-1190}{\includegraphics[scale=0.5]{figures/orcid.jpg}}}$, 
M. Deleuil\inst{26} $^{\href{https://orcid.org/0000-0001-6036-0225}{\includegraphics[scale=0.5]{figures/orcid.jpg}}}$, 
A. Deline\inst{8}, 
L. Delrez\inst{27,28,29} $^{\href{https://orcid.org/0000-0001-6108-4808}{\includegraphics[scale=0.5]{figures/orcid.jpg}}}$, 
O. D. S. Demangeon\inst{17,18} $^{\href{https://orcid.org/0000-0001-7918-0355}{\includegraphics[scale=0.5]{figures/orcid.jpg}}}$, 
B.-O. Demory\inst{10,12} $^{\href{https://orcid.org/0000-0002-9355-5165}{\includegraphics[scale=0.5]{figures/orcid.jpg}}}$, 
A. Derekas\inst{30}, 
B. Edwards\inst{31}, 
D. Ehrenreich\inst{8,32} $^{\href{https://orcid.org/0000-0001-9704-5405}{\includegraphics[scale=0.5]{figures/orcid.jpg}}}$, 
A. Erikson\inst{24}, 
J. Farinato\inst{6} $^{\href{https://orcid.org/0000-0002-5840-8362}{\includegraphics[scale=0.5]{figures/orcid.jpg}}}$, 
A. Fortier\inst{12,10} $^{\href{https://orcid.org/0000-0001-8450-3374}{\includegraphics[scale=0.5]{figures/orcid.jpg}}}$, 
L. Fossati\inst{2} $^{\href{https://orcid.org/0000-0003-4426-9530}{\includegraphics[scale=0.5]{figures/orcid.jpg}}}$, 
M. Fridlund\inst{33,34} $^{\href{https://orcid.org/0000-0002-0855-8426}{\includegraphics[scale=0.5]{figures/orcid.jpg}}}$, 
D. Gandolfi\inst{35} $^{\href{https://orcid.org/0000-0001-8627-9628}{\includegraphics[scale=0.5]{figures/orcid.jpg}}}$, 
K. Gazeas\inst{36}, 
M. Gillon\inst{27} $^{\href{https://orcid.org/0000-0003-1462-7739}{\includegraphics[scale=0.5]{figures/orcid.jpg}}}$, 
M. Güdel\inst{37}, 
Ch. Helling\inst{2,38}, 
K. G. Isaak\inst{7} $^{\href{https://orcid.org/0000-0001-8585-1717}{\includegraphics[scale=0.5]{figures/orcid.jpg}}}$, 
L. L. Kiss\inst{39,40}, 
J. Korth\inst{41} $^{\href{https://orcid.org/0000-0002-0076-6239}{\includegraphics[scale=0.5]{figures/orcid.jpg}}}$, 
K. W. F. Lam\inst{24} $^{\href{https://orcid.org/0000-0002-9910-6088}{\includegraphics[scale=0.5]{figures/orcid.jpg}}}$, 
J. Laskar\inst{42} $^{\href{https://orcid.org/0000-0003-2634-789X}{\includegraphics[scale=0.5]{figures/orcid.jpg}}}$, 
A. Lecavelier des Etangs\inst{43} $^{\href{https://orcid.org/0000-0002-5637-5253}{\includegraphics[scale=0.5]{figures/orcid.jpg}}}$, 
D. Magrin\inst{6} $^{\href{https://orcid.org/0000-0003-0312-313X}{\includegraphics[scale=0.5]{figures/orcid.jpg}}}$, 
B. Merín\inst{44} $^{\href{https://orcid.org/0000-0002-8555-3012}{\includegraphics[scale=0.5]{figures/orcid.jpg}}}$, 
C. Mordasini\inst{12,10}, 
V. Nascimbeni\inst{6} $^{\href{https://orcid.org/0000-0001-9770-1214}{\includegraphics[scale=0.5]{figures/orcid.jpg}}}$, 
G. Olofsson\inst{19} $^{\href{https://orcid.org/0000-0003-3747-7120}{\includegraphics[scale=0.5]{figures/orcid.jpg}}}$, 
R. Ottensamer\inst{37}, 
I. Pagano\inst{1} $^{\href{https://orcid.org/0000-0001-9573-4928}{\includegraphics[scale=0.5]{figures/orcid.jpg}}}$, 
E. Pallé\inst{13,14} $^{\href{https://orcid.org/0000-0003-0987-1593}{\includegraphics[scale=0.5]{figures/orcid.jpg}}}$, 
G. Peter\inst{20} $^{\href{https://orcid.org/0000-0001-6101-2513}{\includegraphics[scale=0.5]{figures/orcid.jpg}}}$, 
D. Piazza\inst{12}, 
G. Piotto\inst{6,5} $^{\href{https://orcid.org/0000-0002-9937-6387}{\includegraphics[scale=0.5]{figures/orcid.jpg}}}$, 
D. Pollacco\inst{9}, 
D. Queloz\inst{45,46} $^{\href{https://orcid.org/0000-0002-3012-0316}{\includegraphics[scale=0.5]{figures/orcid.jpg}}}$, 
R. Ragazzoni\inst{6,5} $^{\href{https://orcid.org/0000-0002-7697-5555}{\includegraphics[scale=0.5]{figures/orcid.jpg}}}$, 
N. Rando\inst{7}, 
H. Rauer\inst{24,47} $^{\href{https://orcid.org/0000-0002-6510-1828}{\includegraphics[scale=0.5]{figures/orcid.jpg}}}$, 
I. Ribas\inst{48,49} $^{\href{https://orcid.org/0000-0002-6689-0312}{\includegraphics[scale=0.5]{figures/orcid.jpg}}}$, 
N. C. Santos\inst{17,18} $^{\href{https://orcid.org/0000-0003-4422-2919}{\includegraphics[scale=0.5]{figures/orcid.jpg}}}$, 
D. Ségransan\inst{8} $^{\href{https://orcid.org/0000-0003-2355-8034}{\includegraphics[scale=0.5]{figures/orcid.jpg}}}$, 
A. E. Simon\inst{12,10} $^{\href{https://orcid.org/0000-0001-9773-2600}{\includegraphics[scale=0.5]{figures/orcid.jpg}}}$, 
A. M. S. Smith\inst{24} $^{\href{https://orcid.org/0000-0002-2386-4341}{\includegraphics[scale=0.5]{figures/orcid.jpg}}}$, 
S. G. Sousa\inst{17} $^{\href{https://orcid.org/0000-0001-9047-2965}{\includegraphics[scale=0.5]{figures/orcid.jpg}}}$, 
M. Stalport\inst{28,27}, 
S. Sulis\inst{26} $^{\href{https://orcid.org/0000-0001-8783-526X}{\includegraphics[scale=0.5]{figures/orcid.jpg}}}$, 
Gy. M. Szabó\inst{30,50} $^{\href{https://orcid.org/0000-0002-0606-7930}{\includegraphics[scale=0.5]{figures/orcid.jpg}}}$, 
S. Udry\inst{8} $^{\href{https://orcid.org/0000-0001-7576-6236}{\includegraphics[scale=0.5]{figures/orcid.jpg}}}$, 
V. Van Grootel\inst{28} $^{\href{https://orcid.org/0000-0003-2144-4316}{\includegraphics[scale=0.5]{figures/orcid.jpg}}}$, 
J. Venturini\inst{8} $^{\href{https://orcid.org/0000-0001-9527-2903}{\includegraphics[scale=0.5]{figures/orcid.jpg}}}$, 
E. Villaver\inst{13,14}, 
N. A. Walton\inst{51} $^{\href{https://orcid.org/0000-0003-3983-8778}{\includegraphics[scale=0.5]{figures/orcid.jpg}}}$
}

\institute{
\label{inst:1} INAF, Osservatorio Astrofisico di Catania, Via S. Sofia 78, 95123 Catania, Italy \and
\label{inst:2} Space Research Institute, Austrian Academy of Sciences, Schmiedlstrasse 6, A-8042 Graz, Austria \and
\label{inst:3} INAF, Osservatorio Astrofisico di Torino, Via Osservatorio, 20, I-10025 Pino Torinese To, Italy \and
\label{inst:4} Astrophysics Group, Lennard Jones Building, Keele University, Staffordshire, ST5 5BG, United Kingdom \and
\label{inst:5} Dipartimento di Fisica e Astronomia "Galileo Galilei", Università degli Studi di Padova, Vicolo dell'Osservatorio 3, 35122 Padova, Italy \and
\label{inst:6} INAF, Osservatorio Astronomico di Padova, Vicolo dell'Osservatorio 5, 35122 Padova, Italy \and
\label{inst:7} European Space Agency (ESA), European Space Research and Technology Centre (ESTEC), Keplerlaan 1, 2201 AZ Noordwijk, The Netherlands \and
\label{inst:8} Observatoire astronomique de l'Université de Genève, Chemin Pegasi 51, 1290 Versoix, Switzerland \and
\label{inst:9} Department of Physics, University of Warwick, Gibbet Hill Road, Coventry CV4 7AL, United Kingdom \and
\label{inst:10} Center for Space and Habitability, University of Bern, Gesellschaftsstrasse 6, 3012 Bern, Switzerland \and
\label{inst:11} SUPA, School of Physics and Astronomy, Kelvin Building, University of Glasgow, Glasgow, G12 8QQ, Scotland, UK \and
\label{inst:12} Weltraumforschung und Planetologie, Physikalisches Institut, University of Bern, Gesellschaftsstrasse 6, 3012 Bern, Switzerland \and
\label{inst:13} Instituto de Astrofísica de Canarias, Vía Láctea s/n, 38200 La Laguna, Tenerife, Spain \and
\label{inst:14} Departamento de Astrofísica, Universidad de La Laguna, Astrofísico Francisco Sanchez s/n, 38206 La Laguna, Tenerife, Spain \and
\label{inst:15} Admatis, 5. Kandó Kálmán Street, 3534 Miskolc, Hungary \and
\label{inst:16} Depto. de Astrofísica, Centro de Astrobiología (CSIC-INTA), ESAC campus, 28692 Villanueva de la Cañada (Madrid), Spain \and
\label{inst:17} Instituto de Astrofisica e Ciencias do Espaco, Universidade do Porto, CAUP, Rua das Estrelas, 4150-762 Porto, Portugal \and
\label{inst:18} Departamento de Fisica e Astronomia, Faculdade de Ciencias, Universidade do Porto, Rua do Campo Alegre, 4169-007 Porto, Portugal \and
\label{inst:19} Department of Astronomy, Stockholm University, AlbaNova University Center, 10691 Stockholm, Sweden \and
\label{inst:20} Institute of Optical Sensor Systems, German Aerospace Center (DLR), Rutherfordstrasse 2, 12489 Berlin, Germany \and
\label{inst:21} Space Research and Planetology, University of Bern, Gesellschaftsstrasse 6, 3012 Bern, Switzerland \and
\label{inst:22} Centre for Exoplanet Science, SUPA School of Physics and Astronomy, University of St Andrews, North Haugh, St Andrews KY16 9SS, UK \and
\label{inst:23} CFisUC, Departamento de F\'{i}sica, Universidade de Coimbra, 3004-516 Coimbra, Portugal \and
\label{inst:24} Institute of Planetary Research, German Aerospace Center (DLR), Rutherfordstrasse 2, 12489 Berlin, Germany \and
\label{inst:25} Centre for Mathematical Sciences, Lund University, Box 118, 221 00 Lund, Sweden \and
\label{inst:26} Aix Marseille Univ, CNRS, CNES, LAM, 38 rue Frédéric Joliot-Curie, 13388 Marseille, France \and
\label{inst:27} Astrobiology Research Unit, Université de Liège, Allée du 6 Août 19C, B-4000 Liège, Belgium \and
\label{inst:28} Space sciences, Technologies and Astrophysics Research (STAR) Institute, Université de Liège, Allée du 6 Août 19C, 4000 Liège, Belgium \and
\label{inst:29} Institute of Astronomy, KU Leuven, Celestijnenlaan 200D, 3001 Leuven, Belgium \and
\label{inst:30} ELTE Gothard Astrophysical Observatory, 9700 Szombathely, Szent Imre h. u. 112, Hungary \and
\label{inst:31} SRON Netherlands Institute for Space Research, Niels Bohrweg 4, 2333 CA Leiden, Netherlands \and
\label{inst:32} Centre Vie dans l’Univers, Faculté des sciences, Université de Genève, Quai Ernest-Ansermet 30, 1211 Genève 4, Switzerland \and
\label{inst:33} Leiden Observatory, University of Leiden, PO Box 9513, 2300 RA Leiden, The Netherlands \and
\label{inst:34} Department of Space, Earth and Environment, Chalmers University of Technology, Onsala Space Observatory, 439 92 Onsala, Sweden \and
\label{inst:35} Dipartimento di Fisica, Università degli Studi di Torino, via Pietro Giuria 1, I-10125, Torino, Italy \and
\label{inst:36} National and Kapodistrian University of Athens, Department of Physics, University Campus, Zografos GR-157 84, Athens, Greece \and
\label{inst:37} Department of Astrophysics, University of Vienna, Türkenschanzstrasse 17, 1180 Vienna, Austria \and
\label{inst:38} Institute for Theoretical Physics and Computational Physics, Graz University of Technology, Petersgasse 16, 8010 Graz, Austria \and
\label{inst:39} Konkoly Observatory, Research Centre for Astronomy and Earth Sciences, 1121 Budapest, Konkoly Thege Miklós út 15-17, Hungary \and
\label{inst:40} ELTE E\"otv\"os Lor\'and University, Institute of Physics, P\'azm\'any P\'eter s\'et\'any 1/A, 1117 Budapest, Hungary \and
\label{inst:41} Lund Observatory, Division of Astrophysics, Department of Physics, Lund University, Box 118, 22100 Lund, Sweden \and
\label{inst:42} IMCCE, UMR8028 CNRS, Observatoire de Paris, PSL Univ., Sorbonne Univ., 77 av. Denfert-Rochereau, 75014 Paris, France \and
\label{inst:43} Institut d'astrophysique de Paris, UMR7095 CNRS, Université Pierre \& Marie Curie, 98bis blvd. Arago, 75014 Paris, France \and
\label{inst:44} European Space Agency, ESA - European Space Astronomy Centre, Camino Bajo del Castillo s/n, 28692 Villanueva de la Cañada, Madrid, Spain \and
\label{inst:45} ETH Zurich, Department of Physics, Wolfgang-Pauli-Strasse 2, CH-8093 Zurich, Switzerland \and
\label{inst:46} Cavendish Laboratory, JJ Thomson Avenue, Cambridge CB3 0HE, UK \and
\label{inst:47} Institut fuer Geologische Wissenschaften, Freie Universitaet Berlin, Maltheserstrasse 74-100,12249 Berlin, Germany \and
\label{inst:48} Institut de Ciencies de l'Espai (ICE, CSIC), Campus UAB, Can Magrans s/n, 08193 Bellaterra, Spain \and
\label{inst:49} Institut d'Estudis Espacials de Catalunya (IEEC), 08860 Castelldefels (Barcelona), Spain \and
\label{inst:50} HUN-REN-ELTE Exoplanet Research Group, Szent Imre h. u. 112., Szombathely, H-9700, Hungary \and
\label{inst:51} Institute of Astronomy, University of Cambridge, Madingley Road, Cambridge, CB3 0HA, United Kingdom
}


 
  \abstract{Hot Jupiters are giant planets subject to intense stellar radiation. The physical and chemical properties of their atmosphere makes them the most amenable targets for the atmospheric characterization.}
  {In this paper we analyze the photometry collected during the secondary eclipses of the hot Jupiter \wtb by \cheops, \tess and \Spitzer. Our aim is to characterize the atmosphere of the planet by measuring the secondary eclipse depth in several passbands and constrain the planetary dayside spectrum.}
  {We update the radius and the ephemeris of \wtb by analyzing the transit photometry collected by \cheops and \tess. We also analyze the \cheops, \tess and \Spitzer photometry of the occultations of the planet measuring the eclipse depth at different wavelengths.}
  {Our update of the stellar and planetary properties is consistent with previous works. The analysis of the occultations returns an eclipse depth of 92$\pm$21 ppm in the \cheops passband, 83$\pm$27 ppm for \tess and $>$2000 ppm in the IRAC 1-2-4 \Spitzer passbands. Using the eclipse depths in the \Spitzer bands we propose a set of likely emission spectra which constrain the emission contribution in the \cheops and \tess passbands to approximately a few dozens of parts per million. This allowed us to measure a geometric albedo of 0.21$\pm$0.07 in the \cheops passband, while the \tess data lead to a 95\% upper limit of $\sim$0.2.}
  {\wtb belongs to the group of ultra-hot Jupiters which are characterized by low Bond albedo (<0.3$\pm$0.1), as predicted by different atmospheric models. On the other hand, it unexpectedly seems to efficiently recirculate the absorbed stellar energy, unlike similar highly irradiated planets. To explain this inconsistency, we propose that other energy recirculation mechanisms may be at play other than advection (for example, dissociation and recombination of H$_2$). Another possibility is that the observations in different bandpasses probe different atmospheric layers, making the atmospheric analysis difficult without an appropriate modeling of the thermal emission spectrum of \wtb, which is not feasible with the limited spectroscopic data available to date.}
  


\authorrunning{Scandariato et al.}


   \maketitle
%







\nolinenumbers

\section{Introduction}

\acp{HJ}, giant exoplanets similar to our Solar System's Jupiter in terms of mass and size, are characterized by orbital periods as short as ten days or less. They are thus subject to intense stellar irradiation, which increases their equilibrium temperature by thousands of kelvins compared to their cooler prototype. Moreover, \acp{HJ} are expected to be tidally locked \citep{Leger2009}, such that they rotates nearly synchronously with their orbits. By consequence, one planetary hemisphere constantly faces the host star, while the other experiences an eternal night. According to theoretical models, this leads to a unusual atmospheric dynamics, including vertical stratification, strong winds, and super-rotating equatorial jets \citep{Perryman,Fortney2021}.

Two key parameters that are thought to determine \ac{HJ} climate are Bond albedo and recirculation efficiency \citep{seager2010exoplanets,Cowan2011,Heng2017}. The Bond albedo quantifies the fraction of incoming stellar radiation that a planet reflects back into space. It sets the energy balance of the planetary atmosphere, impacts the overall climate, and influences circulation patterns in the atmosphere. By studying the albedo, we gain insights into the planet's atmospheric properties, such as cloud cover, atmospheric composition, and scattering behavior. A high albedo suggests reflective surfaces \citep[e.g., silicate clouds,][]{Sudarsky2000}, while a low albedo implies absorption.

The recirculation efficiency refers instead to how efficiently a planet redistributes the absorbed energy from its dayside to its nightside. A high recirculation efficiency implies effective heat redistribution, leading to more uniform temperatures across the planet. Conversely, a low efficiency results in stark day–night temperature contrasts. By constraining the recirculation efficiency, we can understand atmospheric circulation patterns, including jet streams, winds, and other heat transport mechanisms \citep[see e.g.][]{Cowan2011}.

\wtb is a prototypical \ac{HJ}, first reported by \citet{Pollacco2008}. Its close orbit, its large radius and the high effective temperature of its host star place \wtb in the group of the strongly irradiated \acp{HJ}. It is thus an extremely interesting target for albedo and recirculation measurements that would enhance our understanding of the planetary climate and inform theoretical atmospheric models.

For this reason, \wtb was observed for the guaranteed time observing (GTO) program of the \cheops mission \citep{Benz2021,2024arXiv240601716F}. This program was aimed at the detection of \acp{HJ} secondary eclipses, which can directly constrain the reflectivity of a planetary atmosphere. Since the beginning of the mission, this program has proved \cheops's capacity to retrieve faint eclipse signals of the order of tens to hundreds parts per million \citep[see for example][]{Lendl2020,Brandeker2022}. In some cases, when data coming from other facilities operating at different wavelengths were available, \cheops optical observations have helped in constraining the climate properties of several \acp{HJ} \citep[see for example][]{Scandariato2022,Parviainen2022,Singh2022,Hoyer2023,Pagano2024}.

We here present the atmospheric characterization of \wtb using \cheops observations and other space photometry public data: \tess in the optical and \Spitzer in the infrared. Firstly, we revise the stellar parameters in Sect.~\ref{sec:spectroscopy}. We present the \cheops, \tess and \Spitzer photometric datasets in Sect.~\ref{sec:observations} and their analysis in Sect.\ref{sec:dataanalysis}; there, we update the orbital parameters, measure the planetary radius and the secondary eclipse depths. In Sect.~\ref{sec:results}, we combine the derived system parameters to constrain the albedo and recirculation efficiency of \wtb and discuss the implications on its climate. We conclude with our final remarks in Sect.~\ref{sec:conclusion}.

\section{Stellar radius, mass and age}\label{sec:spectroscopy}

The atmospheric stellar parameters were taken from \citet{wasp3} since there were no additional higher quality spectra available in public spectra archives.

To determine the stellar radius of \wtA, we used broadband photometry from \textit{Gaia} $G$, $G_\mathrm{BP}$, and $G_\mathrm{RP}$, 2MASS $J$, $H$, and $K$, and \textit{WISE} $W1$ and $W2$ \citep{Skrutskie2006,Wright2010,GaiaCollaboration2022} and the stellar spectral parameters listed in Table~\ref{tab:parameters} in a MCMC modified infrared flux method \citep{Blackwell1977,Schanche2020} framework. We produced synthetic photometry from constructed spectral energy distributions (SEDs) built using stellar atmospheric models from the ATLAS catalog \citep{Castelli2003} and fitted them to the observations to derive the stellar bolometric flux. We converted this to the stellar effective temperature and angular diameter, and finally the stellar radius using the offset-corrected \textit{Gaia} parallax \citep{Lindegren2021}.

The effective temperature, the metallicity, and the radius along with their uncertainties constitute the basic input set to derive the isochronal mass $M_{\star}$ and age $t_{\star}$ of the star. To this end, we employed the isochrone placement algorithm \citep{bonfanti2015,bonfanti2016} to interpolate the input parameters within two different sets of grids of isochrones and tracks generated using two different stellar evolutionary codes, namely PARSEC\footnote{\textsl{PA}dova \& T\textsl{R}ieste \textsl{S}tellar \textsl{E}volutionary \textsl{C}ode: \url{http://stev.oapd.inaf.it/cgi-bin/cmd}} v1.2S \citep{marigo2017} and CLES \citep{scuflaire2008}. For each set of grids we computed a mass and an age estimate, and after merging the two respective pairs of outcome following the statistical treatment outlined in \citet{bonfanti2021}, we obtained $M_{\star}=1.236\pm0.040\,M_{\oplus}$ and $t_{\star}=1.5_{-0.8}^{+0.9}$ Gyr.

\begin{table*}
\caption{Stellar and system parameters.}             
\label{tab:parameters}      
\centering          
\begin{tabular}{l c c c r}     
\hline\hline       
Parameter & Symbol & Units & Value & Ref.\\
\hline
V mag & & & 10.63 & \citet{Tycho2000}\\
Spectral Type & & & F7V & \citet{Street2007}\\
Effective temperature & $\rm T_{eff}$ & K & $6440\pm120$ & \citet{wasp3}\\
Surface gravity & $\log g$ & $\log c.g.s.$ & $4.49\pm0.08$ & \citet{wasp3}\\
Metallicity & [Fe/H] & --- & $-0.02\pm0.08$ & \citet{wasp3}\\
Projected rotational velocity & $v\sin{i}$ & km/s & $13.4\pm1.5$ & \citet{wasp3}\\
Stellar radius & $\rm R_\star$ & $\rm R_\sun$ & $1.335\pm0.010$ & this work\\
Stellar mass & $\rm M_\star$ & $\rm M_\sun$ & $1.236\pm0.040$ & this work\\
Stellar age & $t_\star$ & Gyr & $1.5_{-0.8}^{+0.9}$ & this work\\
\hline                  
\end{tabular}
\end{table*}

\section{Observations and data reduction}\label{sec:observations}

\subsection{TESS observations}\label{sec:TESSobs}

Observations of \wtA by the TESS mission \citep{2015JATIS...1a4003R} with a cadence of 120\,s are available for 5 sectors (26, 40, 53, 54, 74) covering 54 transits and 57 secondary eclipses. These observations span the date range 2020-06-09 to 2024-01-30. 

We used the pre-search data conditioning SAP fluxes (PDCSAP\_FLUX) available from the TESS Science Processing Operations Centre (SPOC) and downloaded using the software package  {\sc lightkurve~2.0}  \citep{2018ascl.soft12013L}.\footnote{\url{https://docs.lightkurve.org/}}
Our analysis improves on the results of \citet{Wong2021}, who analyzed only sector 26, the only one available at the time of their writing.
 
\subsection{CHEOPS observations}\label{sec:CHEOPSobs}
We observed 12 complete secondary eclipses of \wtb as part of the guaranteed time observing (GTO) program ID-016  ``Occultations''. \wtA was also included in the GTO  program ID-052 ``Short-period EBLM and hot-Jupiter systems''. 
This is a ``filler'' program that makes use of gaps between other GTO programs. The 4 observations of a few hours each from the filler program do not cover complete transits by themselves, but do cover the full orbital phase range of the transit when taken together. Observations from both programs used an exposure time of 60\,s. 
The log of the observations is shown in Table~\ref{tab:obslog}.

The data used here were processed using version 13 of the \cheops \ac{DRP} \citep{2020A&A...635A..24H}. We used \acp{LC} generated from the \cheops imagettes with the point-spread function photometry package PIPE\footnote{\url{https://github.com/alphapsa/PIPE}}
developed specifically for \cheops \citep[Brandeker et al., in prep.;][]{2021A&A...654A.159S, 2021A&A...651L..12M}. Previous studies have shown that PIPE \acp{LC} give results that are consistent with \acp{LC} computed using synthetic aperture photometry by the \ac{DRP} but typically with improved precision because the PIPE \acp{LC} are less affected by instrumental noise \citep{Brandeker2022, 2023AJ....165..134O, 2023A&A...672A..24K}.
 
 \begin{table*}
    \centering
    \caption{Log of \cheops\ observations.   } 
    \label{tab:obslog}
    \begin{tabular}{@{}lllrrrl}
\hline
&\multicolumn{1}{l}{File key} & 
\multicolumn{1}{l}{Event\tablefootmark{a}} & 
\multicolumn{1}{l}{Start date} & 
\multicolumn{1}{l}{Dur.\tablefootmark{b}} & 
\multicolumn{1}{l}{Eff.\tablefootmark{c}} & 
\multicolumn{1}{l}{Decorrelation} \\ 
&\multicolumn{1}{l}{} & 
\multicolumn{1}{l}{} & 
\multicolumn{1}{c}{ [UTC]} & 
\multicolumn{1}{c}{ [h]} & 
\multicolumn{1}{c}{ [\%]} & 
\multicolumn{1}{l}{ parameters\tablefootmark{d}} \\ 
\hline
\noalign{\smallskip}
  1 & CH\_PR100016\_TG014001\_V0200 & Ecl. & 2021-07-03T07:32:18 & 10.45 & 66 & $x$, $y$, $\sin\phi$, $\sin2\phi$, $\sin3\phi$, $\cos3\phi$ \\
  2 & CH\_PR100016\_TG014002\_V0200 & Ecl. & 2021-07-05T02:27:59 & 10.50 & 65 & $y$, $\cos\phi$ \\
  3 & CH\_PR100016\_TG014003\_V0200 & Ecl. & 2021-07-06T23:13:58 & 11.91 & 65 & {\tt bg}, $t$, $\cos\phi$, $\cos2\phi$ \\
  4 & CH\_PR100016\_TG014004\_V0200 & Ecl. & 2021-07-08T19:34:58 & 10.42 & 65 & $x$, $y$, $\sin\phi$, $\sin2\phi$, $\sin3\phi$ \\
  5 & CH\_PR100016\_TG014005\_V0200 & Ecl. & 2021-07-10T17:11:38 & 10.24 & 65 & $\sin\phi$, $\sin2\phi$, $\sin3\phi$\\
  6 & CH\_PR100016\_TG014006\_V0200 & Ecl. & 2021-07-12T12:09:58 & 10.45 & 65 & {\tt bg}, $\cos2\phi$, $\cos3\phi$ \\
  7 & CH\_PR100016\_TG014007\_V0200 & Ecl. & 2021-07-19T22:35:58 & 10.49 & 67 & $y$, $t$, $\sin\phi$, $\sin3\phi$ \\
  8 & CH\_PR120052\_TG002201\_V0200 & Tr.  & 2022-06-22T02:27:39 &  3.17 & 67 & $\sin\phi$, $\cos\phi$ \\
  9 & CH\_PR100016\_TG014008\_V0200 & Ecl. & 2022-06-22T21:43:39 & 10.49 & 66 & {\tt bg},  $x$, $\sin\phi$, $\cos2\phi$, $\sin3\phi$ \\
 10 & CH\_PR120052\_TG002202\_V0200 & Tr.  & 2022-06-23T23:37:38 &  2.57 & 79 & $\cos\phi$,$\sin2\phi$, $\cos2\phi$ \\
 11 & CH\_PR120052\_TG002203\_V0200 & Tr.  & 2022-06-25T18:41:00 &  3.17 & 59 & $y$, $\cos2\phi$ \\
 12 & CH\_PR100016\_TG014009\_V0200 & Ecl. & 2022-06-26T14:37:39 & 10.62 & 66 & $y$, $\sin\phi$, $\cos3\phi$ \\
 13 & CH\_PR100016\_TG014010\_V0200 & Ecl. & 2022-07-09T12:18:59 & 10.59 & 65 & $t$, $\sin\phi$, $\sin3\phi$ \\
 14 & CH\_PR100016\_TG014011\_V0200 & Ecl. & 2022-07-13T04:06:59 & 12.06 & 67 & $x$, $y$, $t$, $\sin\phi$ \\
 15 & CH\_PR120052\_TG002204\_V0200 & Tr.  & 2022-07-14T05:59:58 &  2.68 & 78 & $y$, $\cos\phi$, $\sin3\phi$ \\
 16 & CH\_PR100016\_TG014012\_V0200 & Ecl. & 2022-07-15T00:13:59 & 10.82 & 70 & $x$, {\tt bg}, $\sin\phi$ \\
\noalign{\smallskip}
\hline
\end{tabular}
\tablefoot{
        \tablefoottext{a}{Either a secondary eclipse (Ecl.) or a transit (Tr.).}
        \tablefoottext{b}{The duration of the observing interval.}
        \tablefoottext{c}{The fraction of the observing interval covered by valid observations of the target.}
        \tablefoottext{d}{This column reports the variables used for the decorrelation against the instrumental noise: spacecraft roll angle, $\phi$, time, $t$, image background level, {\tt bg}, and the centroid of the target PSF on the detector ($x$,$y$).}
}
\end{table*}

\subsection{Spitzer observations}

The \textit{Spitzer Space Telescope}\ observed four eclipses \acp{LC} of
\wtb\ from three visits (Table \ref{tab:spitzer_obs}).  All
observations were carried out using the Infrared Array Camera (IRAC)
instrument \citep{FazioEtal2004apjsIRAC}, sampling three different
bands.  The 2008 observations were obtained during the cryogenic
mission, in full-array mode, with a cadence of 12.0~s per frame.  The
2009 and 2016 observations were obtained during the warm mission, in
sub-array mode, with a cadence of 2.0~s per frame.

\begin{table}
\centering
\caption{{\Spitzer} secondary eclipse observations of \wtb.} 
\label{tab:spitzer_obs}
\begin{tabular}{ccl}
\hline
\multicolumn{1}{l}{Band ($\lambda$)}  & 
\multicolumn{1}{l}{Date} & 
PI (program) \\
\hline
\noalign{\smallskip}
IRAC 2 (4.5 {\microns})  &  2008-09-18  &   P. Wheatley (50759) \\
IRAC 4 (8.0 {\microns})  &  2008-09-18  &   P. Wheatley (50759) \\
IRAC 1 (3.6 {\microns})  &  2009-10-26  &   H. Knutson (60021) \\
IRAC 2 (4.5 {\microns})  &  2016-11-10  &   D. Deming (13044) \\
\hline
\end{tabular}
\end{table}

\section{Data analysis}\label{sec:dataanalysis}

\subsection{TESS light curve analysis}\label{sec:tesslc}

We performed the analysis of the \tess\ \acp{LC} using PlanetModel model in \pycheops\ \citep{2022MNRAS.514...77M}.\footnote{\url{https://github.com/pmaxted/pycheops}}
This model uses the qpower2 algorithm \citep{2019A&A...622A..33M} to compute the \acp{LC} for the transit assuming the power-2 limb-darkening law, $I_{\lambda}(\mu) = 1 - c\left(1-\mu^{\alpha}\right)$, where $\mu$ is the cosine of the angle between the surface normal and the line of sight.
The flux from the planet is calculated using a Lambertian phase function and the secondary eclipse due to the occultation of this reflected flux is computed assuming a uniform disc for the planet.
There is no evidence for any orbital eccentricity in the orbit of \wtb from either radial velocity measurements nor the phase of the secondary eclipse, so we also assumed that the planet has  a circular orbit, as is typical for short-period hot-Jupiter systems \citep{2017A&A...602A.107B}.
We only used data within one transit duration of the times of mid-transit or mid-eclipse, resulting in 18\,812 observations being used for this analysis.
The flux values for each transit or secondary eclipse were divided by a robust straight-line fit \citep{siegelslopes} to the flux values either side of the primary or secondary eclipse. 
There is no evidence of variability in the \tess \ac{LC} between the transits on the timescales expected for stellar rotation. The amplitude of any such variability is less than 500\,ppm, i.e. \wtA is not a magnetically active star so we do not expect any significant variability in the transits due to the planet occulting star spots.    

\begin{figure*}
\begin{center}
	\includegraphics[width=0.9\textwidth]{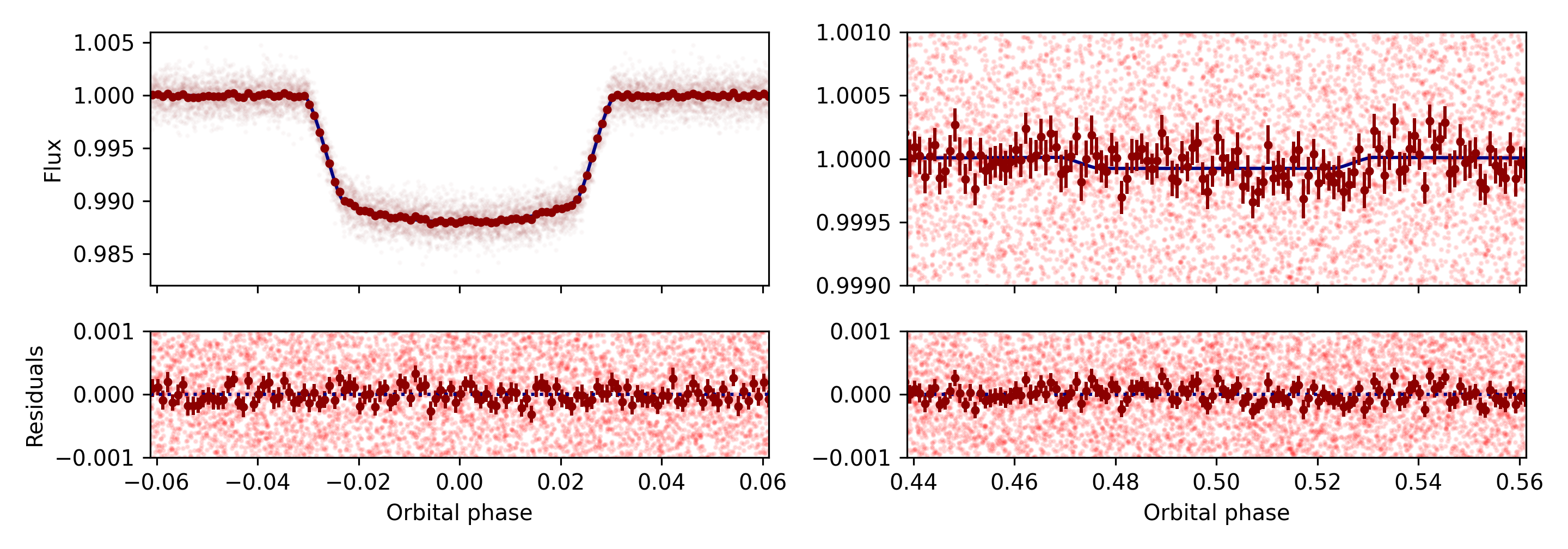}
\end{center}
    \caption{The phase-folded \tess \ac{LC} of \wtA (pale red points) centered on the planetary transit (left column) and secondary eclipses (right column). Our best fit model is shown with a dark blue line. The dark red points show the mean value of the observations and residuals in bins of 0.001 phase units (160\,s).}
    \label{fig:tesslcfit}
\end{figure*}

\begin{figure*}
\begin{center}
	\includegraphics[width=0.75\textwidth]{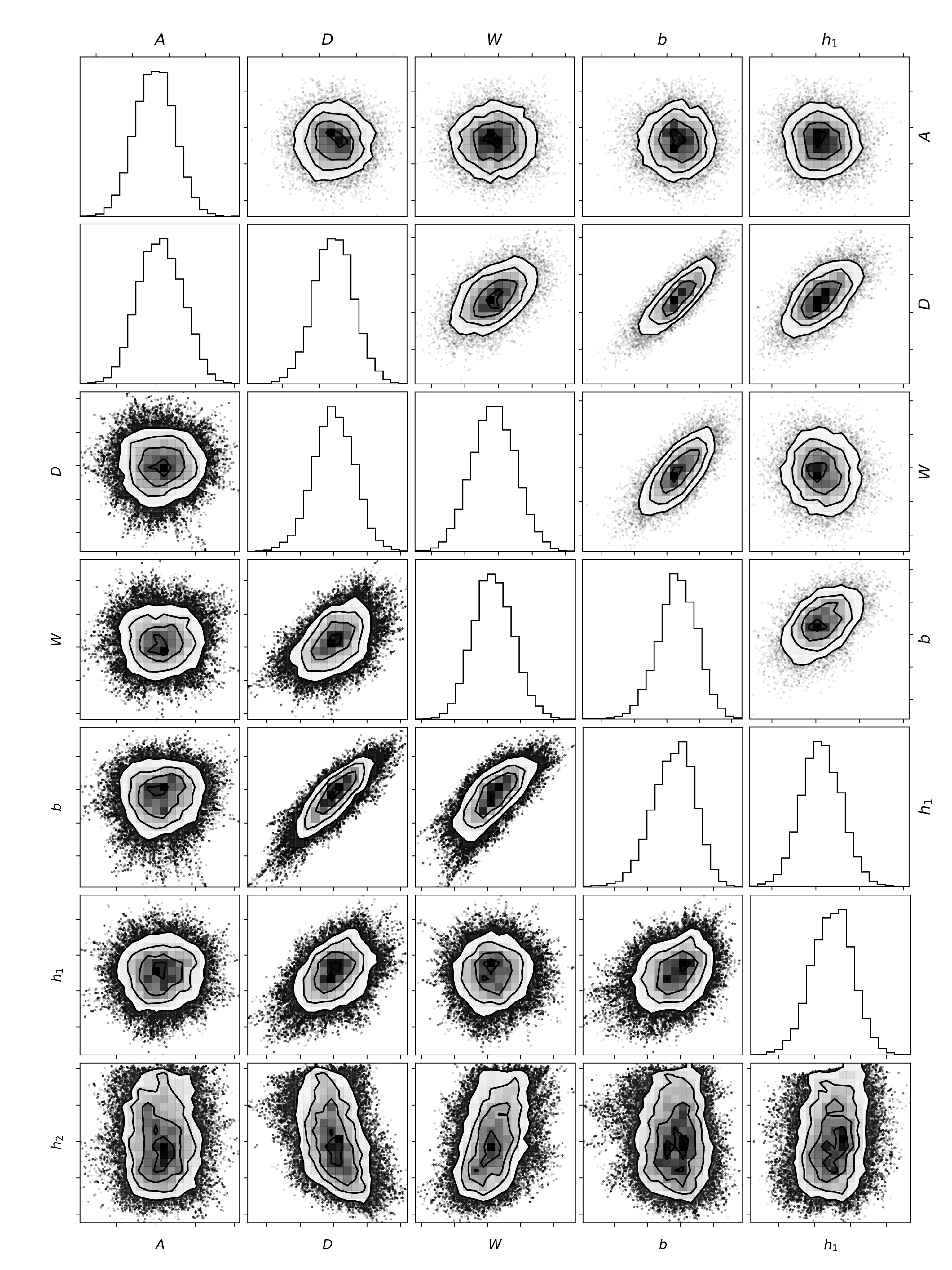}
\end{center}
    \caption{Parameter correlation plots for selected parameters from our analysis of the \cheops (lower-left) and \tess (upper-right) \acp{LC} of \wtA.}
    \label{fig:lcfit_corner}
\end{figure*}

We used the \cheops exposure time calculator\footnote{\url{https://cheops.unige.ch/pht2/exposure-time-calculator/}} to estimate that the granulation noise on a time scale of 3~hours is 15~ppm. This is an order of magnitude smaller than the white noise and a factor of a few less than the eclipse depth, so we assumed a Gaussian white-noise model with equal standard deviations $\sigma_w$ for each observation. 
The free parameters in the fit are: the time of mid-transit, $T_0$;  
the orbital period, $P$; the transit impact parameter, $b= a\cos(i)/R_{\star}$ where $a$ is the orbital semi-major axis and $i$ is the orbital inclination; the transit depth parameter, $D= (R_{\rm pl}/R_{\star})^2 = k^2$; the transit width, $W=(R_{\star}/a)\sqrt{(1+k)^2 - b^2}/\pi$; the limb-darkening parameter $h_1= I_{\lambda}(\frac{1}{2}) = 1 - c(1-2^{-\alpha})$; the planet's geometric albedo assuming zero thermal emission, $A_{g,0}$; the log of the standard error per observation, $\log \sigma_w$. We assume uniform priors on $\cos i$, $\log k$, $\log a/R_{\star}$. The limb-darkening parameter $h_2 = I_{\lambda}(\frac{1}{2}) - I_{\lambda}(0) = c2^{-\alpha}$ has a negligible effect on the quality of the fit and so was fixed at the value $h_2=0.44$, appropriate for a star of this type \citep{2018A+A...616A..39M}. 
All other parameters have broad uniform priors. To sample the \ac{PPD} of the model parameters we used \texttt{emcee} \citep{Foreman2013} with 64 walkers and 256 steps after 512 burn-in steps. Convergence of the sampler was checked by visual inspection of the sample values for each parameter and each walker as a function of step number. The mean and standard error for each parameter calculated from the samples \ac{PPD} are given in Table~\ref{tab:tessfit} and the best fit to the \acp{LC} is shown in Fig.~\ref{fig:tesslcfit}. Correlations between selected parameters from the \ac{PPD} are shown in Fig.~\ref{fig:lcfit_corner}. The parameter of interest is the eclipse depth
\begin{equation}
    \delta_{\rm ecl}= A_{g,0}\left(\frac{R_{\rm pl}}{a}\right)^2,\label{eq:ageo}
\end{equation}
so we also quote this value in Table~\ref{tab:tessfit}. We also tried a model with zero albedo and pure thermal emission from the planet and found that there was a negligible change in the results, i.e. the value of \decl derived is insensitive to the details of the model used to describe the flux from the planet.


 \begin{table}
    \centering
    \caption{Results from our fit to \tess \acp{LC} of \wtA.} 
    \label{tab:tessfit}
    \begin{tabular}{lrl}
\hline
\multicolumn{1}{l}{Parameter\tablefootmark{a} [Unit]} &  \multicolumn{1}{l}{Value} \\
\hline
\noalign{\smallskip}
 $T_0$ [BJD$_{\rm TDB}$ ]       & $ 2459606.789477 \pm 0.000052 $\\
 $P$ [days]         & $   1.8468354 \pm   0.0000002 $ \\
 $D$         & $    0.011043 \pm     0.000054 $ \\
 $W$ [phase]        & $    0.06118 \pm      0.00012 $ \\
 $b$         & $       0.510 \pm       0.015 $ \\
 $A_{g,0}$   & $       0.195 \pm        0.063$ \\
 $h_1$       & $      0.828 \pm         0.007$ \\
 $h_2$\tablefootmark{b}  & $ =  0.44$ \\
 $\ln \sigma_w$ & $     -6.736 \pm    0.005 $ \\
    \noalign{\smallskip}
\hline
\multicolumn{3}{l}{Derived parameters} \\
\hline
 $k=R_{\rm pl}/R_{\star}$   &$ 0.10509 \pm 0.00026 $ \\
 $a/R_{\star}$   &$ 5.099 \pm      0.047 $ \\  
 \decl  [ppm]  & $          83 \pm      27 $\\
 $ \sigma_{\rm w} $ [ppm] &$ 1188 \pm 6 $& \\ 
 $\rm T_{irr}=T_{eff}\sqrt{R_\star/a}$  [K]  &  $2850 \pm 50$  &\\ 
\noalign{\smallskip}
\hline
\end{tabular}
\tablefoot{
        \tablefoottext{a}{$T_0$ is the time of mid-transit, $P$ is the orbital period, $D=(R_{\rm pl}/R_{\star})^2$ is the transit depth,  $W$ is the transit width, $b$ is the transit impact parameter, $A_{g,0}$ is the planet's geometric albedo assuming zero thermal emission,  $h_1$ and $h_2$ are the parameters of the power-2 limb-darkening law,  $\sigma_w$ is the standard error per observation, $a$ is the orbital semi-major axis, and \decl is the eclipse depth.}
        \tablefoottext{b}{Fixed parameter.}}
\end{table}

\subsection{CHEOPS light curve analysis}\label{cheopslc}

We analyzed the \cheops\acp{LC} of \wtA\ produced using PIPE \ac{PSF} photometry using \pycheops\ version 1.1.8. The model and assumptions are the same as for the analysis of the \tess\ photometry (Sect.~\ref{sec:tesslc}).

This version of \pycheops\ is fundamentally the same as the version described in \citet{2022MNRAS.514...77M}, but includes a number of functions that were introduced in \pycheops\ version 1.1.0 that make it easier to analyze \cheops \acp{LC}. 
For example, \pycheops\ can now create Dataset objects containing \acp{LC} and metadata for a single visit directly from PIPE output files, and includes a new method, MultiVisit.fit\_planet(), that can be used to analyze multiple Dataset objects covering transits and secondary eclipses using the PlanetModel model described above. 

We use a linear model of the form $c (1+\sum_k a_k d_k(t))$ to describe the instrumental noise in the \ac{LC} for each Dataset object. 
The complete model for the \ac{LC} is a PlanetModel \ac{LC} multiplied by this linear instrumental noise model. 
The basis vectors $d_k(t)$ are generated from the house-keeping data provided in the Dataset, e.g. $\sin\phi, \cos\phi, \sin2\phi$, etc., where $\phi$ is the spacecraft roll angle. 
To select basis vectors that give a significant improvement to the quality of the fit, we compute the Bayes factor,
\[B_k = e^{-(a_k/\sigma_k)^2/2}\,\sigma_0/\sigma_k.\]
The parameters $a_k$ (``detrending coefficients'') are assumed to have Gaussian priors with mean 0 and standard deviation $\sigma_k$. 
For a model where $a_k$ is included as a free parameter, $a_k$ in this equation is the best-fit (maximum likelihood) value and $\sigma_k$ is its standard error. 
If we set $\sigma_0$ to the standard deviation of the residuals from a model with no instrumental noise model, we find that free parameters with $B_k > 0.5$ give a negligible improvement to the quality of the fit, are poorly constrained, have a best-fit value consistent with 0, and have a negligible impact on the results obtained. 
We therefore only include detrending coefficients as free parameters in the instrumental noise model if their Bayes factor is $B_k<0.5$.  Detrending coefficients should be selected one-by-one, recomputing the Bayes factor for each new model. This can be done automatically in \pycheops\ version 1.1.0 using the new feature Dataset.select\_detrend(). The basis vectors selected for the instrumental noise model for each visit are listed in Table~\ref{tab:obslog}. We excluded the coefficient for the basis vector $f(t) = t^2$ because this tends to be degenerate with the model for the secondary eclipse, and including this term is not justified for the short visits that cover the transits. 

To model the residual noise in the \acp{LC}, we assume a white-noise model where the variance on the flux measurement $f_i$ is $\sigma_i^2+\sigma_w^2$, where $\sigma_i$ is the standard error estimate provided with the data that quantifies known noise sources (photon-counting noise, read-out noise, etc.), and the extra white-noise contribution $\sigma_w$ is assumed to be the same for all observations and for all visits. These errors are assumed to be normally distributed and independent. \cheops is able to detect variability due to granulation or solar-like pulsations in solar-type stars brighter than $V\approx 6.5$ \citep{2023A&A...670A..24S}. \wtA is much fainter than this limit (V=10.6) so correlated noise due to granulation or solar-like pulsations will be undetectable in the \cheops data. There is also very little correlated noise due to the instrument or data reduction apart from the trends captured by the instrument noise model described above \citep{2024arXiv240601716F}.

All the Datasets listed in Table~\ref{tab:obslog} were analyzed together using MultiVisit.fit\_planet(). 
The observed flux values for each Dataset were first corrected for instrumental noise using the best-fit instrumental noise model prior to this joint analysis (``unwrap=True'' option in MultiVisit.fit\_planet()). 
The orbital period was fixed at the value $P = 1.84683542$\,d taken from the analysis of the \tess \ac{LC}. 
MultiVisit.fit\_planet() includes the option to implicitly include trends correlated with space-craft roll angle using a harmonic series with $n_{\rm roll}$ terms. 
We tried models including values of  $n_{\rm roll}$ up to 4, but found that none of these models significantly improved the quality of the fit and the change in the results for the parameters of interest is negligible, so we did not use this option for the results presented here.

Previous studies using \cheops \acp{LC} have used a variant of the \scalpels\ algorithm  \citep{2021MNRAS.505.1699C} to generate basis vectors from the \ac{PSF} autocorrelation function that can be used to generate a linear model for instrumental noise associated with changes in the shape of the PSF \citep{2022MNRAS.511.1043W}. 
We calculated \scalpels\ basis vectors for all our observations of \wtA with \cheops and tested whether they provide a significant improvement to the quality of the fit if we include them in the instrumental noise model. 
This was straightforward to do using the new extra\_decorr\_vectors functionality that was added to the fitting routines in \pycheops\ for version 1.1.0. 
We find that the quality of the fit is not significantly improved for any of the visits if we include these extra basis functions, and the results we obtain are indistinguishable from the results presented here if we do include the few \scalpels\ basis functions that provide a marginal improvement to the quality of the fit. 
We therefore decided, for simplicity, to present results here that do not include \scalpels\ basis vectors in the instrumental noise model for any of the visits. 

\begin{figure*}
\begin{center}
	\includegraphics[width=0.9\textwidth]{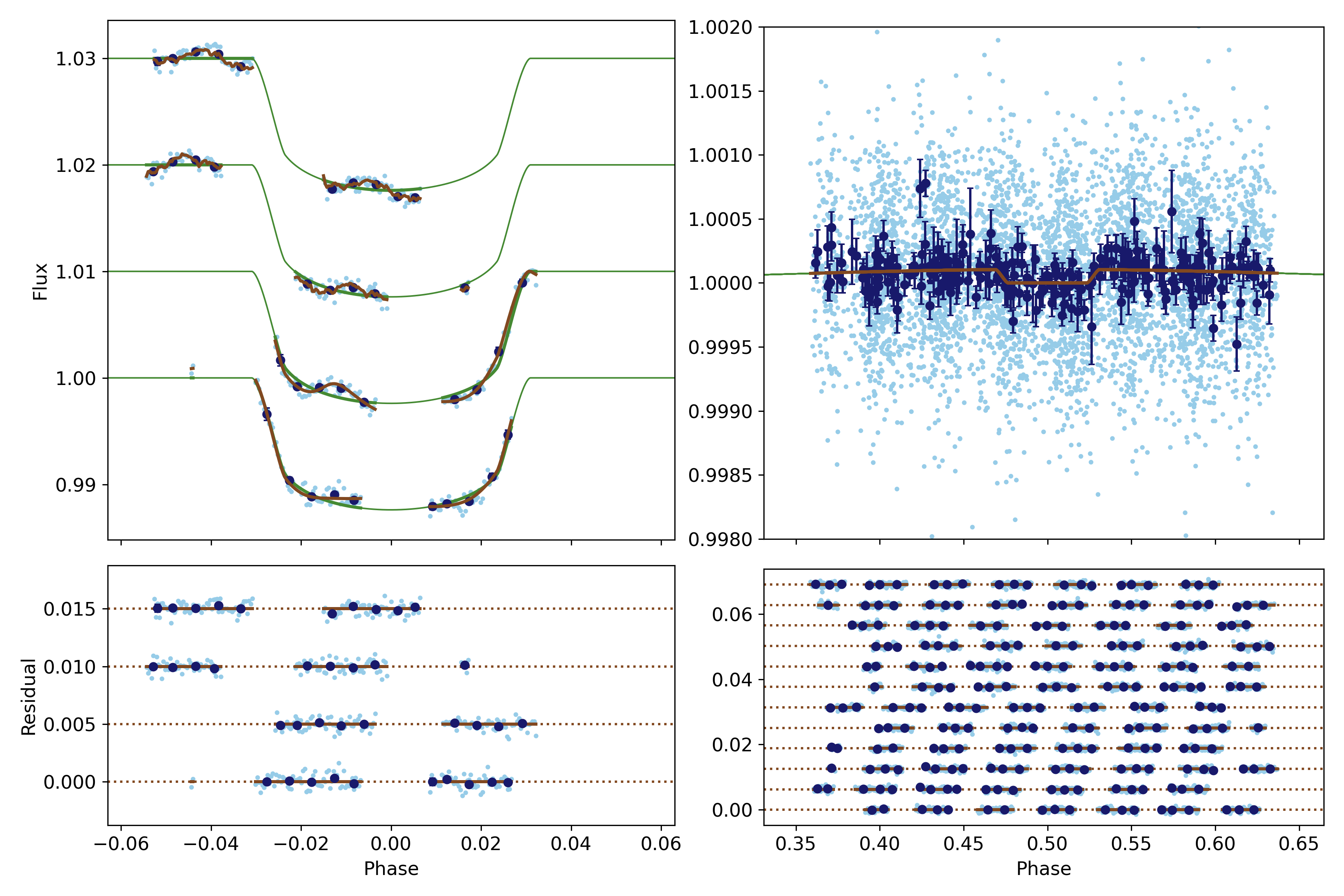}
\end{center}
    \caption{The \cheops \ac{LC} of \wtA and our best-fit model for the transits and eclipses. Upper left panel: Observed transit \acp{LC} are displayed in cyan offset vertically by multiples of 0.01 units. The dark blue points are the data points binned over 0.01 phase units. The full model including instrumental trends is shown in brown and the transit model without trends is shown in green. Lower-left panel: Residuals from the best-fit model to the transit \acp{LC} offset vertically by multiples of 0.005 units. Upper-right panel: secondary eclipse \acp{LC} after removal of instrumental noise from our best-fit model, plotted without any vertical offset. Lower-right panel: Residuals from the best-fit model to the secondary eclipses offset vertically by multiples of 0.005 units.}
    \label{fig:cheopslcfit}
\end{figure*}

We assumed uniform priors on $\cos i$, $\log k$, $\log a/R_{\star}$ across the full allowed range of these parameters. 
All other parameters have broad uniform priors around their best-fit values from a previous fit to the CHEOPS data. We included $h_2$ as a free parameter to the \cheops \ac{LC} because the parameter is constrained by the data and the value obtained is reasonable for a star of this type.
To sample the \ac{PPD} of the model parameters we used \texttt{emcee} with 256 walkers and 1024 steps after 2048 burn-in steps. 
Convergence of the sampler was checked by visual inspection of the sample values for each parameter and each walker as a function of step number. 
The mean and standard error for each parameter calculated from the samples \ac{PPD} are given in Table~\ref{tab:cheopsfit} and the best fit to the \ac{LC} is shown in Fig.~\ref{fig:cheopslcfit}. 
Correlations between selected parameters from the \ac{PPD} are shown in Fig.~\ref{fig:lcfit_corner}.

 \begin{table}
    \centering
    \caption{Results from our fit to \cheops \ac{LC} of \wtA. Detrending coefficients are labeled using the notation $df/dq_i$ for quantity $q$ in Dataset $i$, as listed in Table~\ref{tab:obslog}. }
    \label{tab:cheopsfit}
    \begin{tabular}{lrl}
\hline
\multicolumn{1}{l}{Parameter [Unit]} &  Value \\
\hline
\noalign{\smallskip}
   $D$           &$ 0.01095 \pm 0.00016 $ \\
   $W$ [phase]          &$ 0.06134 \pm 0.00033 $ \\
   $b$           &$   0.487 \pm 0.033 $ \\
   $A_{g,0}$     &$   0.222 \pm 0.053 $ \\
   $T_0$ [BJD$_{\rm TDB}$]        &$ 2459606.78941 \pm 0.00010 $  \\
   $h_1$         &$ 0.748 \pm 0.012$ \\
   $h_2$         &$ 0.53  \pm 0.15$ \\
   $\ln\sigma_w$ &$-8.010 \pm 0.024 $ \\
   $df/dx_1$     &$ 0.00049 \pm 0.00014 $ \\
   $df/dy_1$     &$-0.00081 \pm 0.00012 $ \\
   $df/dt_1$     &$ 0.00056 \pm 0.00021 $ \\
   $df/dy_2$     &$-0.00035 \pm 0.00011 $ \\
   $df/d{\tt bg}_3$    &$-0.00064 \pm 0.00026 $ \\
   $df/dt_3$     &$ 0.00067 \pm 0.00018 $ \\
   $df/dx_4$     &$ 0.00041 \pm 0.00014 $ \\
   $df/dy_4$     &$-0.00089 \pm 0.00012 $ \\
   $df/d{\tt bg}_6$    &$-0.00081 \pm 0.00021 $ \\
   $df/dy_7$     &$-0.00052 \pm 0.00010 $ \\
   $df/dt_7$     &$ 0.00094 \pm 0.00020 $ \\
   $df/d{\tt bg}_9$    &$ 0.00120 \pm 0.00018 $ \\
   $df/dx_9$     &$ 0.00052 \pm 0.00015 $ \\
   $df/dy_{11}$     &$-0.00103 \pm 0.00024 $ \\
   $df/dy_{12}$     &$ 0.00077 \pm 0.00013 $ \\
   $df/dt_{13}$     &$-0.00091 \pm 0.00020 $ \\
   $df/dx_{14}$     &$ 0.00099 \pm 0.00014 $ \\
   $df/dy_{14}$     &$-0.00131 \pm 0.00011 $ \\
   $df/dt_{14}$     &$-0.00042 \pm 0.00017 $ \\
   $df/dy_{15}$     &$-0.00126 \pm 0.00023 $ \\
   $df/d{\tt bg}_{16}$    &$ 0.00052 \pm 0.00020 $ \\
   $df/dx_{16}$     &$ 0.00067 \pm 0.00017 $ \\
\noalign{\smallskip}
\hline
\multicolumn{3}{l}{Derived parameters} \\
\hline
 \decl [ppm]     &$ 92 \pm 21 $ \\
 $k=R_{\rm pl}/R_{\star}$   &$ 0.10464  \pm 0.00075 $ \\
 $ \sigma_{\rm w} $ [ppm] &$ 332 \pm 8 $& \\ 
\noalign{\smallskip}
\hline
\end{tabular}
\end{table}

\subsection{Masses and radii}
 We computed the mass and radius of \wtb using the average values of the parameters $k$, $\sin i$ and $a/R_{\star}$ from Tables~\ref{tab:tessfit} and \ref{tab:cheopsfit} and their standard errors computed using the ``combine'' algorithm described in Appendix~A of \citet{2022MNRAS.514...77M}. The semi-amplitude of the star's spectroscopic orbit used to compute the planet's mass, $K_{\rm rv}$, is the weighted mean of the values derived by \cite{2008MNRAS.385.1576P} and \cite{2013AJ....146..147M}. 
 Note that we did not use the value of $R_{\star}$ from Table~\ref{tab:parameters} in this calculation because this quantity follows directly from the mean stellar density that can be computed from $P$ and $a/R_{\star}$. The results are given in Table~\ref{tab:massradius}. The agreement between the radius derived from the mean stellar density ($1.332 \pm 0.018\,R_{\odot}$) and the independent estimate given in Table~\ref{tab:parameters} ($1.335 \pm 0.010\,R_{\odot}$) is better than 1$\sigma$.

\begin{table}
\centering
\caption{Mass and radius estimates for \wtA and \wtb as inferred from the transit observables.}
\label{tab:massradius}
\begin{tabular}{llrr}
\hline\hline
\multicolumn{1}{@{}l}{Parameter}   & \multicolumn{1}{l}{Units}    & \multicolumn{1}{l}{Value} & \multicolumn{1}{l}{Error} \\
\hline
\noalign{\smallskip}
\multicolumn{3}{@{}l}{Input values} \\
\noalign{\smallskip}
P               &             [d]    &$ 1.8468354 $ &                 \\
$M_\star$       & [$M_{\odot}$]      &$     1.236 $ & $\pm\,  0.040 $ \\
$K_{\rm RV}$    & [m\,s$^{-1}$]      &$       259 $ & $\pm\,      7 $ \\
$e$             &                    &            0 (fixed) &   \\
$\sin i$        &                    &$    0.9951 $ & $\pm\, 0.0004 $ \\
$R_{\rm pl}/R_{\star}$ &             &$   0.10503 $ & $\pm\,0.00027 $ \\
$a/R_{\star}$   &                    &$     5.108 $ & $\pm\,  0.043 $ \\
\noalign{\smallskip}
\hline
\multicolumn{3}{@{}l}{Derived quantities} \\
\hline
\noalign{\smallskip}
$R_\star$       & [$R_{\odot}$]      &$      1.332 $ & $\pm\, 0.018 $ \\
$M_{\rm pl}$    & [$M_{\rm Jup}$]    &$      1.811 $ & $\pm\, 0.063 $ \\
$R_{\rm pl}$    & [$R_{\rm Jup}$]    &$      1.392 $ & $\pm\, 0.019 $ \\
$g_{\rm pl}$    &   [m\,s$^{-2}$]    &$       24.2 $ & $\pm\,   0.8 $ \\
$\rho_{\rm pl}$ &    [g\,cm$^{-3}$]  &$      0.891 $ & $\pm\, 0.035 $ \\
$T_4-T_1$       &  [d]               &$      0.114 $ & $\pm\, 0.002 $ \\  
$T_3-T_2$       &  [d]               &$      0.085 $ & $\pm\, 0.002 $ \\ 
$\rm T_{irr}=T_{eff}\sqrt{\frac{R_\star}{a}}$ & [K]  &$      2850 $ & $\pm\, 50 $ \\ 
\noalign{\smallskip}
\hline
\end{tabular}
\tablefoot{
R$_{\rm Jup}$=69911\,km is the volume-average radius of Jupiter. 
$T_4-T_1$ and $T_3-T_2$ are the time intervals between the contact points of the primary eclipse (transit). 
}
\end{table}

\subsection{Spitzer light curve analysis}

We analyzed the {\Spitzer} data using the \ac{POET} pipeline
\citep{StevensonEtal2010natGJ436b,
  StevensonEtal2012apjSpitzerHD149026b,
  StevensonEtal2012apjGJ346nonPlanets, CampoEtal2011apjWASP12b,
  NymeyerEtal2011apjWASP18b, CubillosEtal2013apjWASP8b,
  CubillosEtal2014apjTrES1}.  The \ac{POET} pipeline consists of two main
stages, first it extracts raw \acp{LC} from the 2D time-series
images, and then it models the \acp{LC} to constrain the eclipse
properties.

Stage one of the \ac{POET} analysis starts by reading the {\Spitzer}
\ac{BCD}, produced by the {\Spitzer} pipeline
version 18.25.0 for the 2008 observation and version 19.2.0 for the
2009 and 2016 observations. The \ac{BCD} data provides the science
images, uncertainty images, and bad-pixel masks.  Next, \ac{POET} flags bad
pixels from the Spitzer \ac{BCD} files using the bad-pixel masks and
performing a sigma-rejection routine for outlier pixel values.  As the
2008 observation consists of an Instrument Engineering Request data,
there are no associated masks files to each science image.  Therefore,
we performed a more careful sigma-rejection step to flag pixels
showing transient fluctuations away from the median of the neighboring
ten frames.
Then, \ac{POET} estimates the target center position on the detector by
fitting a two-dimensional Gaussian function.  Finally, \ac{POET} obtains
raw \acp{LC} by applying interpolated aperture photometry, testing
several circular apertures with radii ranging from 1.5 to 4.5 pixels,
in 0.25 pixel increments.

In stage two, \ac{POET} models the raw \acp{LC} by simultaneously
fitting the astrophysical signal (i.e., the eclipse) and the
instrumental systematics.  To model the astrophysical signal we used a
\citet{MandelAgol2002apjLightcurves} eclipse model.  The free fitting
parameters of this model are the stellar flux, the eclipse mid-point
epoch, duration, depth, and ingress duration (setting the egress duration equal to the ingress duration). We adopt uniform prior for all parameters.
The IRAC detectors typically exhibit temporal and intra-pixel
sensitivity variations \citep{CharbonneauEtal2005apjTrES1}.  To model
the temporal instrumental systematics we tested several parametric
``ramp'' models as a function of time (linear, quadratic, exponential
ramps).  To model the intra-pixel systematics we employed the
Bi-Linearly Interpolated Sub-pixel Sensitivity (BLISS) map model
\citep{StevensonEtal2012apjSpitzerHD149026b}.

The final \ac{LC} model is then composed of the multiplicative
factors of the astrophysical, ramp, and intra-pixel models.  Using the
\textsc{mc3} statistical package \citep{CubillosEtal2017apjRednoise},
\ac{POET} determines parameter's best-fitting values (via a Trust Region
Reflective optimization) and their Bayesian posterior distributions
\citep[via the differential-evolution Markov-chain Monte Carlo
  sampler][]{terBraak2008SnookerDEMC}.  The Markov chain uses the
Gelman-Rubin statistic \citep{GelmanRubin1992stascGRstatistics} to be
within 1\% of unity to assess parameter convergence.

We analyzed the \acp{LC} in two steps.  First we modeled each
independent observation to determine the optimal aperture photometry
radii and optimal systematic models.  Then we performed a joint-fit
analysis of all four observations to determine the final values.
We selected the optimal systematics model by minimizing the Bayesian
Information Criterion \citep[BIC,][]{Schwarz1978anstaBIC} from a set of
fits to a given dataset.  Here we tested fits with and without a BLISS
model, and we tested fits with each of the different ramp models:
linear, quadratic, exponential, or no ramp.
We selected the optimal aperture photometry by minimizing the standard
deviation of the residual between the raw \ac{LC} and their
respective best-fitting model.  We also paid attention to the
time-correlated noise by computing the Allan deviation \citep[also
  known as time-averaging method,][]{Allan1966ieeepTimeAveraging},
finding no significant deviation for any of the datasets.  Table
\ref{tab:spitzer_results} lists the preferred aperture radii and
systematics models for each individual observation.

\begin{figure*}
\begin{center}
\includegraphics[width=\textwidth]{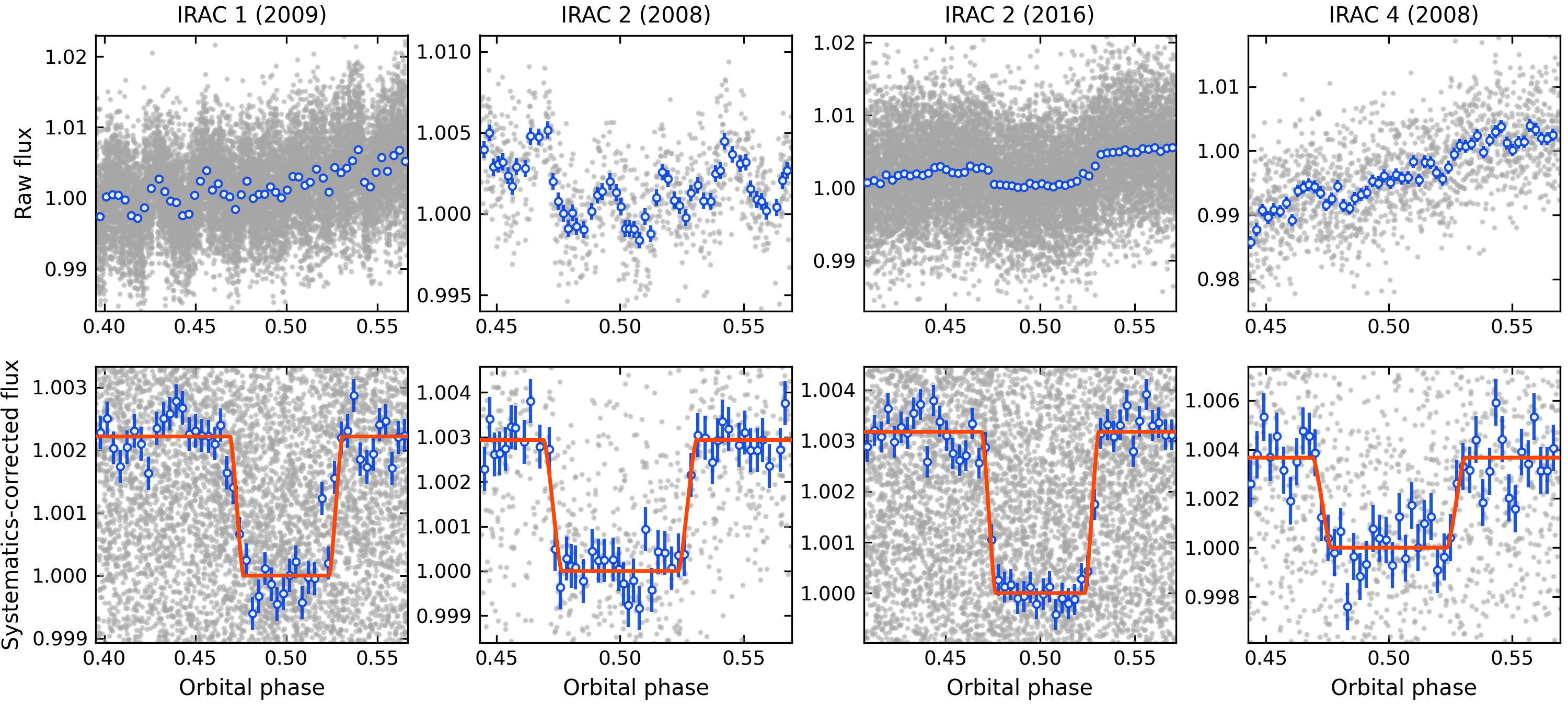}
\end{center}
\caption{Normalized {\Spitzer} \acp{LC} of \wtA after the raw
  aperture-photometry extraction (top panels) and after removing
  instrumental systematics from the joint \ac{LC} fit (bottom
  panels).  The gray markers show the flux of the individual frames,
  whereas the blue markers show the binned fluxes and their
  uncertainties.  The red curves at the bottom panels show the
  best-fit eclipse model for each observation.}
\label{fig:spitzer_lightcurves}
\end{figure*}

To obtain final eclipse depths we ran a joint fit combining all four
events.  In this fit, we shared the mid-eclipse epoch, duration, and
ingress-time parameters among all events. The combined analysis thus
enables a more robust estimation of the eclipse \ac{LC} shape of
the individual observations.
Figure \ref{fig:spitzer_lightcurves} shows the systematics-corrected
\ac{LC} data and joint best-fitting model.  Table
\ref{tab:spitzer_results} presents the retrieved values for the
astrophysical parameters, derived from the posterior distribution
(median and central 68\% percentile).

The {\Spitzer} transit depths analyzed with \ac{POET} are consistent with those of  \citet{Rostron2014} within $1 \sigma$, albeit we consistently obtain deeper occultation depths.
From the occultation depths we can estimate a planetary brightness temperature at each band.  As expected, 
we found a similar trend as \citet{Rostron2014}, that is a deeper occultation depth at 4.5 {\microns} than at 3.6 or 8.0 {\microns} (Table \ref{tab:spitzer_results}).  To first order, the {\Spitzer} observations do not deviate much from a blackbody spectrum for the planet.  In the following section we will perform a more in-depth physical interpretation of the observations.



We also compared our results to the population-level analysis of
Spitzer secondary analyses by
\citet{Deming2023}, who used more
modern pixel-level decorrelation method (PLD) to detrend the telescope
systematics \citep{DemingEtal2015apjHATP20bPLD}(see Fig.~\ref{fig:retrievals} for a graphical representation). They reported depths
of $0.229 \pm 0.011$\% at 3.6~{\micron}, $0.273 \pm 0.016$\% at
4.5~{\micron} (2008), and $0.290 +/- 0.012$\% at 4.5~{\micron} (2016).
While our 3.6~{\microns} occultation depths are consistent within the uncertainty, at
4.5~{\microns} we obtained depths that are $\sim$1.5 and 2$\sigma$
deeper (2008 and 2016 epochs, respectively).  This is intriguing, given
that both the BLISS and PLD methods are among the most robust and
accurate noise-retrieval techniques for Spitzer time-series
observations \citep{IngallsEtal2016ajSpitzerRepeatability}.
We thus conjecture that the discrepancies arise from other
assumptions made during the light-curve analysis.  Since
\citet{Deming2023} analyzed the entire
population of Spitzer occultations for hot Jupiters, they did not
provide the specific details of the \wtb analyses.  Therefore we
are forced to draw our conclusions based on their general analysis
description (Appendix A).  We identified two significant differences
between our analysis approaches.  One is that
\citet{Deming2023} adopted a quadratic
ramp for all planets with equilibrium temperature greater than 2000~K
(this includes \wtb).  From our individual-epoch analyses at
4.5~{\microns} we found occultation depth differences of the order of
0.01\% between our fits with a quadratic ramp and the selected ramp
(Table \ref{tab:spitzer_results}).  Another difference in the
approaches is that \citet{Deming2023}
analyzed the occultations individually, keeping the eclipse shape
parameters fixed at values found in the literature.
Instead, we obtained our final values from a joint fit of all occultations, where
we simultaneously fit for the (shared) eclipse shape parameters.
By sharing parameters in our joint-fit, we estimated the eclipse midpoint, duration, and ingress/egress times with a precision comparable to that of the transit-shape parameters.  Moreover, our eclipse values are consistent within 1$\sigma$ of those expected from an orbit derived from the transit parameters assuming a circular orbit (see tables \ref{tab:massradius} and \ref{tab:spitzer_results}).
Comparing the 4.5~{\microns} occultation depths between our individual
and joint-fit analysis we see differences on the order of
0.002--0.005\%.  While admittedly speculative, given the limitations
on the analysis details, occultation depth variations of this magnitude
can bring the \citet{Deming2023} and
our results closer to a 1$\sigma$ agreement.

{\renewcommand{\arraystretch}{1.3}
\begin{table*}
\centering
\caption{{\Spitzer} \ac{LC} observations of \wtA.} 
\label{tab:spitzer_results}
\begin{tabular}{@{\extracolsep{-0.1cm}} lcccc}
\hline
\hline
Parameter   & IRAC 1 (2009) &  IRAC 2 (2008) & IRAC 2 (2016) & IRAC 4 (2008) \\
\hline
Aperture radius (pixels) & 2.00  & 2.25  &  2.25 & 2.75 \\
BLISS model      & yes         &    yes &  yes &  no \\
Ramp model       & quadratic   &   linear & none &  exponential \\
\decl (\%) &  $0.224^{+0.014}_{-0.013}$ &  $0.295^{+0.015}_{-0.015}$ & $0.320^{+0.014}_{-0.015}$  & $0.369^{+0.049}_{-0.051}$ \\
$T_{\rm bright}$ (K) &  $2361 \pm 66$ &  $2460 \pm 72$ & $2575 \pm 67$  & $2390 \pm 220$ \\
Mid-eclipse epoch$^{\dagger}$  &  $0.49995^{+0.00028}_{-0.00031}$ & ---  & ---  & --- \\
Eclipse duration$^\dagger$ ($T_4-T_1$)  &  $0.0627^{+0.0022}_{-0.0016}$  & ---  & ---  & --- \\
Ingress time$^\dagger$ ($T_2-T_1$) &  $0.0082^{+0.0021}_{-0.0015}$  & ---  & ---  & --- \\
\hline
\end{tabular}
\begin{tablenotes}
  \item $^\dagger$ Free parameter shared between all four {\Spitzer}
    observations. The units are in orbital phase relative to the
    transit epoch and period from Table \ref{tab:massradius}.
\end{tablenotes}
\end{table*}
}

\section{Results}\label{sec:results}

\subsection{Atmospheric modeling}\label{sec:atmpsphericModeling}

To convert the observed eclipse depth into an estimate of the geometric albedo of a planet in a given passband it is important to quantify the flux thermally radiated by the planet at the same wavelengths. To this purpose, we performed various spectral retrievals of the \Spitzer dataset, where the contribution from atmospheric reflection is negligible, with different model assumptions to determine a range of possible models compatible with the infrared observations. In this analysis, we exclude the eclipse depth measurement in the K$_S$ band of \citet{Zhao2012} as it has been shown to be hardly reconciled to atmospheric models \citep{Rostron2014}.

To derive physically based priors for one retrieval run, we performed a forward model for \wtb with the orbital and planetary parameters as listed in Table~\ref{tab:parameters}. For this purpose, we used the 3D general circulation model (GCM) \texttt{expeRT/MITgcm} \citep{Carone2020,Schneider2022} with equilibrium chemistry. We chose the canonical opacity sources for a solar metallicity atmosphere as listed in \citep[][Tab.1]{Schneider2022}, including TiO. The dayside atmosphere composition of this simulation was used to constrain part of the retrieval pipeline, where we averaged the abundances over the whole dayside.

The retrieval models are defined using the TauREx code \citep{Alrefaie2021,Waldmann2015a,Waldmann2015b}, under different assumptions (from the simplest to the most complex one). The number of parameters defines the complexity of the model and, given the low number of input parameters (three wavelength points from the Spitzer dataset). We performed five different retrievals (we list them by increasing the number of parameters), changing the P-T profile and the chemical composition of the atmosphere:

\begin{enumerate}
    \item Isothermal fit: we left only the radius and the equilibrium temperature as free parameters (2 parameters);
    \item 3PT: we fit three-point P-T profile and the radius (5 free parameters). The atmosphere is assumed to be made only by H and He and solar abundances;
    \item 4PT: we fit a four-point P-T profile and the radius (7 free parameters). We assumed the same chemistry of  the 3PT model;
    \item 4PT with TiO: we fit a four-point P-T profile and the radius fixing the TiO abundance as a function of the pressure from the 3D \ac{GCM} model (7 free parameters). We used the TiO opacities from the ExoMol group \citep{buldyreva2022};
    \item 4PT with equilibrium chemistry: we fit a four-point P-T profile, the radius and the C/O ratio assuming equilibrium chemistry with Fastchem (8 free parameters).
\end{enumerate}

All the retrievals have been performed using Nested Sampling with the \texttt{multinest} \citep{feroz2009} through the \texttt{pymultinest} interface \citep{buchner2014} with 1000 live points and uniform prior distributions for all parameters. The atmospheric grid is defined between $10^{-4}$\,Pa and $10^6$\,Pa and, then the retrieved radius corresponds to the radius of the planet at $10^6$Pa. We used the \texttt{phoenix} stellar model \citep{hauschildt1997,hauschildt1999} of \wtA for all the configurations.

\begin{figure*}
    \centering
    \includegraphics[width=0.45\textwidth]{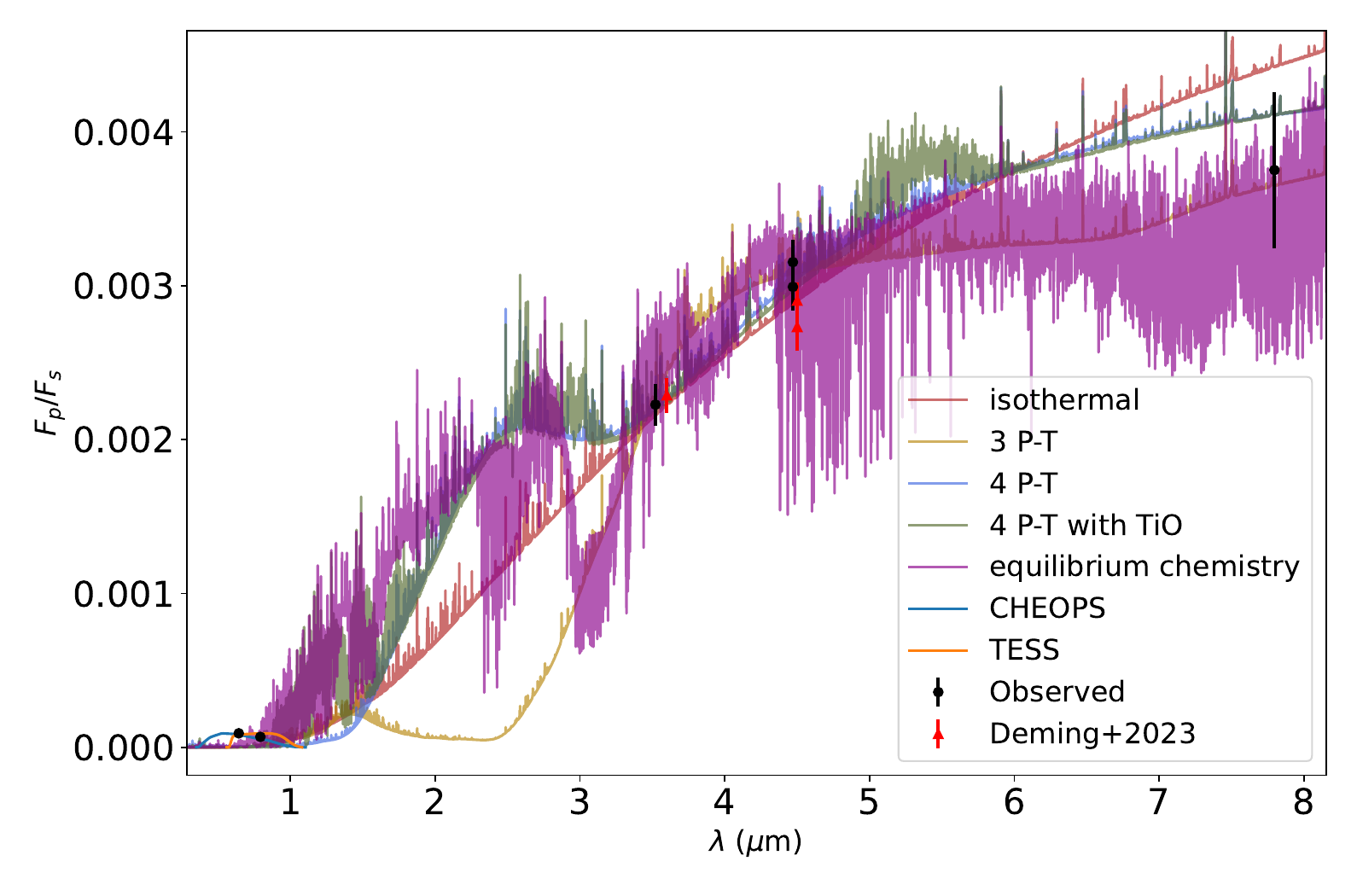}
    \includegraphics[width=0.45\textwidth]{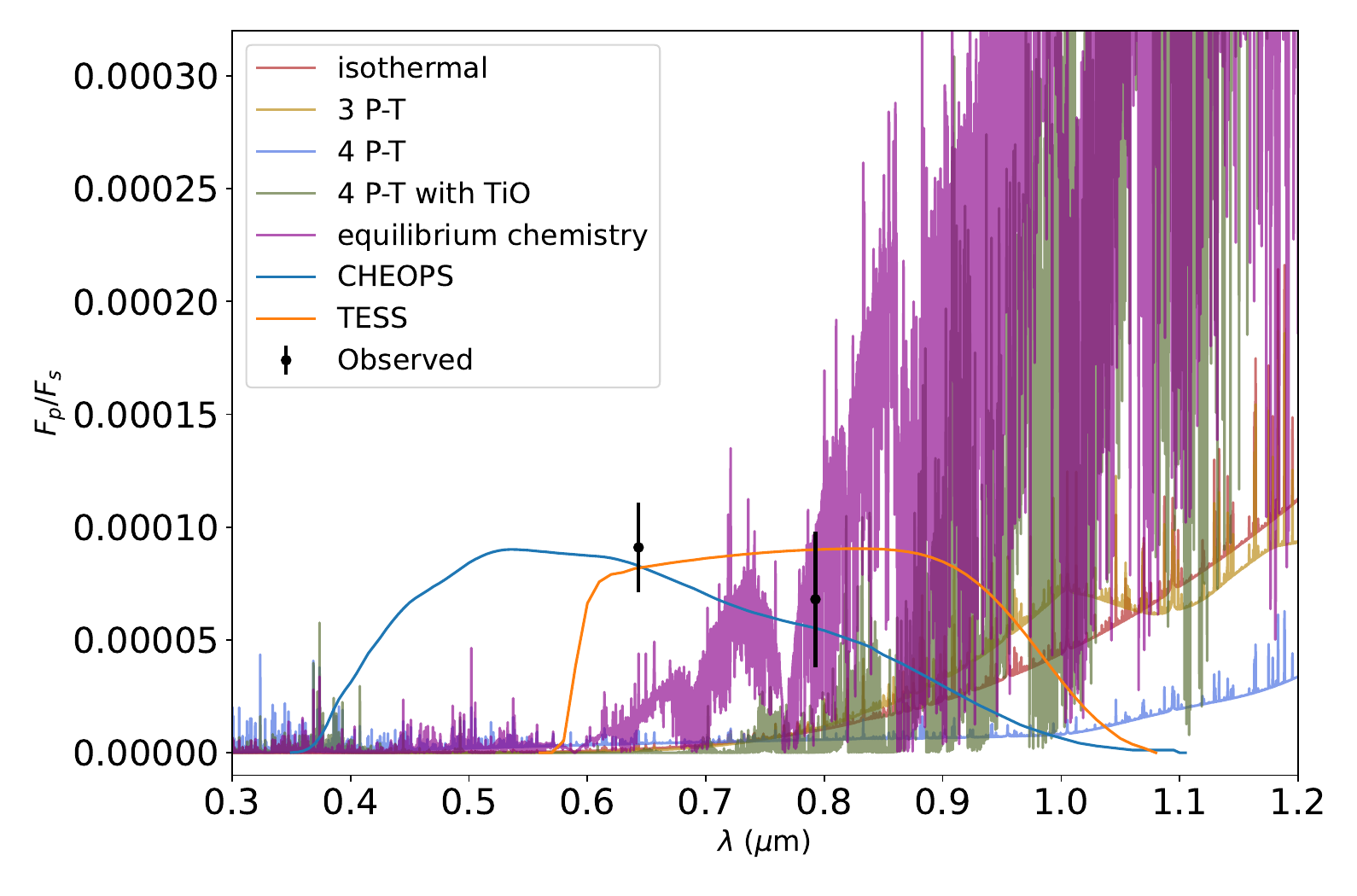}
    \caption{Emission spectra with different model assumptions. On the left are displayed all the best-fit models we performed in this work. The black points represent the observations with \tess, \cheops, and \Spitzer, while red points show the measurements by \citet{Deming2023}. In the right panel, we show a zoom on the optical wavelength, together with the \tess and \cheops observations and bandpasses. \label{fig:retrievals}}
\end{figure*}

The best-fit models for all these scenarios are shown in Fig.~\ref{fig:retrievals}. The atmospheric models with the chemical species within them were computed using the cross sections from the ExoMol database \citep{Tennyson2013,Tennyson2020}\footnote{\url{https://www.exomol.com/data/molecules/}} while the equilibrium chemistry was computed with the FastChem \footnote{\url{https://ascl.net/1804.025}} \citep{Stock2022} through the \texttt{taurex-fastchem} plugin\footnote{\url{https://pypi.org/project/taurex-fastchem}}.

\begin{figure}
    \centering
    \includegraphics[width=0.45\textwidth]{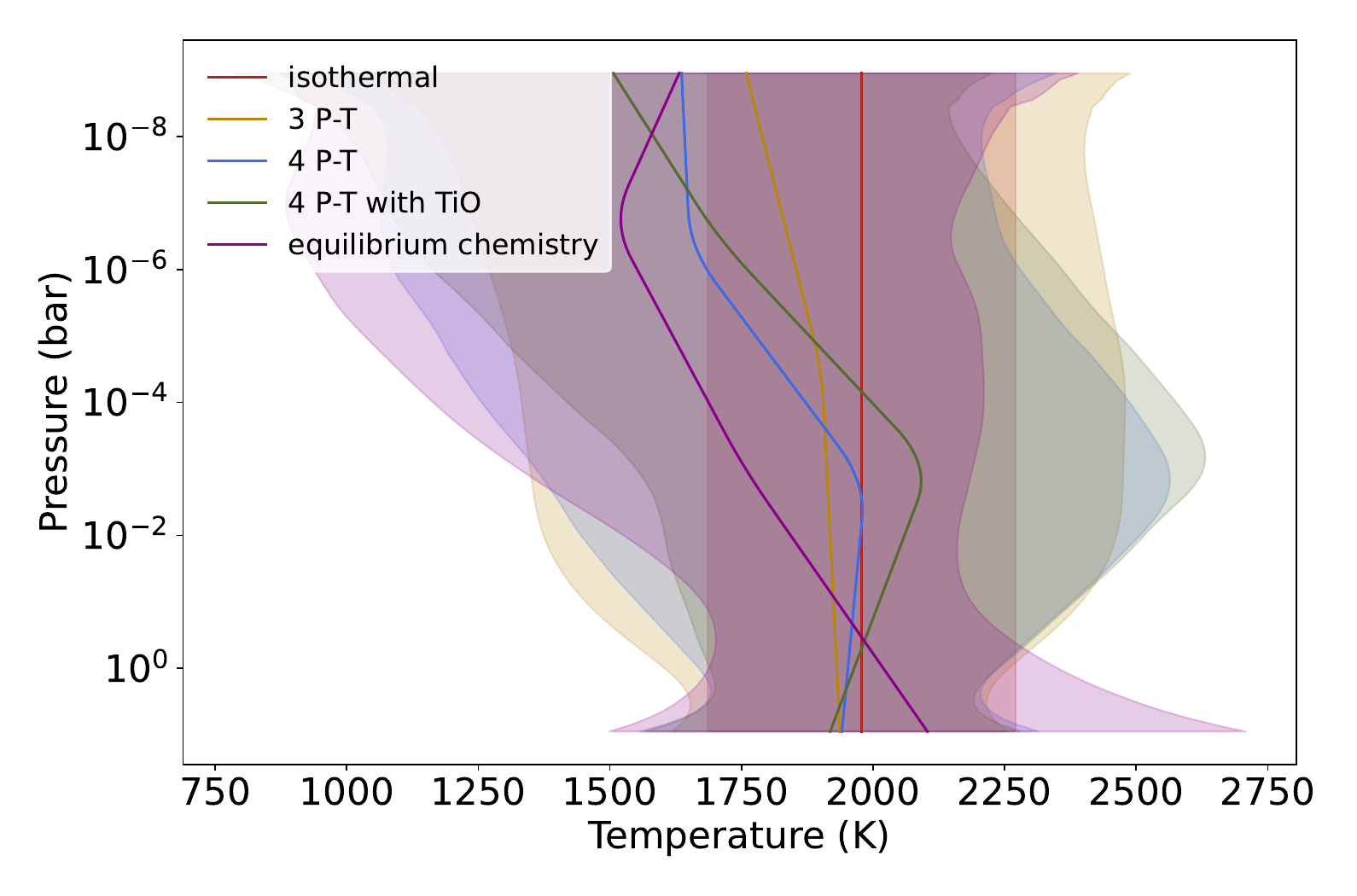}
    \caption{Retrieved temperature-pressure profiles for all the atmospheric configurations. \label{fig:tp_profiles}}
\end{figure}

The temperature-pressure profiles retrieved with all the configurations are consistent with themselves (see Fig \ref{fig:tp_profiles}). The presence of TiO causes a slight thermal inversion and it causes an increase of the optical thickness in the CHEOPS and TESS wavelength ranges as shown in Fig.~\ref{fig:tau_level}, preventing sensitive optical band missions from probing low regions of the atmosphere. The level that can be probed with the different instruments is reported in Tab \ref{tab:level_probed}.

All these models are equivalent from a statistical point of view: their Bayesian Evidence $\log{\mathcal{E}_{min}} = 23.8$ for the case of the isothermal fit with two parameters while $\log{\mathcal{E}_{max}} = 24.4$ for the case of 4PT with Fastchem. Given that the Bayes factor among any of the models is $< 3$, we cannot exclude any possibility in the computation of the thermal spectrum. Considering the Occam's principle, the simplest model gives already a solid understanding of the dayside brightness temperature $T_{\rm d} = 1980_{-240}^{340}$K and a radius $R_p = 1.25\pm0.19 R_{\mathrm{Jup}}$ (consistent with the radius reported in Table~\ref{tab:massradius}). We report all the thermal fluxes integrated into the \cheops and the \tess bandpass for all the scenarios in Tab \ref{tab:optical_fluxes}.

{\renewcommand{\arraystretch}{1.3}
\begin{table}
\centering
\caption{Thermal fluxes for the observed CHEOPS and TESS observations and the respective best-fit models from the TauREx retrievals assuming different forward models. The error bars are computed from their 1$\sigma$ distributions.}
\label{tab:optical_fluxes}
\begin{tabular}{lrr}
\hline\hline
\multicolumn{1}{@{}l}{Model}   & \multicolumn{1}{l}{CHEOPS $F_p/F_s$}    & \multicolumn{1}{l}{TESS $F_p/F_s$}\\
   &  [ppm]    &  [ppm]\\
\hline
\noalign{\smallskip}
isothermal             & $2\pm20$	& $12\pm47$           \\
3 P-T                  & $1\pm20$	& $14\pm53$           \\
4 P-T                  & $4\pm22$	& $6\pm60$            \\
4 P-T with TiO         & $0.2\pm36$	& $10\pm80$           \\
4 P-T with eq. chemistry  & $17\pm14$	& $97\pm38$           \\
\noalign{\smallskip}
\hline
\end{tabular}
\end{table}
}

{\renewcommand{\arraystretch}{1.3}
\begin{table*}
\centering
\caption{Maximum contribution level to the optical depth at the wavelength ranges of CHEOPS, TESS and Spitzer.}
\label{tab:level_probed}
\begin{tabular}{lrrr}
\hline\hline
\multicolumn{1}{@{}l}{Model}   & \multicolumn{1}{l}{CHEOPS level (bar)}    & \multicolumn{1}{l}{TESS level (bar)} & \multicolumn{1}{l}{Spitzer level (bar)}\\
\hline
\noalign{\smallskip}
isothermal             & 8.9	& 8.9  & 1.32       \\
3 P-T                  & 7.12	& 8.2   &   0.6     \\
4 P-T                  & 8.9	& 8.9  &  1.76        \\
4 P-T with TiO         & $1.8\cdot10^{-5}$	& $1.5\cdot10^{-5}$	  &    1.39    \\
4 P-T with eq. chemistry  & 0.15 &    0.27  &   $1.8\cdot10^{-2}$   \\
\noalign{\smallskip}
\hline
\end{tabular}
\end{table*}
}

\subsection{Planetary albedo and heat recirculation}

\begin{figure*}
    \centering
    \includegraphics[width=\linewidth]{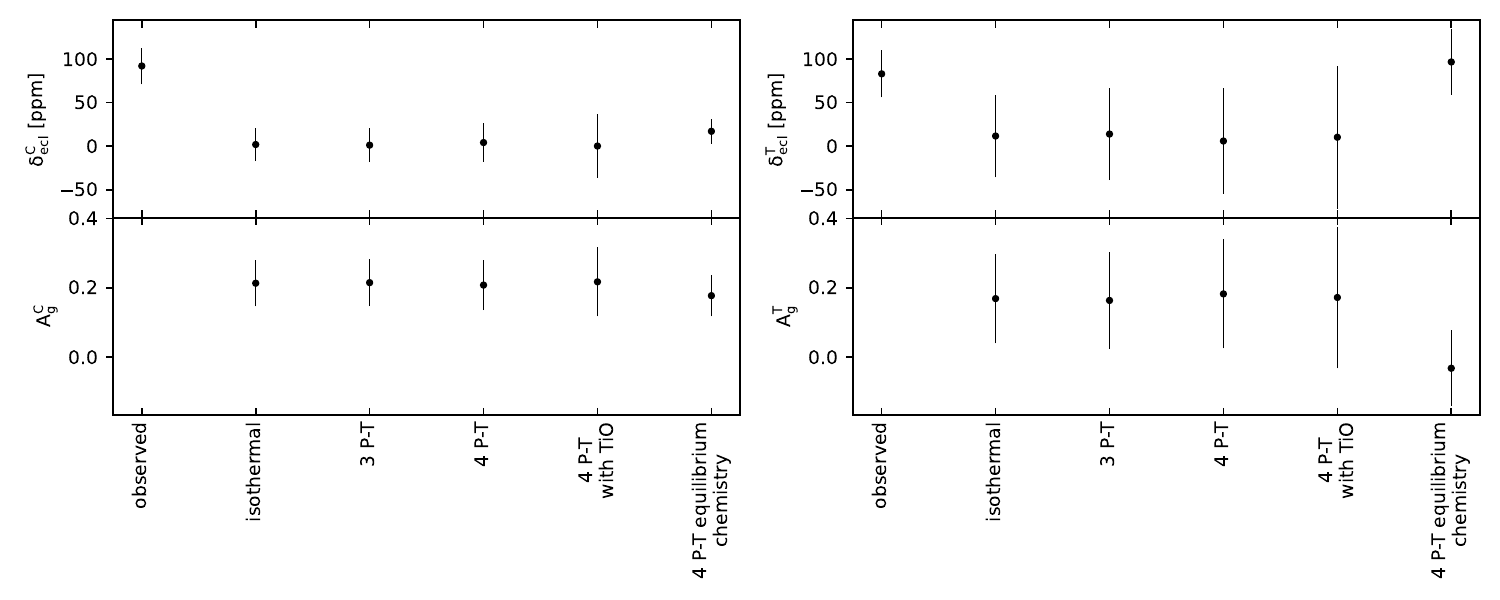}
    \caption{Measured eclipse depth in the \cheops ($\rm\delta^C_{ecl}$, top left panel) and \tess ($\rm\delta^T_{ecl}$, top right panel) passbands together with the theoretical predictions discussed in the text. The bottom panels show the corresponding geometric albedos after correction for thermal emission.}\label{fig:ageo}
\end{figure*}
\begin{figure}
    \centering
    \includegraphics[width=\linewidth]{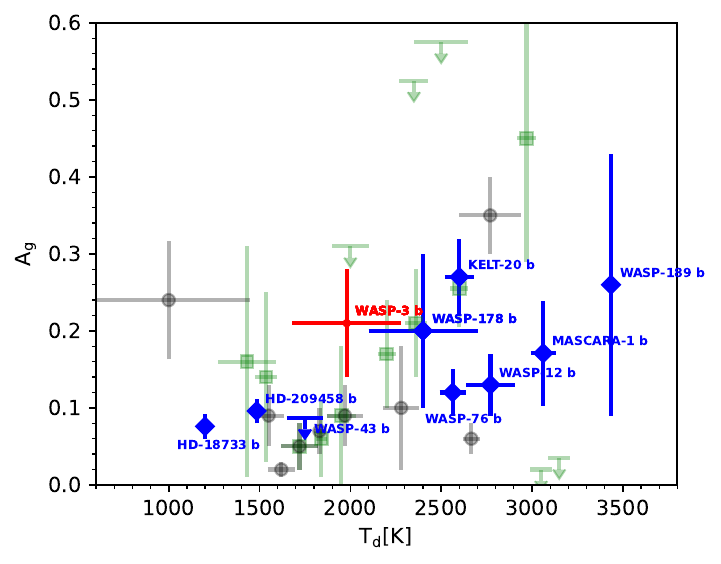}
    \caption{Relationship between \ageo and \tday, adapted from Fig.~10 in \citet{Wong2021} and including our analysis of \wtb (in red). The green squares indicate the systems from the first and second year of the \tess\ primary mission. The black circles indicate the Kepler-/CoRoT-band geometric albedos for the targets that were observed by those missions. Blue symbols are other targets from the \cheops GTO program \citep{Lendl2020, Hooton2021, Brandeker2022,Singh2022,Scandariato2022, 2023A&A...672A..24K,Pagano2024,Akinsami2024,Demangeon2024}.}\label{fig:wongPlot}
\end{figure}

While the irradiation temperature $\rm T_{irr}=T_{eff}\sqrt{R_\star/a}$ is a useful quantity to compare exoplanets because it only depends on the stellar irradiation, during secondary eclipse we measure the dayside effective temperature \tday that also factors in heat redistribution and albedo. The nominal thermal contribution to the \cheops and \tess eclipse depths is of the order of a few ppm regardless of the thermal model, with uncertainties on average of $\sim$20~ppm for \cheops and $\sim$60~ppm for \tess (Fig.~\ref{fig:ageo}, top panels). We subtracted these estimates to the measured eclipse depths and propagated the uncertainties in Eq.~\ref{eq:ageo} to derive an unbiased estimate of the geometric albedo \ageo of \wtb  (Fig.~\ref{fig:ageo}, bottom panels). The geometric albedo \rm $\rm A_{\rm g}^{\rm C}$ measured in the \cheops passband is $\sim$0.21$\pm$0.07, while the uncertainties derived for the \tess passband do not allow a statistically significant measurement of the corresponding geometric albedos $\rm A_{\rm g}^{\rm T}$ but only lead to a 95\% upper limit of about 0.2. Our result thus follows the general trend of \ageo\ increasing with $T_{\rm d}$ indicated by \citet{Wong2021} (Fig.~\ref{fig:wongPlot}).

Since the measurements in the case of \tess do not lead to a statistically significant measurement of \ageo, in the following we will only discuss the case of \cheops, dropping the \lq\lq C\rq\rq\ superscript for simplicity.

By definition, the geometric albedo \ageo\ quantifies the incident light reflected back to the star at a given wavelength. The spherical albedo \asph\ is obtained from \ageo by integration at all angles. The relationship between the two is parameterized  by the phase integral $q$: \asph=$q$\ageo\ \citep[see for example][]{Seager2010}. Exoplanetary atmospheres have 1<$q$<1.5 depending on their scattering law \citep{Pollack1986,Burrows2010} and can be observationally derived from the analysis of the phase curve. Unfortunately no phase curve of \wtb is available to date with enough precision, hence we could not place any constraint on $q$. In the following we thus consider the two limiting scenarios $A_S^{\rm min}=A_g$ and $A_S^{\rm max}=1.5A_g$.

The Bond albedo \abond\ can be obtained as the average of \asph\ weighted over the incident stellar spectrum:
\begin{equation}
    A_{\rm B}=\frac{\int_0^\infty A_{\rm S}(\lambda)I_\star(\lambda)d\lambda}{\int_0^\infty I_\star(\lambda)d\lambda}.\label{eq:abond}
\end{equation}

\begin{figure}
    \centering
    \includegraphics[width=\linewidth]{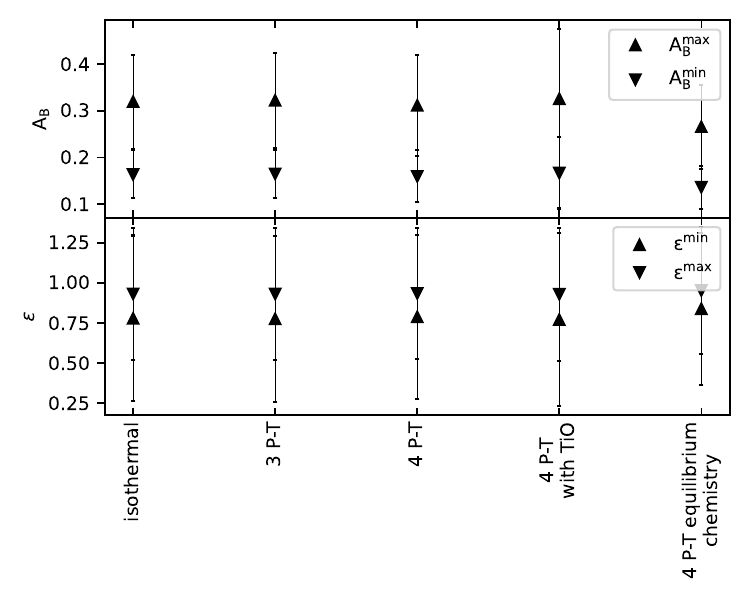}
    \caption{Maximum and minimum Bond albedo \abond of \wtb depending on the assumed thermal model (see text) and corresponding minimum and maximum recirculation efficiency (top and bottom panel respectively).}\label{fig:abond}
\end{figure}

The conversion into \abond\ thus relies on the measurement of \asph\ across the stellar spectrum, but we only have this information integrated in the \cheops passband. Following \citet{Schwartz2015}, we explore the limit case of minimum Bond albedo $A_{\rm B}^{\rm min}$ obtained through Eq.~\ref{eq:abond} assuming \asph=$A_{\rm S}^{\rm min}$ in the spectral range covered by \cheops and \asph=0 otherwise. The opposite limit case assumes \asph$(\lambda)=A_{\rm S}^{\rm max}$ at all wavelengths, that leads to $A_{\rm B}^{\rm max}=A_{\rm S}^{\rm max}$. To compute the integrals in Eq.~\ref{eq:abond}, we used the synthetic spectrum in the BT-Settl library corresponding to the parameters of \wtA\ \citep{Allard2012}. Our results indicate that \abond has a weak dependence on the assumed atmospheric model: regardless of the scenario, \abond increases from $\sim$0.16$\pm$0.06 to $\sim$0.3$\pm$0.1 with increasing $q$  (Fig.~\ref{fig:abond}, top panel).


\citet{Cowan2011} discuss a simple atmospheric energy budget and relate the dayside effective temperature \tday to the Bond albedo \abond (the fraction of incident stellar light not absorbed by the planet) and the recirculation efficiency $\rm\epsilon$ (the homogeneity by which the absorbed stellar energy is distributed across the planetary surface):
\begin{equation}
    T_{\rm d}=T_{\rm eff}\sqrt{\frac{R_\star}{a}}\left(1-A_{\rm B}\right)^{1/4}\left(\frac{2}{3}-\frac{5}{12}\epsilon\right)^{1/4}\label{eq:td}
\end{equation}
Inverting Eq.~\ref{eq:td} yields:
\begin{equation}
    \epsilon=\frac{1}{5}\left[8-\left(\frac{T_{\rm d}}{T_{\rm eff}}\right)^4\left(\frac{a}{R_\star}\right)^2\frac{12}{1-A_{\rm B}}\right].\label{eq:epsilon}
\end{equation}
Equation \ref{eq:epsilon} indicates that $\epsilon$ decreases as \abond increases. This is due to the fact that for a fixed \tday the model allows for a more efficient energy recirculation from day- to nightside (higher $\rm\epsilon$) when more incident stellar light is absorbed by the atmosphere (low \abond). Plugging in the \tday and \abond estimates discussed above and propagating the corresponding uncertainties we derived that the assumptions of minimum and maximum \asph lead to $\rm\epsilon$ predictions compared with the 1$\sigma$ uncertainties (Fig.~\ref{fig:abond}).

Figure~\ref{fig:abond} also shows that the $\epsilon$ estimates according to different thermal models are all consistent with each other, the differences being much smaller than the $1\sigma$ uncertainties. They all agree in indicating an almost complete energy recirculation ($\rm\epsilon\lesssim1$). This result would lead to a perfectly isothermal planetary surface, an unrealistic scenario which we discuss in Sect.~\ref{sec:conclusion}.

\section{Discussion and conclusion}\label{sec:conclusion}

With the aim of characterizing the climate of the \ac{HJ} \wtb, in this paper we have analyzed the \cheops, \tess and \Spitzer photometry of the secondary eclipses of the planet. We found that the planetary atmosphere has a maximum Bond albedo of 0.3$\pm$0.1 and likely has an efficient energy redistribution. \wtb is thus as reflective as similar \acp{UHJ} but does not fit the general expectation of low $\epsilon$ proposed by \citet{Cowan2011}.


From a theoretical perspective, the temperature contrast between the day and nightside of \acp{HJ} is governed by the presence of strong advective latitudinal winds in the planetary atmospheres \citep[][and references therein]{Komacek2016}. The efficiency by which energy is redistributed from day to nightside decreases as $T_{\rm irr}$ (see Table~\ref{tab:massradius}) increases, because the timescale of irradation becomes shorter than the advective timescale \citep[see also][]{Perez2013,Schwartz2017,Parmentier2018,Baeyens2021}. More specifically, the day--night temperature contrast is highest for ultra-hot planets with $T_{\rm 
 irr}\geq 2000$~K \citep{Helling2023}. 
Accordingly, \wtb is in the ultra-hot temperature regime where heat recirculation is expected to be low, in contrast with our finding.

Similarly, \citet{Zhang2018} and \citet{May2021} collected observational evidence that the phase curves offset of \acp{HJ} decreases with increasing $\rm T_{irr}$, hinting to the fact that the presence of zonal winds is less efficient in redistributing the energy in the atmosphere of highly irradiated \acp{HJ} like \wtb.

Furthermore, for \acp{UHJ} similar to \wtb, a clear dayside that is too hot for cloud formation and a cloudy nightside is predicted from 3D climate models \citep[e.g.][]{2019A&A...631A..79H,Helling2023,Roman2021,Parmentier2021}. The backwarming effect due to clouds will warm up the nightside and thus reduce the day-to-nightside temperature gradient, reducing the horizontal wind speeds and thus the efficiency of horizontal heat transfer 
even more 
\citep{Roman2021,Parmentier2021}.
\citet{Roman2021} 
predicts low day side albedos $\leq 0.15$, consistent with our findings on \wtb.



Theoretical predictions from 3D climate models and observed phase curve offsets consistently suggest that highly irradiated \acp{UHJ} like \wtb should be characterized by low values of the recirculation efficiency $\rm\epsilon$. Conversely, our analysis suggests that the atmosphere of \wtb efficiently redistributes energy from dayside to nightside, leaving the possibility that other mechanisms for energy redistribution than advective winds are at play. One possibility that is emerging in the last years is that for planets with $T_{\rm irr}\geq2500~K$ the temperature of the dayside is high enough to dissociate the H$_2$ molecules. This ionized gas then recombines crossing the limbs and reaching the nightside, where lower temperatures allow the molecular recombination and the consequent energy release. This scenario is supported by several studies \citep{Bell2018,Tan2019,Mansfield2020,Helling2021,Helling2023} and is consistent with the irradiation temperature of \wtb (Table~\ref{tab:tessfit}). 

The conundrum of a large recirculation efficiency can be an artifact of either the  of the model depicted in Eqs.~\ref{eq:td}--\ref{eq:epsilon} or our assumed physical parameters. For example, $\epsilon$ has a strong dependence on the brightness temperature: by assuming a higher T$_{\rm d}$=2400~K (statistically consistent within 2$\sigma$ with the estimate discussed in Sect.~\ref{sec:atmpsphericModeling}) $\epsilon$ decreases to a lower value of $\sim$0.4, with an uncertainty that makes it consistent with 0 within 1$\sigma$.

\begin{figure*}
    \centering
    \includegraphics[width=0.45\textwidth]{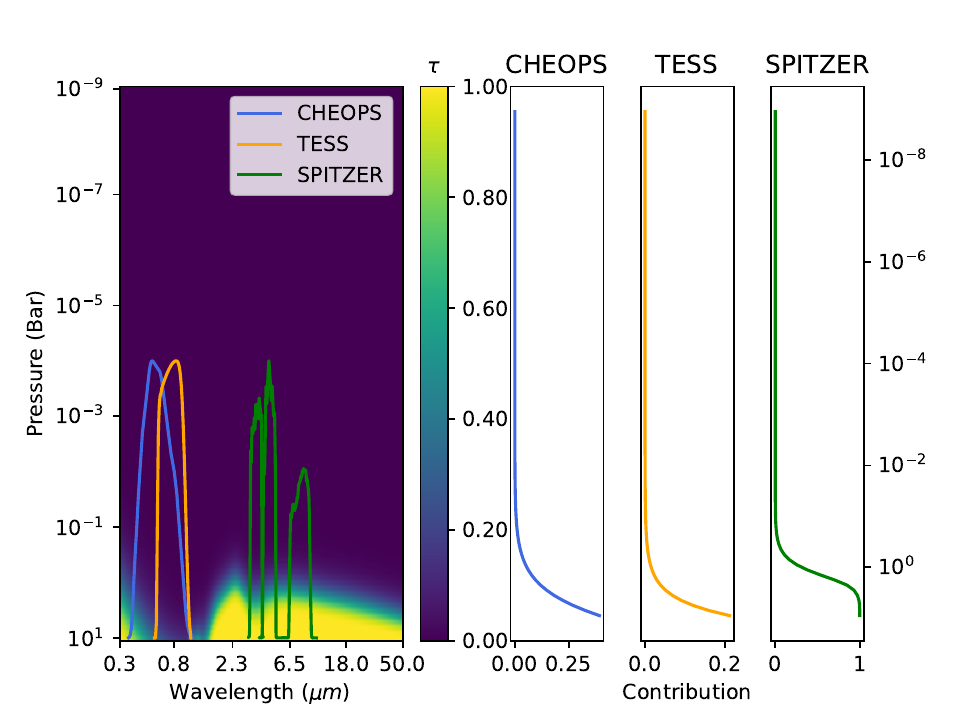}
    \includegraphics[width=0.45\textwidth]{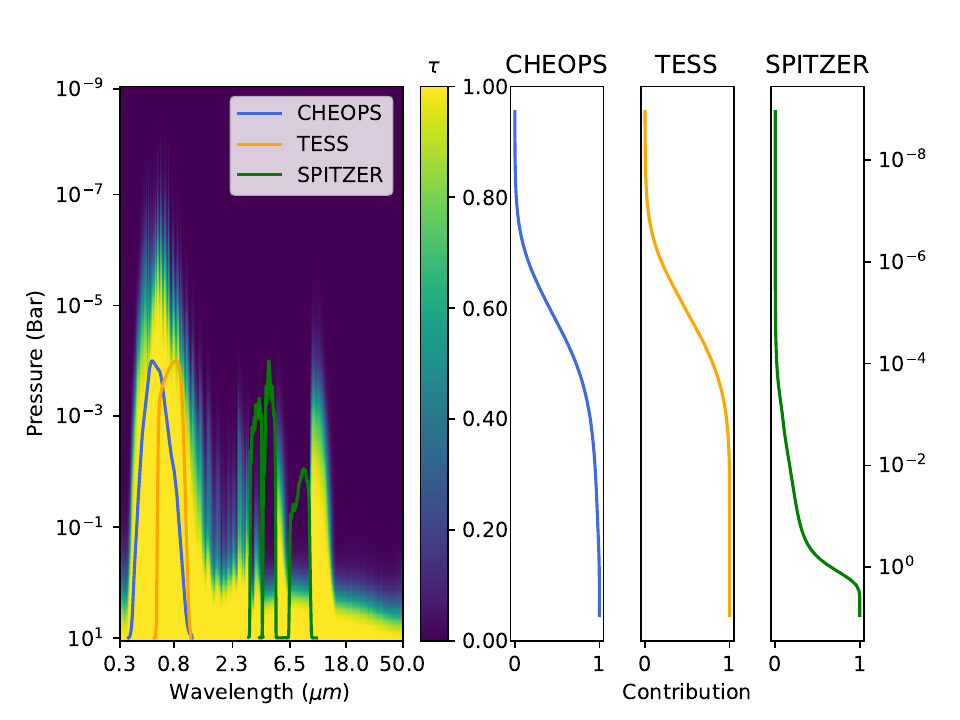}
    \caption{Optical depth ($\tau$) in the wavelength range of \cheops, \tess, and \Spitzer space telescopes at every pressure level for the simulation labeled 4P-T (left) and 4 P-T with TiO (right). In the left panel of each figure is depicted the optical depth at each atmospheric pressure level. In the right panels there are the cumulative contribution functions at each atmospheric level for the \cheops, \tess and \Spitzer space missions. In the presence of optical absorbers, \cheops and \tess can probe at a higher level than \Spitzer compared to a case without optical absorbers.}
    \label{fig:tau_level}
\end{figure*}

One explanation for our underestimation of the dayside brightness temperature is the presence of optical absorbers in the atmosphere of \wtb. For example, in Fig.~\ref{fig:tau_level} we show that if TiO is present in the atmosphere then \cheops and \tess probe higher and hotter atmospheric layers than \Spitzer. This can potentially increase \tday enough to reconcile $\epsilon$ with the expectations. Recently, \citet{Roth2024} revisited the question of TiO cold-trapping from the gas phase in hot Jupiters \citep{Parmentier2013}. These authors predicted that there is a transition at $T_{eq} = 1800$~K from hot Jupiters with low abundances of TiO in the gas phase towards ultra-hot Jupiters with significant abundances of TiO. A transition in dayside brightness at 1800~K was also inferred by a statistical analysis of Spitzer eclipse observation \citep{Deming2023}. The slight thermal inversion inferred from atmosphere retrieval appears to confirm that \wtb lies in 
a transition region where TiO abundances may be affected by cloud formation. However, neither \citet{Roth2024} nor \cite{Deming2023} performed microphysical modeling that studies the  TiO abundances affected by cloud formation. Microphysical cloud models, like  DRIFT (e.g., \citealt{Helling2006}) provide further insight. \citet[][Fig.15]{Helling2023} and also \citet[][Fig. A12]{Helling2019} demonstrate that the TiO abundances are affected substantially by condensation processes into e.g. TiO$_2$ as it can indeed occur on the dayside of planets in this temperature transition region. At the same time, aerosols of highly diverse composition may form in the upper atmosphere that reduce also the oxygen abundance - if the temperature is cold enough to allow thermal stability of these particles. 
If TiO is abundant enough to force a strong upper atmosphere temperature inversion on the day side of a hot Jupiter, then the resulting high temperatures inhibit aerosol formation \citep[][Fig.15]{Helling2023}. Thus, the formation of upper atmosphere temperature inversions, TiO abundances and absence/presence of high altitude aerosols go hand in hand. Answering the question if aerosols form on the dayside of \wtb and contribute to the observed albedo requires, however, a dedicated microphysical nucleation model and better constraints of the TiO abundances for this planet. In any case, \wtb emerges as a highly interesting target for further in-depth characterization to address, for example,  TiO nucleation, cloud dissipation and the nature of high altitude aerosols on the dayside of hot Jupiters.

Moreover, in the assumption of different instruments probing different atmospheric layers, in order to have a comprehensive view it is necessary to model how the physical and chemical conditions vary with atmospheric height. This dependence can potentially break the assumptions of the analytic model proposed by \citet{Cowan2011}. In this regard, the unexpected high $\epsilon$ can be an artifact of adapting a simple atmospheric model to a stratified atmosphere where complex processes are at play. In conclusion, \wtb\ is an interesting test case for more in depth future investigations, aiming at validating and constraining current atmospheric models for \acp{UHJ}.

\begin{acknowledgements}
CHEOPS is an ESA mission in partnership with Switzerland with important contributions to the payload and the ground segment from Austria, Belgium, France, Germany, Hungary, Italy, Portugal, Spain, Sweden, and the United Kingdom. The CHEOPS Consortium would like to gratefully acknowledge the support received by all the agencies, offices, universities, and industries involved. Their flexibility and willingness to explore new approaches were essential to the success of this mission. CHEOPS data analyzed in this article will be made available in the CHEOPS mission archive (\url{https://cheops.unige.ch/archive_browser/}).
GSc and CSt acknowledge the long waited \lq\lq $\rm M\Tilde{a}$\rq\rq\ project and are speechlessly thankful to all the colleagues who have wonderfully run the project so far on their behalf (ACa, MLu and co-workers).
LBo, GBr, VNa, IPa, GPi, RRa, GSc, VSi, and TZi acknowledge support from CHEOPS ASI-INAF agreement n. 2019-29-HH.0. 
LCa  and ChH acknowledge the European Union H2020-MSCA-ITN-2019 under Grant Agreement no. 860470 (CHAMELEON), and the HPC facilities of the Vienna Science Cluster (VSC project 72245).
P.E.C. is funded by the Austrian Science Fund (FWF) Erwin Schroedinger Fellowship, program J4595-N. 
PM acknowledges support from STFC research grant number ST/R000638/1. 
TZi acknowledges NVIDIA Academic Hardware Grant Program for the use of the Titan V GPU card and the Italian MUR Departments of Excellence grant 2023-2027 “Quantum Frontiers”. 
MNG is the ESA CHEOPS Project Scientist and Mission Representative, and as such also responsible for the Guest Observers (GO) Programme. MNG does not relay proprietary information between the GO and Guaranteed Time Observation (GTO) Programmes, and does not decide on the definition and target selection of the GTO Programme. 
ML acknowledges support of the Swiss National Science Foundation under grant number PCEFP2\_194576. 
TWi acknowledges support from the UKSA and the University of Warwick. 
YAl acknowledges support from the Swiss National Science Foundation (SNSF) under grant 200020\_192038. 
We acknowledge financial support from the Agencia Estatal de Investigación of the Ministerio de Ciencia e Innovación MCIN/AEI/10.13039/501100011033 and the ERDF “A way of making Europe” through projects PID2019-107061GB-C61, PID2019-107061GB-C66, PID2021-125627OB-C31, and PID2021-125627OB-C32, from the Centre of Excellence “Severo Ochoa” award to the Instituto de Astrofísica de Canarias (CEX2019-000920-S), from the Centre of Excellence “María de Maeztu” award to the Institut de Ciències de l’Espai (CEX2020-001058-M), and from the Generalitat de Catalunya/CERCA programme. 
We acknowledge financial support from the Agencia Estatal de Investigación of the Ministerio de Ciencia e Innovación MCIN/AEI/10.13039/501100011033 and the ERDF “A way of making Europe” through projects PID2019-107061GB-C61, PID2019-107061GB-C66, PID2021-125627OB-C31, and PID2021-125627OB-C32, from the Centre of Excellence “Severo Ochoa'' award to the Instituto de Astrofísica de Canarias (CEX2019-000920-S), from the Centre of Excellence “María de Maeztu” award to the Institut de Ciències de l’Espai (CEX2020-001058-M), and from the Generalitat de Catalunya/CERCA programme. 
S.C.C.B. acknowledges support from FCT through FCT contracts nr. IF/01312/2014/CP1215/CT0004. 
ABr was supported by the SNSA. 
C.B. acknowledges support from the Swiss Space Office through the ESA PRODEX program. 
This work has been carried out within the framework of the NCCR PlanetS supported by the Swiss National Science Foundation under grants 51NF40\_182901 and 51NF40\_205606. 
ACC acknowledges support from STFC consolidated grant number ST/V000861/1, and UKSA grant number ST/X002217/1. 
ACMC acknowledges support from the FCT, Portugal, through the CFisUC projects UIDB/04564/2020 and UIDP/04564/2020, with DOI identifiers 10.54499/UIDB/04564/2020 and 10.54499/UIDP/04564/2020, respectively. 
A.C., A.D., B.E., K.G., and J.K. acknowledge their role as ESA-appointed CHEOPS Science Team Members. 
This project was supported by the CNES. 
The Belgian participation to CHEOPS has been supported by the Belgian Federal Science Policy Office (BELSPO) in the framework of the PRODEX Program, and by the University of Liège through an ARC grant for Concerted Research Actions financed by the Wallonia-Brussels Federation. 
L.D. thanks the Belgian Federal Science Policy Office (BELSPO) for the provision of financial support in the framework of the PRODEX Programme of the European Space Agency (ESA) under contract number 4000142531. 
This work was supported by FCT - Funda\c{c}\~{a}o para a Ci\^{e}ncia e a Tecnologia through national funds and by FEDER through COMPETE2020 through the research grants UIDB/04434/2020, UIDP/04434/2020, 2022.06962.PTDC. 
O.D.S.D. is supported in the form of work contract (DL 57/2016/CP1364/CT0004) funded by national funds through FCT. 
B.-O. D. acknowledges support from the Swiss State Secretariat for Education, Research and Innovation (SERI) under contract number MB22.00046. 
This project has received funding from the Swiss National Science Foundation for project 200021\_200726. It has also been carried out within the framework of the National Centre of Competence in Research PlanetS supported by the Swiss National Science Foundation under grant 51NF40\_205606. The authors acknowledge the financial support of the SNSF. 
MF and CMP gratefully acknowledge the support of the Swedish National Space Agency (DNR 65/19, 174/18). 
DG gratefully acknowledges financial support from the CRT foundation under Grant No. 2018.2323 “Gaseousor rocky? Unveiling the nature of small worlds”. 
M.G. is an F.R.S.-FNRS Senior Research Associate. 
CHe acknowledges support from the European Union H2020-MSCA-ITN-2019 under Grant Agreement no. 860470 (CHAMELEON). 
KGI is the ESA CHEOPS Project Scientist and is responsible for the ESA CHEOPS Guest Observers Programme. She does not participate in, or contribute to, the definition of the Guaranteed Time Programme of the CHEOPS mission through which observations described in this paper have been taken, nor to any aspect of target selection for the programme. 
K.W.F.L. was supported by Deutsche Forschungsgemeinschaft grants RA714/14-1 within the DFG Schwerpunkt SPP 1992, Exploring the Diversity of Extrasolar Planets. 
This work was granted access to the HPC resources of MesoPSL financed by the Region Ile de France and the project Equip@Meso (reference ANR-10-EQPX-29-01) of the programme Investissements d'Avenir supervised by the Agence Nationale pour la Recherche. 
This work was also partially supported by a grant from the Simons Foundation (PI Queloz, grant number 327127). 
NCSa acknowledges funding by the European Union (ERC, FIERCE, 101052347). Views and opinions expressed are however those of the author(s) only and do not necessarily reflect those of the European Union or the European Research Council. Neither the European Union nor the granting authority can be held responsible for them. 
A. S. acknowledges support from the Swiss Space Office through the ESA PRODEX program. 
S.G.S. acknowledge support from FCT through FCT contract nr. CEECIND/00826/2018 and POPH/FSE (EC). 
The Portuguese team thanks the Portuguese Space Agency for the provision of financial support in the framework of the PRODEX Programme of the European Space Agency (ESA) under contract number 4000142255. 
GyMSz acknowledges the support of the Hungarian National Research, Development and Innovation Office (NKFIH) grant K-125015, a PRODEX Experiment Agreement No. 4000137122, the Lend\"ulet LP2018-7/2021 grant of the Hungarian Academy of Science and the support of the city of Szombathely. 
V.V.G. is an F.R.S-FNRS Research Associate. 
JV acknowledges support from the Swiss National Science Foundation (SNSF) under grant PZ00P2\_208945. 
NAW acknowledges UKSA grant ST/R004838/1. 
E.V. acknowledges support from the 'DISCOBOLO' project funded by the Spanish Ministerio de Ciencia, Innovación y Universidades under grant PID2021-127289NB-I00.
\end{acknowledgements}

\bibliographystyle{aa} 
\bibliography{references.bib}

\begin{thebibliography}{111}
\expandafter\ifx\csname natexlab\endcsname\relax\def\natexlab#1{#1}\fi

\bibitem[{{Akinsanmi} {et~al.}(2024){Akinsanmi}, {Barros}, {Lendl}, {Carone}, {Cubillos}, {Bekkelien}, {Fortier}, {Flor{\'e}n}, {Collier Cameron}, {Bou{\'e}}, {Bruno}, {Demory}, {Brandeker}, {Sousa}, {Wilson}, {Deline}, {Bonfanti}, {Scandariato}, {Hooton}, {Correia}, {Demangeon}, {Smith}, {Singh}, {Alibert}, {Alonso}, {Asquier}, {B{\'a}rczy}, {Barrado Navascues}, {Baumjohann}, {Beck}, {Beck}, {Benz}, {Billot}, {Bonfils}, {Borsato}, {Broeg}, {Buder}, {Charnoz}, {Csizmadia}, {Davies}, {Deleuil}, {Delrez}, {Ehrenreich}, {Erikson}, {Farinato}, {Fossati}, {Fridlund}, {Gandolfi}, {Gillon}, {G{\"u}del}, {G{\"u}nther}, {Heitzmann}, {Helling}, {Hoyer}, {Isaak}, {Kiss}, {Lam}, {Laskar}, {Lecavelier des Etangs}, {Magrin}, {Maxted}, {Mecina}, {Mordasini}, {Nascimbeni}, {Olofsson}, {Ottensamer}, {Pagano}, {Pall{\'e}}, {Peter}, {Piazza}, {Piotto}, {Pollacco}, {Queloz}, {Ragazzoni}, {Rando}, {Rauer}, {Ribas}, {Santos}, {S{\'e}gransan}, {Simon}, {Stalport}, {Szab{\'o}}, {Thomas}, {Udry}, {Van Grootel}, {Venturini},
  {Villaver}, \& {Walton}}]{Akinsami2024}
{Akinsanmi}, B., {Barros}, S.~C.~C., {Lendl}, M., {et~al.} 2024, \aap, 685, A63

\bibitem[{{Al-Refaie} {et~al.}(2021){Al-Refaie}, {Changeat}, {Waldmann}, \& {Tinetti}}]{Alrefaie2021}
{Al-Refaie}, A.~F., {Changeat}, Q., {Waldmann}, I.~P., \& {Tinetti}, G. 2021, \apj, 917, 37

\bibitem[{{Allan}(1966)}]{Allan1966ieeepTimeAveraging}
{Allan}, D.~W. 1966, IEEE Proceedings, 54, 221

\bibitem[{{Allard} {et~al.}(2012){Allard}, {Homeier}, \& {Freytag}}]{Allard2012}
{Allard}, F., {Homeier}, D., \& {Freytag}, B. 2012, Philosophical Transactions of the Royal Society of London Series A, 370, 2765

\bibitem[{{Baeyens} {et~al.}(2021){Baeyens}, {Decin}, {Carone}, {Venot}, {Ag{\'u}ndez}, \& {Molli{\`e}re}}]{Baeyens2021}
{Baeyens}, R., {Decin}, L., {Carone}, L., {et~al.} 2021, \mnras, 505, 5603

\bibitem[{Bell \& Cowan(2018)}]{Bell2018}
Bell, T.~J. \& Cowan, N.~B. 2018, The Astrophysical Journal, 857, L20

\bibitem[{{Benz} {et~al.}(2021){Benz}, {Broeg}, {Fortier}, {Rando}, {Beck}, {Beck}, {Queloz}, {Ehrenreich}, {Maxted}, {Isaak}, {Billot}, {Alibert}, {Alonso}, {Ant{\'o}nio}, {Asquier}, {Bandy}, {B{\'a}rczy}, {Barrado}, {Barros}, {Baumjohann}, {Bekkelien}, {Bergomi}, {Biondi}, {Bonfils}, {Borsato}, {Brandeker}, {Busch}, {Cabrera}, {Cessa}, {Charnoz}, {Chazelas}, {Collier Cameron}, {Corral Van Damme}, {Cortes}, {Davies}, {Deleuil}, {Deline}, {Delrez}, {Demangeon}, {Demory}, {Erikson}, {Farinato}, {Fossati}, {Fridlund}, {Futyan}, {Gandolfi}, {Garcia Munoz}, {Gillon}, {Guterman}, {Gutierrez}, {Hasiba}, {Heng}, {Hernandez}, {Hoyer}, {Kiss}, {Kovacs}, {Kuntzer}, {Laskar}, {Lecavelier des Etangs}, {Lendl}, {L{\'o}pez}, {Lora}, {Lovis}, {L{\"u}ftinger}, {Magrin}, {Malvasio}, {Marafatto}, {Michaelis}, {de Miguel}, {Modrego}, {Munari}, {Nascimbeni}, {Olofsson}, {Ottacher}, {Ottensamer}, {Pagano}, {Palacios}, {Pall{\'e}}, {Peter}, {Piazza}, {Piotto}, {Pizarro}, {Pollaco}, {Ragazzoni}, {Ratti}, {Rauer}, {Ribas}, {Rieder},
  {Rohlfs}, {Safa}, {Salatti}, {Santos}, {Scandariato}, {S{\'e}gransan}, {Simon}, {Smith}, {Sordet}, {Sousa}, {Steller}, {Szab{\'o}}, {Szoke}, {Thomas}, {Tschentscher}, {Udry}, {Van Grootel}, {Viotto}, {Walter}, {Walton}, {Wildi}, \& {Wolter}}]{Benz2021}
{Benz}, W., {Broeg}, C., {Fortier}, A., {et~al.} 2021, Experimental Astronomy, 51, 109

\bibitem[{{Blackwell} \& {Shallis}(1977)}]{Blackwell1977}
{Blackwell}, D.~E. \& {Shallis}, M.~J. 1977, \mnras, 180, 177

\bibitem[{{Bonfanti} {et~al.}(2021){Bonfanti}, {Delrez}, {Hooton}, {Wilson}, {Fossati}, {Alibert}, {Hoyer}, {Mustill}, {Osborn}, {Adibekyan}, {Gandolfi}, {Salmon}, {Sousa}, {Tuson}, {Van Grootel}, {Cabrera}, {Nascimbeni}, {Maxted}, {Barros}, {Billot}, {Bonfils}, {Borsato}, {Broeg}, {Davies}, {Deleuil}, {Demangeon}, {Fridlund}, {Lacedelli}, {Lendl}, {Persson}, {Santos}, {Scandariato}, {Szab{\'o}}, {Collier Cameron}, {Udry}, {Benz}, {Beck}, {Ehrenreich}, {Fortier}, {Isaak}, {Queloz}, {Alonso}, {Asquier}, {Bandy}, {B{\'a}rczy}, {Barrado}, {Barrag{\'a}n}, {Baumjohann}, {Beck}, {Bekkelien}, {Bergomi}, {Brandeker}, {Busch}, {Cessa}, {Charnoz}, {Chazelas}, {Corral Van Damme}, {Demory}, {Erikson}, {Farinato}, {Futyan}, {Garcia Mu{\~n}oz}, {Gillon}, {Guedel}, {Guterman}, {Hasiba}, {Heng}, {Hernandez}, {Kiss}, {Kuntzer}, {Laskar}, {Lecavelier des Etangs}, {Lovis}, {Magrin}, {Malvasio}, {Marafatto}, {Michaelis}, {Munari}, {Olofsson}, {Ottacher}, {Ottensamer}, {Pagano}, {Pall{\'e}}, {Peter}, {Piazza}, {Piotto},
  {Pollacco}, {Ragazzoni}, {Rando}, {Ratti}, {Rauer}, {Ribas}, {Rieder}, {Rohlfs}, {Safa}, {Salatti}, {S{\'e}gransan}, {Simon}, {Smith}, {Sordet}, {Steller}, {Thomas}, {Tschentscher}, {Van Eylen}, {Viotto}, {Walter}, {Walton}, {Wildi}, \& {Wolter}}]{bonfanti2021}
{Bonfanti}, A., {Delrez}, L., {Hooton}, M.~J., {et~al.} 2021, \aap, 646, A157

\bibitem[{{Bonfanti} {et~al.}(2016){Bonfanti}, {Ortolani}, \& {Nascimbeni}}]{bonfanti2016}
{Bonfanti}, A., {Ortolani}, S., \& {Nascimbeni}, V. 2016, \aap, 585, A5

\bibitem[{{Bonfanti} {et~al.}(2015){Bonfanti}, {Ortolani}, {Piotto}, \& {Nascimbeni}}]{bonfanti2015}
{Bonfanti}, A., {Ortolani}, S., {Piotto}, G., \& {Nascimbeni}, V. 2015, \aap, 575, A18

\bibitem[{{Bonomo} {et~al.}(2017){Bonomo}, {Desidera}, {Benatti}, {Borsa}, {Crespi}, {Damasso}, {Lanza}, {Sozzetti}, {Lodato}, {Marzari}, {Boccato}, {Claudi}, {Cosentino}, {Covino}, {Gratton}, {Maggio}, {Micela}, {Molinari}, {Pagano}, {Piotto}, {Poretti}, {Smareglia}, {Affer}, {Biazzo}, {Bignamini}, {Esposito}, {Giacobbe}, {H{\'e}brard}, {Malavolta}, {Maldonado}, {Mancini}, {Martinez Fiorenzano}, {Masiero}, {Nascimbeni}, {Pedani}, {Rainer}, \& {Scandariato}}]{2017A&A...602A.107B}
{Bonomo}, A.~S., {Desidera}, S., {Benatti}, S., {et~al.} 2017, \aap, 602, A107

\bibitem[{{Brandeker} {et~al.}(2022){Brandeker}, {Heng}, {Lendl}, {Patel}, {Morris}, {Broeg}, {Guterman}, {Beck}, {Maxted}, {Demangeon}, {Delrez}, {Demory}, {Kitzmann}, {Santos}, {Singh}, {Alibert}, {Alonso}, {Anglada}, {B{\'a}rczy}, {Barrado y Navascues}, {Barros}, {Baumjohann}, {Beck}, {Benz}, {Billot}, {Bonfils}, {Bruno}, {Cabrera}, {Charnoz}, {Collier Cameron}, {Corral van Damme}, {Csizmadia}, {Davies}, {Deleuil}, {Deline}, {Ehrenreich}, {Erikson}, {Farinato}, {Fortier}, {Fossati}, {Fridlund}, {Gandolfi}, {Gillon}, {G{\"u}del}, {Hoyer}, {Isaak}, {Kiss}, {Laskar}, {Lecavelier des Etangs}, {Lovis}, {Luntzer}, {Magrin}, {Nascimbeni}, {Olofsson}, {Ottensamer}, {Pagano}, {Pall{\'e}}, {Peter}, {Piotto}, {Pollacco}, {Queloz}, {Ragazzoni}, {Rando}, {Rauer}, {Ribas}, {Scandariato}, {S{\'e}gransan}, {Simon}, {Smith}, {Sousa}, {Steller}, {Szab{\'o}}, {Thomas}, {Udry}, {Van Grootel}, {Walton}, \& {Wolter}}]{Brandeker2022}
{Brandeker}, A., {Heng}, K., {Lendl}, M., {et~al.} 2022, \aap, 659, L4

\bibitem[{{Buchner} {et~al.}(2014){Buchner}, {Georgakakis}, {Nandra}, {Hsu}, {Rangel}, {Brightman}, {Merloni}, {Salvato}, {Donley}, \& {Kocevski}}]{buchner2014}
{Buchner}, J., {Georgakakis}, A., {Nandra}, K., {et~al.} 2014, \aap, 564, A125

\bibitem[{Buldyreva {et~al.}(2022)Buldyreva, Yurchenko, \& Tennyson}]{buldyreva2022}
Buldyreva, J., Yurchenko, S.~N., \& Tennyson, J. 2022, RAS Techniques and Instruments, 1, 43

\bibitem[{{Burrows} \& {Orton}(2010)}]{Burrows2010}
{Burrows}, A. \& {Orton}, G. 2010, in Exoplanets, ed. S.~{Seager}, 419--440

\bibitem[{{Campo} {et~al.}(2011){Campo}, {Harrington}, {Hardy}, {Stevenson}, {Nymeyer}, {Ragozzine}, {Lust}, {Anderson}, {Collier-Cameron}, {Blecic}, {Britt}, {Bowman}, {Wheatley}, {Loredo}, {Deming}, {Hebb}, {Hellier}, {Maxted}, {Pollaco}, \& {West}}]{CampoEtal2011apjWASP12b}
{Campo}, C.~J., {Harrington}, J., {Hardy}, R.~A., {et~al.} 2011, \apj, 727, 125

\bibitem[{{Carone} {et~al.}(2020){Carone}, {Baeyens}, {Molli{\`e}re}, {Barth}, {Vazan}, {Decin}, {Sarkis}, {Venot}, \& {Henning}}]{Carone2020}
{Carone}, L., {Baeyens}, R., {Molli{\`e}re}, P., {et~al.} 2020, \mnras, 496, 3582

\bibitem[{{Castelli} \& {Kurucz}(2003)}]{Castelli2003}
{Castelli}, F. \& {Kurucz}, R.~L. 2003, in IAU Symposium, Vol. 210, Modelling of Stellar Atmospheres, ed. N.~{Piskunov}, W.~W. {Weiss}, \& D.~F. {Gray}, A20

\bibitem[{{Charbonneau} {et~al.}(2005){Charbonneau}, {Allen}, {Megeath}, {Torres}, {Alonso}, {Brown}, {Gilliland}, {Latham}, {Mandushev}, {O'Donovan}, \& {Sozzetti}}]{CharbonneauEtal2005apjTrES1}
{Charbonneau}, D., {Allen}, L.~E., {Megeath}, S.~T., {et~al.} 2005, \apj, 626, 523

\bibitem[{{Collier Cameron} {et~al.}(2021){Collier Cameron}, {Ford}, {Shahaf}, {Aigrain}, {Dumusque}, {Haywood}, {Mortier}, {Phillips}, {Buchhave}, {Cecconi}, {Cegla}, {Cosentino}, {Cr{\'e}tignier}, {Ghedina}, {Gonz{\'a}lez}, {Latham}, {Lodi}, {L{\'o}pez-Morales}, {Micela}, {Molinari}, {Pepe}, {Piotto}, {Poretti}, {Queloz}, {Juan}, {S{\'e}gransan}, {Sozzetti}, {Szentgyorgyi}, {Thompson}, {Udry}, \& {Watson}}]{2021MNRAS.505.1699C}
{Collier Cameron}, A., {Ford}, E.~B., {Shahaf}, S., {et~al.} 2021, \mnras, 505, 1699

\bibitem[{{Cowan} \& {Agol}(2011)}]{Cowan2011}
{Cowan}, N.~B. \& {Agol}, E. 2011, \apj, 729, 54

\bibitem[{{Cubillos} {et~al.}(2017){Cubillos}, {Harrington}, {Loredo}, {Lust}, {Blecic}, \& {Stemm}}]{CubillosEtal2017apjRednoise}
{Cubillos}, P., {Harrington}, J., {Loredo}, T.~J., {et~al.} 2017, \aj, 153, 3

\bibitem[{{Cubillos} {et~al.}(2014){Cubillos}, {Harrington}, {Madhusudhan}, {Foster}, {Lust}, {Hardy}, \& {Bowman}}]{CubillosEtal2014apjTrES1}
{Cubillos}, P., {Harrington}, J., {Madhusudhan}, N., {et~al.} 2014, \apj, 797, 42

\bibitem[{{Cubillos} {et~al.}(2013){Cubillos}, {Harrington}, {Madhusudhan}, {Stevenson}, {Hardy}, {Blecic}, {Anderson}, {Hardin}, \& {Campo}}]{CubillosEtal2013apjWASP8b}
{Cubillos}, P., {Harrington}, J., {Madhusudhan}, N., {et~al.} 2013, \apj, 768, 42

\bibitem[{{Demangeon} {et~al.}(2024){Demangeon}, {Cubillos}, {Singh}, {Wilson}, {Carone}, {Bekkelien}, {Deline}, {Ehrenreich}, {Maxted}, {Demory}, {Zingales}, {Lendl}, {Bonfanti}, {Sousa}, {Brandeker}, {Alibert}, {Alonso}, {Asquier}, {B{\'a}rczy}, {Navascues}, {Barros}, {Baumjohann}, {Beck}, {Beck}, {Benz}, {Billot}, {Biondi}, {Borsato}, {Broeg}, {Buder}, {Cameron}, {Csizmadia}, {Davies}, {Deleuil}, {Delrez}, {Erikson}, {Fortier}, {Fossati}, {Fridlund}, {Gandolfi}, {Gillon}, {G{\"u}del}, {G{\"u}nther}, {Heitzmann}, {Helling}, {Hoyer}, {Isaak}, {Kiss}, {Lam}, {Laskar}, {des Etangs}, {Magrin}, {Mecina}, {Mordasini}, {Nascimbeni}, {Olofsson}, {Ottensamer}, {Pagano}, {Pall{\'e}}, {Peter}, {Piotto}, {Pollacco}, {Queloz}, {Ragazzoni}, {Rando}, {Rauer}, {Ribas}, {Rieder}, {Salmon}, {Santos}, {Scandariato}, {S{\'e}gransan}, {Simon}, {Smith}, {Stalport}, {Szab{\'o}}, {Thomas}, {Udry}, {Van Grootel}, {Venturini}, {Villaver}, \& {Walton}}]{Demangeon2024}
{Demangeon}, O.~D.~S., {Cubillos}, P.~E., {Singh}, V., {et~al.} 2024, \aap, 684, A27

\bibitem[{{Deming} {et~al.}(2015){Deming}, {Knutson}, {Kammer}, {Fulton}, {Ingalls}, {Carey}, {Burrows}, {Fortney}, {Todorov}, {Agol}, {Cowan}, {Desert}, {Fraine}, {Langton}, {Morley}, \& {Showman}}]{DemingEtal2015apjHATP20bPLD}
{Deming}, D., {Knutson}, H., {Kammer}, J., {et~al.} 2015, \apj, 805, 132

\bibitem[{{Deming} {et~al.}(2023){Deming}, {Line}, {Knutson}, {Crossfield}, {Kempton}, {Komacek}, {Wallack}, \& {Fu}}]{Deming2023}
{Deming}, D., {Line}, M.~R., {Knutson}, H.~A., {et~al.} 2023, \aj, 165, 104

\bibitem[{{Fazio} {et~al.}(2004){Fazio}, {Hora}, {Allen}, {Ashby}, {Barmby}, {Deutsch}, {Huang}, {Kleiner}, {Marengo}, {Megeath}, {Melnick}, {Pahre}, {Patten}, {Polizotti}, {Smith}, {Taylor}, {Wang}, {Willner}, {Hoffmann}, {Pipher}, {Forrest}, {McMurty}, {McCreight}, {McKelvey}, {McMurray}, {Koch}, {Moseley}, {Arendt}, {Mentzell}, {Marx}, {Losch}, {Mayman}, {Eichhorn}, {Krebs}, {Jhabvala}, {Gezari}, {Fixsen}, {Flores}, {Shakoorzadeh}, {Jungo}, {Hakun}, {Workman}, {Karpati}, {Kichak}, {Whitley}, {Mann}, {Tollestrup}, {Eisenhardt}, {Stern}, {Gorjian}, {Bhattacharya}, {Carey}, {Nelson}, {Glaccum}, {Lacy}, {Lowrance}, {Laine}, {Reach}, {Stauffer}, {Surace}, {Wilson}, {Wright}, {Hoffman}, {Domingo}, \& {Cohen}}]{FazioEtal2004apjsIRAC}
{Fazio}, G.~G., {Hora}, J.~L., {Allen}, L.~E., {et~al.} 2004, \apjs, 154, 10

\bibitem[{{Feroz} {et~al.}(2009){Feroz}, {Hobson}, \& {Bridges}}]{feroz2009}
{Feroz}, F., {Hobson}, M.~P., \& {Bridges}, M. 2009, \mnras, 398, 1601

\bibitem[{{Foreman-Mackey} {et~al.}(2013){Foreman-Mackey}, {Hogg}, {Lang}, \& {Goodman}}]{Foreman2013}
{Foreman-Mackey}, D., {Hogg}, D.~W., {Lang}, D., \& {Goodman}, J. 2013, \pasp, 125, 306

\bibitem[{{Fortier} {et~al.}(2024){Fortier}, {Simon}, {Broeg}, {Olofsson}, {Deline}, {Wilson}, {Maxted}, {Brandeker}, {Collier Cameron}, {Beck}, {Bekkelien}, {Billot}, {Bonfanti}, {Bruno}, {Cabrera}, {Delrez}, {Demory}, {Futyan}, {Flor{\'e}n}, {G{\"u}nther}, {Heitzmann}, {Hoyer}, {Isaak}, {Sousa}, {Stalport}, {Turin}, {Verhoeve}, {Akinsanmi}, {Alibert}, {Alonso}, {B{\'a}nhidi}, {B{\'a}rczy}, {Barrado}, {Barros}, {Baumjohann}, {Baycroft}, {Beck}, {Benz}, {B{\'\i}r{\'o}}, {B{\'o}di}, {Bonfils}, {Borsato}, {Charnoz}, {Cseh}, {Csizmadia}, {Cs{\'a}nyi}, {Cubillos}, {Davies}, {Davis}, {Deleuil}, {Demangeon}, {Derekas}, {Dransfield}, {Ducrot}, {Ehrenreich}, {Erikson}, {Fari{\~n}a}, {Fossati}, {Fridlund}, {Gandolfi}, {Garai}, {Garcia}, {Gillon}, {G{\'o}mez Maqueo Chew}, {G{\'o}mez-Mu{\~n}oz}, {Granata}, {G{\"u}del}, {Guterman}, {Heged{\"u}s}, {Helling}, {Jehin}, {Kalup}, {Kilkenny}, {Kiss}, {Kriskovics}, {Lam}, {Laskar}, {Lecavelier des Etangs}, {Lendl}, {Lopez Pina}, {Luntzer}, {Magrin}, {Miller}, {Modrego
  Contreras}, {Mordasini}, {Munari}, {Murray}, {Nascimbeni}, {Ottacher}, {Ottensamer}, {Pagano}, {P{\'a}l}, {Pall{\'e}}, {Pasetti}, {Pedersen}, {Peter}, {Petrucci}, {Piotto}, {Pizarro-Rubio}, {Pollacco}, {Pribulla}, {Queloz}, {Ragazzoni}, {Rando}, {Rauer}, {Ribas}, {Sabin}, {Santos}, {Scandariato}, {Schanche}, {Schroffenegger}, {Scutt}, {Sebastian}, {S{\'e}gransan}, {Seli}, {Smith}, {Southworth}, {Standing}, {Szab{\'o}}, {Szak{\'a}ts}, {Thomas}, {Timmermans}, {Triaud}, {Udry}, {Van Grootel}, {Venturini}, {Villaver}, {Vink{\'o}}, {Walton}, {Wells}, \& {Wolter}}]{2024arXiv240601716F}
{Fortier}, A., {Simon}, A.~E., {Broeg}, C., {et~al.} 2024, arXiv e-prints, arXiv:2406.01716

\bibitem[{{Fortney} {et~al.}(2021){Fortney}, {Dawson}, \& {Komacek}}]{Fortney2021}
{Fortney}, J.~J., {Dawson}, R.~I., \& {Komacek}, T.~D. 2021, Journal of Geophysical Research (Planets), 126, e06629

\bibitem[{{Gaia Collaboration} {et~al.}(2022){Gaia Collaboration}, {Vallenari}, {Brown}, {Prusti}, {de Bruijne}, {Arenou}, {Babusiaux}, {Biermann}, {Creevey}, {Ducourant}, {Evans}, {Eyer}, {Guerra}, {Hutton}, {Jordi}, {Klioner}, {Lammers}, {Lindegren}, {Luri}, {Mignard}, {Panem}, {Pourbaix}, {Randich}, {Sartoretti}, {Soubiran}, {Tanga}, {Walton}, {Bailer-Jones}, {Bastian}, {Drimmel}, {Jansen}, {Katz}, {Lattanzi}, {van Leeuwen}, {Bakker}, {Cacciari}, {Casta{\~n}eda}, {De Angeli}, {Fabricius}, {Fouesneau}, {Fr{\'e}mat}, {Galluccio}, {Guerrier}, {Heiter}, {Masana}, {Messineo}, {Mowlavi}, {Nicolas}, {Nienartowicz}, {Pailler}, {Panuzzo}, {Riclet}, {Roux}, {Seabroke}, {Sordo{\o}rcit}, {Th{\'e}venin}, {Gracia-Abril}, {Portell}, {Teyssier}, {Altmann}, {Andrae}, {Audard}, {Bellas-Velidis}, {Benson}, {Berthier}, {Blomme}, {Burgess}, {Busonero}, {Busso}, {C{\'a}novas}, {Carry}, {Cellino}, {Cheek}, {Clementini}, {Damerdji}, {Davidson}, {de Teodoro}, {Nu{\~n}ez Campos}, {Delchambre}, {Dell'Oro}, {Esquej},
  {Fern{\'a}ndez-Hern{\'a}ndez}, {Fraile}, {Garabato}, {Garc{\'\i}a-Lario}, {Gosset}, {Haigron}, {Halbwachs}, {Hambly}, {Harrison}, {Hern{\'a}ndez}, {Hestroffer}, {Hodgkin}, {Holl}, {Jan{\ss}en}, {Jevardat de Fombelle}, {Jordan}, {Krone-Martins}, {Lanzafame}, {L{\"o}ffler}, {Marchal}, {Marrese}, {Moitinho}, {Muinonen}, {Osborne}, {Pancino}, {Pauwels}, {Recio-Blanco}, {Reyl{\'e}}, {Riello}, {Rimoldini}, {Roegiers}, {Rybizki}, {Sarro}, {Siopis}, {Smith}, {Sozzetti}, {Utrilla}, {van Leeuwen}, {Abbas}, {{\'A}brah{\'a}m}, {Abreu Aramburu}, {Aerts}, {Aguado}, {Ajaj}, {Aldea-Montero}, {Altavilla}, {{\'A}lvarez}, {Alves}, {Anders}, {Anderson}, {Anglada Varela}, {Antoja}, {Baines}, {Baker}, {Balaguer-N{\'u}{\~n}ez}, {Balbinot}, {Balog}, {Barache}, {Barbato}, {Barros}, {Barstow}, {Bartolom{\'e}}, {Bassilana}, {Bauchet}, {Becciani}, {Bellazzini}, {Berihuete}, {Bernet}, {Bertone}, {Bianchi}, {Binnenfeld}, {Blanco-Cuaresma}, {Blazere}, {Boch}, {Bombrun}, {Bossini}, {Bouquillon}, {Bragaglia}, {Bramante}, {Breedt},
  {Bressan}, {Brouillet}, {Brugaletta}, {Bucciarelli}, {Burlacu}, {Butkevich}, {Buzzi}, {Caffau}, {Cancelliere}, {Cantat-Gaudin}, {Carballo}, {Carlucci}, {Carnerero}, {Carrasco}, {Casamiquela}, {Castellani}, {Castro-Ginard}, {Chaoul}, {Charlot}, {Chemin}, {Chiaramida}, {Chiavassa}, {Chornay}, {Comoretto}, {Contursi}, {Cooper}, {Cornez}, {Cowell}, {Crifo}, {Cropper}, {Crosta}, {Crowley}, {Dafonte}, {Dapergolas}, {David}, {David}, {de Laverny}, {De Luise}, {De March}, {De Ridder}, {de Souza}, {de Torres}, {del Peloso}, {del Pozo}, {Delbo}, {Delgado}, {Delisle}, {Demouchy}, {Dharmawardena}, {Di Matteo}, {Diakite}, {Diener}, {Distefano}, {Dolding}, {Edvardsson}, {Enke}, {Fabre}, {Fabrizio}, {Faigler}, {Fedorets}, {Fernique}, {Fienga}, {Figueras}, {Fournier}, {Fouron}, {Fragkoudi}, {Gai}, {Garcia-Gutierrez}, {Garcia-Reinaldos}, {Garc{\'\i}a-Torres}, {Garofalo}, {Gavel}, {Gavras}, {Gerlach}, {Geyer}, {Giacobbe}, {Gilmore}, {Girona}, {Giuffrida}, {Gomel}, {Gomez}, {Gonz{\'a}lez-N{\'u}{\~n}ez},
  {Gonz{\'a}lez-Santamar{\'\i}a}, {Gonz{\'a}lez-Vidal}, {Granvik}, {Guillout}, {Guiraud}, {Guti{\'e}rrez-S{\'a}nchez}, {Guy}, {Hatzidimitriou}, {Hauser}, {Haywood}, {Helmer}, {Helmi}, {Sarmiento}, {Hidalgo}, {Hilger}, {H{\l}adczuk}, {Hobbs}, {Holland}, {Huckle}, {Jardine}, {Jasniewicz}, {Jean-Antoine Piccolo}, {Jim{\'e}nez-Arranz}, {Jorissen}, {Juaristi Campillo}, {Julbe}, {Karbevska}, {Kervella}, {Khanna}, {Kontizas}, {Kordopatis}, {Korn}, {K{\'o}sp{\'a}l}, {Kostrzewa-Rutkowska}, {Kruszy{\'n}ska}, {Kun}, {Laizeau}, {Lambert}, {Lanza}, {Lasne}, {Le Campion}, {Lebreton}, {Lebzelter}, {Leccia}, {Leclerc}, {Lecoeur-Taibi}, {Liao}, {Licata}, {Lindstr{\o}m}, {Lister}, {Livanou}, {Lobel}, {Lorca}, {Loup}, {Madrero Pardo}, {Magdaleno Romeo}, {Managau}, {Mann}, {Manteiga}, {Marchant}, {Marconi}, {Marcos}, {Marcos Santos}, {Mar{\'\i}n Pina}, {Marinoni}, {Marocco}, {Marshall}, {Polo}, {Mart{\'\i}n-Fleitas}, {Marton}, {Mary}, {Masip}, {Massari}, {Mastrobuono-Battisti}, {Mazeh}, {McMillan}, {Messina}, {Michalik},
  {Millar}, {Mints}, {Molina}, {Molinaro}, {Moln{\'a}r}, {Monari}, {Mongui{\'o}}, {Montegriffo}, {Montero}, {Mor}, {Mora}, {Morbidelli}, {Morel}, {Morris}, {Muraveva}, {Murphy}, {Musella}, {Nagy}, {Noval}, {Oca{\~n}a}, {Ogden}, {Ordenovic}, {Osinde}, {Pagani}, {Pagano}, {Palaversa}, {Palicio}, {Pallas-Quintela}, {Panahi}, {Payne-Wardenaar}, {Pe{\~n}alosa Esteller}, {Penttil{\"a}}, {Pichon}, {Piersimoni}, {Pineau}, {Plachy}, {Plum}, {Poggio}, {Pr{\v{s}}a}, {Pulone}, {Racero}, {Ragaini}, {Rainer}, {Raiteri}, {Rambaux}, {Ramos}, {Ramos-Lerate}, {Re Fiorentin}, {Regibo}, {Richards}, {Rios Diaz}, {Ripepi}, {Riva}, {Rix}, {Rixon}, {Robichon}, {Robin}, {Robin}, {Roelens}, {Rogues}, {Rohrbasser}, {Romero-G{\'o}mez}, {Rowell}, {Royer}, {Ruz Mieres}, {Rybicki}, {Sadowski}, {S{\'a}ez N{\'u}{\~n}ez}, {Sagrist{\`a} Sell{\'e}s}, {Sahlmann}, {Salguero}, {Samaras}, {Sanchez Gimenez}, {Sanna}, {Santove{\~n}a}, {Sarasso}, {Schultheis}, {Sciacca}, {Segol}, {Segovia}, {S{\'e}gransan}, {Semeux}, {Shahaf}, {Siddiqui}, {Siebert},
  {Siltala}, {Silvelo}, {Slezak}, {Slezak}, {Smart}, {Snaith}, {Solano}, {Solitro}, {Souami}, {Souchay}, {Spagna}, {Spina}, {Spoto}, {Steele}, {Steidelm{\"u}ller}, {Stephenson}, {S{\"u}veges}, {Surdej}, {Szabados}, {Szegedi-Elek}, {Taris}, {Taylo}, {Teixeira}, {Tolomei}, {Tonello}, {Torra}, {Torra}, {Torralba Elipe}, {Trabucchi}, {Tsounis}, {Turon}, {Ulla}, {Unger}, {Vaillant}, {van Dillen}, {van Reeven}, {Vanel}, {Vecchiato}, {Viala}, {Vicente}, {Voutsinas}, {Weiler}, {Wevers}, {Wyrzykowski}, {Yoldas}, {Yvard}, {Zhao}, {Zorec}, {Zucker}, \& {Zwitter}}]{GaiaCollaboration2022}
{Gaia Collaboration}, {Vallenari}, A., {Brown}, A.~G.~A., {et~al.} 2022, arXiv e-prints, arXiv:2208.00211

\bibitem[{{Gelman} \& {Rubin}(1992)}]{GelmanRubin1992stascGRstatistics}
{Gelman}, A. \& {Rubin}, D.~B. 1992, Statistical Science, 7, 457

\bibitem[{{Hauschildt} {et~al.}(1999){Hauschildt}, {Allard}, \& {Baron}}]{hauschildt1999}
{Hauschildt}, P.~H., {Allard}, F., \& {Baron}, E. 1999, \apj, 512, 377

\bibitem[{{Hauschildt} {et~al.}(1997){Hauschildt}, {Baron}, \& {Allard}}]{hauschildt1997}
{Hauschildt}, P.~H., {Baron}, E., \& {Allard}, F. 1997, \apj, 483, 390

\bibitem[{{Helling} {et~al.}(2019{\natexlab{a}}){Helling}, {Gourbin}, {Woitke}, \& {Parmentier}}]{Helling2019}
{Helling}, C., {Gourbin}, P., {Woitke}, P., \& {Parmentier}, V. 2019{\natexlab{a}}, \aap, 626, A133

\bibitem[{{Helling} {et~al.}(2019{\natexlab{b}}){Helling}, {Iro}, {Corrales}, {Samra}, {Ohno}, {Alam}, {Steinrueck}, {Lew}, {Molaverdikhani}, {MacDonald}, {Herbort}, {Woitke}, \& {Parmentier}}]{2019A&A...631A..79H}
{Helling}, C., {Iro}, N., {Corrales}, L., {et~al.} 2019{\natexlab{b}}, \aap, 631, A79

\bibitem[{{Helling} {et~al.}(2023){Helling}, {Samra}, {Lewis}, {Calder}, {Hirst}, {Woitke}, {Baeyens}, {Carone}, {Herbort}, \& {Chubb}}]{Helling2023}
{Helling}, C., {Samra}, D., {Lewis}, D., {et~al.} 2023, \aap, 671, A122

\bibitem[{{Helling} \& {Woitke}(2006)}]{Helling2006}
{Helling}, C. \& {Woitke}, P. 2006, \aap, 455, 325

\bibitem[{{Helling} {et~al.}(2021){Helling}, {Worters}, {Samra}, {Molaverdikhani}, \& {Iro}}]{Helling2021}
{Helling}, C., {Worters}, M., {Samra}, D., {Molaverdikhani}, K., \& {Iro}, N. 2021, \aap, 648, A80

\bibitem[{{Heng}(2017)}]{Heng2017}
{Heng}, K. 2017, {Exoplanetary Atmospheres: Theoretical Concepts and Foundations}

\bibitem[{{H{\o}g} {et~al.}(2000){H{\o}g}, {Fabricius}, {Makarov}, {Urban}, {Corbin}, {Wycoff}, {Bastian}, {Schwekendiek}, \& {Wicenec}}]{Tycho2000}
{H{\o}g}, E., {Fabricius}, C., {Makarov}, V.~V., {et~al.} 2000, \aap, 355, L27

\bibitem[{{Hooton} {et~al.}(2022){Hooton}, {Hoyer}, {Kitzmann}, {Morris}, {Smith}, {Collier Cameron}, {Futyan}, {Maxted}, {Queloz}, {Demory}, {Heng}, {Lendl}, {Cabrera}, {Csizmadia}, {Deline}, {Parviainen}, {Salmon}, {Sulis}, {Wilson}, {Bonfanti}, {Brandeker}, {Demangeon}, {Oshagh}, {Persson}, {Scandariato}, {Alibert}, {Alonso}, {Anglada Escud{\'e}}, {B{\'a}rczy}, {Barrado}, {Barros}, {Baumjohann}, {Beck}, {Beck}, {Benz}, {Billot}, {Bonfils}, {Bourrier}, {Broeg}, {Busch}, {Charnoz}, {Davies}, {Deleuil}, {Delrez}, {Ehrenreich}, {Erikson}, {Farinato}, {Fortier}, {Fossati}, {Fridlund}, {Gandolfi}, {Gillon}, {G{\"u}del}, {Isaak}, {Jones}, {Kiss}, {Laskar}, {Lecavelier des Etangs}, {Lovis}, {Luntzer}, {Magrin}, {Nascimbeni}, {Olofsson}, {Ottensamer}, {Pagano}, {Pall{\'e}}, {Peter}, {Piotto}, {Pollacco}, {Ragazzoni}, {Rando}, {Ratti}, {Rauer}, {Ribas}, {Santos}, {S{\'e}gransan}, {Simon}, {Sousa}, {Steller}, {Szab{\'o}}, {Thomas}, {Udry}, {Ulmer}, {Van Grootel}, \& {Walton}}]{Hooton2021}
{Hooton}, M.~J., {Hoyer}, S., {Kitzmann}, D., {et~al.} 2022, \aap, 658, A75

\bibitem[{{Hoyer} {et~al.}(2020){Hoyer}, {Guterman}, {Demangeon}, {Sousa}, {Deleuil}, {Meunier}, \& {Benz}}]{2020A&A...635A..24H}
{Hoyer}, S., {Guterman}, P., {Demangeon}, O., {et~al.} 2020, \aap, 635, A24

\bibitem[{{Hoyer} {et~al.}(2023){Hoyer}, {Jenkins}, {Parmentier}, {Deleuil}, {Scandariato}, {Wilson}, {D{\'\i}az}, {Crossfield}, {Dragomir}, {Kataria}, {Lendl}, {Ramirez}, {Pe{\~n}a Rojas}, \& {Vin{\'e}s}}]{Hoyer2023}
{Hoyer}, S., {Jenkins}, J.~S., {Parmentier}, V., {et~al.} 2023, \aap, 675, A81

\bibitem[{{Ingalls} {et~al.}(2016){Ingalls}, {Krick}, {Carey}, {Stauffer}, {Lowrance}, {Grillmair}, {Buzasi}, {Deming}, {Diamond-Lowe}, {Evans}, {Morello}, {Stevenson}, {Wong}, {Capak}, {Glaccum}, {Laine}, {Surace}, \& {Storrie-Lombardi}}]{IngallsEtal2016ajSpitzerRepeatability}
{Ingalls}, J.~G., {Krick}, J.~E., {Carey}, S.~J., {et~al.} 2016, \aj, 152, 44

\bibitem[{{Komacek} \& {Showman}(2016)}]{Komacek2016}
{Komacek}, T.~D. \& {Showman}, A.~P. 2016, \apj, 821, 16

\bibitem[{{Krenn} {et~al.}(2023){Krenn}, {Lendl}, {Patel}, {Carone}, {Deleuil}, {Sulis}, {Collier Cameron}, {Deline}, {Guterman}, {Queloz}, {Fossati}, {Brandeker}, {Heng}, {Akinsanmi}, {Adibekyan}, {Bonfanti}, {Demangeon}, {Kitzmann}, {Salmon}, {Sousa}, {Wilson}, {Alibert}, {Alonso}, {Anglada}, {B{\'a}rczy}, {Barrado Navascues}, {Barros}, {Baumjohann}, {Beck}, {Beck}, {Benz}, {Billot}, {Blecha}, {Bonfils}, {Borsato}, {Broeg}, {Cabrera}, {Charnoz}, {Corral van Damme}, {Csizmadia}, {Cubillos}, {Davies}, {Delrez}, {Demory}, {Ehrenreich}, {Erikson}, {Farinato}, {Fortier}, {Fridlund}, {Gandolfi}, {Gillon}, {G{\"u}del}, {Hoyer}, {Isaak}, {Kiss}, {Kopp}, {Laskar}, {Lecavelier des Etangs}, {Lovis}, {Magrin}, {Maxted}, {Mordasini}, {Nascimbeni}, {Olofsson}, {Ottensamer}, {Pagano}, {Pall{\'e}}, {Peter}, {Piotto}, {Pollacco}, {Ragazzoni}, {Rando}, {Rauer}, {Ribas}, {Santos}, {Scandariato}, {S{\'e}gransan}, {Simon}, {Smith}, {Steller}, {Szab{\'o}}, {Thomas}, {Udry}, {Ulmer}, {Van Grootel}, {Venturini}, \&
  {Walton}}]{2023A&A...672A..24K}
{Krenn}, A.~F., {Lendl}, M., {Patel}, J.~A., {et~al.} 2023, \aap, 672, A24

\bibitem[{{L{\'e}ger} {et~al.}(2009){L{\'e}ger}, {Rouan}, {Schneider}, {Barge}, {Fridlund}, {Samuel}, {Ollivier}, {Guenther}, {Deleuil}, {Deeg}, {Auvergne}, {Alonso}, {Aigrain}, {Alapini}, {Almenara}, {Baglin}, {Barbieri}, {Bruntt}, {Bord{\'e}}, {Bouchy}, {Cabrera}, {Catala}, {Carone}, {Carpano}, {Csizmadia}, {Dvorak}, {Erikson}, {Ferraz-Mello}, {Foing}, {Fressin}, {Gandolfi}, {Gillon}, {Gondoin}, {Grasset}, {Guillot}, {Hatzes}, {H{\'e}brard}, {Jorda}, {Lammer}, {Llebaria}, {Loeillet}, {Mayor}, {Mazeh}, {Moutou}, {P{\"a}tzold}, {Pont}, {Queloz}, {Rauer}, {Renner}, {Samadi}, {Shporer}, {Sotin}, {Tingley}, {Wuchterl}, {Adda}, {Agogu}, {Appourchaux}, {Ballans}, {Baron}, {Beaufort}, {Bellenger}, {Berlin}, {Bernardi}, {Blouin}, {Baudin}, {Bodin}, {Boisnard}, {Boit}, {Bonneau}, {Borzeix}, {Briet}, {Buey}, {Butler}, {Cailleau}, {Cautain}, {Chabaud}, {Chaintreuil}, {Chiavassa}, {Costes}, {Cuna Parrho}, {de Oliveira Fialho}, {Decaudin}, {Defise}, {Djalal}, {Epstein}, {Exil}, {Faur{\'e}}, {Fenouillet}, {Gaboriaud},
  {Gallic}, {Gamet}, {Gavalda}, {Grolleau}, {Gruneisen}, {Gueguen}, {Guis}, {Guivarc'h}, {Guterman}, {Hallouard}, {Hasiba}, {Heuripeau}, {Huntzinger}, {Hustaix}, {Imad}, {Imbert}, {Johlander}, {Jouret}, {Journoud}, {Karioty}, {Kerjean}, {Lafaille}, {Lafond}, {Lam-Trong}, {Landiech}, {Lapeyrere}, {Larqu{\'e}}, {Laudet}, {Lautier}, {Lecann}, {Lefevre}, {Leruyet}, {Levacher}, {Magnan}, {Mazy}, {Mertens}, {Mesnager}, {Meunier}, {Michel}, {Monjoin}, {Naudet}, {Nguyen-Kim}, {Orcesi}, {Ottacher}, {Perez}, {Peter}, {Plasson}, {Plesseria}, {Pontet}, {Pradines}, {Quentin}, {Reynaud}, {Rolland}, {Rollenhagen}, {Romagnan}, {Russ}, {Schmidt}, {Schwartz}, {Sebbag}, {Sedes}, {Smit}, {Steller}, {Sunter}, {Surace}, {Tello}, {Tiph{\`e}ne}, {Toulouse}, {Ulmer}, {Vandermarcq}, {Vergnault}, {Vuillemin}, \& {Zanatta}}]{Leger2009}
{L{\'e}ger}, A., {Rouan}, D., {Schneider}, J., {et~al.} 2009, \aap, 506, 287

\bibitem[{{Lendl} {et~al.}(2020){Lendl}, {Csizmadia}, {Deline}, {Fossati}, {Kitzmann}, {Heng}, {Hoyer}, {Salmon}, {Benz}, {Broeg}, {Ehrenreich}, {Fortier}, {Queloz}, {Bonfanti}, {Brandeker}, {Collier Cameron}, {Delrez}, {Garcia Mu{\~n}oz}, {Hooton}, {Maxted}, {Morris}, {Van Grootel}, {Wilson}, {Alibert}, {Alonso}, {Asquier}, {Bandy}, {B{\'a}rczy}, {Barrado}, {Barros}, {Baumjohann}, {Beck}, {Beck}, {Bekkelien}, {Bergomi}, {Billot}, {Biondi}, {Bonfils}, {Bourrier}, {Busch}, {Cabrera}, {Cessa}, {Charnoz}, {Chazelas}, {Corral Van Damme}, {Davies}, {Deleuil}, {Demangeon}, {Demory}, {Erikson}, {Farinato}, {Fridlund}, {Futyan}, {Gandolfi}, {Gillon}, {Guterman}, {Hasiba}, {Hernandez}, {Isaak}, {Kiss}, {Kuntzer}, {Lecavelier des Etangs}, {L{\"u}ftinger}, {Laskar}, {Lovis}, {Magrin}, {Malvasio}, {Marafatto}, {Michaelis}, {Munari}, {Nascimbeni}, {Olofsson}, {Ottacher}, {Ottensamer}, {Pagano}, {Pall{\'e}}, {Peter}, {Piazza}, {Piotto}, {Pollacco}, {Ratti}, {Rauer}, {Ragazzoni}, {Rando}, {Ribas}, {Rieder}, {Rohlfs},
  {Safa}, {Santos}, {Scandariato}, {S{\'e}gransan}, {Simon}, {Singh}, {Smith}, {Sordet}, {Sousa}, {Steller}, {Szab{\'o}}, {Thomas}, {Tschentscher}, {Udry}, {Viotto}, {Walter}, {Walton}, {Wildi}, \& {Wolter}}]{Lendl2020}
{Lendl}, M., {Csizmadia}, S., {Deline}, A., {et~al.} 2020, \aap, 643, A94

\bibitem[{{Lightkurve Collaboration} {et~al.}(2018){Lightkurve Collaboration}, {Cardoso}, {Hedges}, {Gully-Santiago}, {Saunders}, {Cody}, {Barclay}, {Hall}, {Sagear}, {Turtelboom}, {Zhang}, {Tzanidakis}, {Mighell}, {Coughlin}, {Bell}, {Berta-Thompson}, {Williams}, {Dotson}, \& {Barentsen}}]{2018ascl.soft12013L}
{Lightkurve Collaboration}, {Cardoso}, J.~V.~d.~M., {Hedges}, C., {et~al.} 2018, {Lightkurve: Kepler and TESS time series analysis in Python}, Astrophysics Source Code Library

\bibitem[{{Lindegren} {et~al.}(2021){Lindegren}, {Bastian}, {Biermann}, {Bombrun}, {de Torres}, {Gerlach}, {Geyer}, {Hern{\'a}ndez}, {Hilger}, {Hobbs}, {Klioner}, {Lammers}, {McMillan}, {Ramos-Lerate}, {Steidelm{\"u}ller}, {Stephenson}, \& {van Leeuwen}}]{Lindegren2021}
{Lindegren}, L., {Bastian}, U., {Biermann}, M., {et~al.} 2021, \aap, 649, A4

\bibitem[{{Maciejewski} {et~al.}(2013){Maciejewski}, {Niedzielski}, {Wolszczan}, {Nowak}, {Neuh{\"a}user}, {Winn}, {Deka}, {Adam{\'o}w}, {G{\'o}recka}, {Fern{\'a}ndez}, {Aceituno}, {Ohlert}, {Errmann}, {Seeliger}, {Dimitrov}, {Latham}, {Esquerdo}, {McKnight}, {Holman}, {Jensen}, {Kramm}, {Pribulla}, {Raetz}, {Schmidt}, {Ginski}, {Mottola}, {Hellmich}, {Adam}, {Gilbert}, {Mugrauer}, {Saral}, {Popov}, \& {Raetz}}]{2013AJ....146..147M}
{Maciejewski}, G., {Niedzielski}, A., {Wolszczan}, A., {et~al.} 2013, \aj, 146, 147

\bibitem[{{Mandel} \& {Agol}(2002)}]{MandelAgol2002apjLightcurves}
{Mandel}, K. \& {Agol}, E. 2002, \apjl, 580, L171

\bibitem[{{Mansfield} {et~al.}(2020){Mansfield}, {Bean}, {Stevenson}, {Komacek}, {Bell}, {Tan}, {Malik}, {Beatty}, {Wong}, {Cowan}, {Dang}, {D{\'e}sert}, {Fortney}, {Gaudi}, {Keating}, {Kempton}, {Kreidberg}, {Line}, {Parmentier}, {Stassun}, {Swain}, \& {Zellem}}]{Mansfield2020}
{Mansfield}, M., {Bean}, J.~L., {Stevenson}, K.~B., {et~al.} 2020, \apjl, 888, L15

\bibitem[{{Marigo} {et~al.}(2017){Marigo}, {Girardi}, {Bressan}, {Rosenfield}, {Aringer}, {Chen}, {Dussin}, {Nanni}, {Pastorelli}, {Rodrigues}, {Trabucchi}, {Bladh}, {Dalcanton}, {Groenewegen}, {Montalb{\'a}n}, \& {Wood}}]{marigo2017}
{Marigo}, P., {Girardi}, L., {Bressan}, A., {et~al.} 2017, \apj, 835, 77

\bibitem[{{Maxted}(2018)}]{2018A+A...616A..39M}
{Maxted}, P.~F.~L. 2018, \aap, 616, A39

\bibitem[{{Maxted} {et~al.}(2022){Maxted}, {Ehrenreich}, {Wilson}, {Alibert}, {Cameron}, {Hoyer}, {Sousa}, {Olofsson}, {Bekkelien}, {Deline}, {Delrez}, {Bonfanti}, {Borsato}, {Alonso}, {Anglada Escud{\'e}}, {Barrado}, {Barros}, {Baumjohann}, {Beck}, {Beck}, {Benz}, {Billot}, {Biondi}, {Bonfils}, {Brandeker}, {Broeg}, {B{\'a}rczy}, {Cabrera}, {Charnoz}, {Corral Van Damme}, {Csizmadia}, {Davies}, {Deleuil}, {Demangeon}, {Demory}, {Erikson}, {Flor{\'e}n}, {Fortier}, {Fossati}, {Fridlund}, {Futyan}, {Gandolfi}, {Gillon}, {Guedel}, {Guterman}, {Heng}, {Isaak}, {Kiss}, {Laskar}, {Lecavelier des Etangs}, {Lendl}, {Lovis}, {Magrin}, {Nascimbeni}, {Ottensamer}, {Pagano}, {Pall{\'e}}, {Peter}, {Piotto}, {Pollacco}, {Pozuelos}, {Queloz}, {Ragazzoni}, {Rando}, {Rauer}, {Reimers}, {Ribas}, {Salmon}, {Santos}, {Scandariato}, {Simon}, {Smith}, {Steller}, {Swayne}, {Szab{\'o}}, {S{\'e}gransan}, {Thomas}, {Udry}, {Van Grootel}, \& {Walton}}]{2022MNRAS.514...77M}
{Maxted}, P.~F.~L., {Ehrenreich}, D., {Wilson}, T.~G., {et~al.} 2022, \mnras, 514, 77

\bibitem[{{Maxted} \& {Gill}(2019)}]{2019A&A...622A..33M}
{Maxted}, P.~F.~L. \& {Gill}, S. 2019, \aap, 622, A33

\bibitem[{{May} {et~al.}(2021){May}, {Komacek}, {Stevenson}, {Kempton}, {Bean}, {Malik}, {Ih}, {Mansfield}, {Savel}, {Deming}, {Desert}, {Feng}, {Fortney}, {Kataria}, {Lewis}, {Morley}, {Rauscher}, \& {Showman}}]{May2021}
{May}, E.~M., {Komacek}, T.~D., {Stevenson}, K.~B., {et~al.} 2021, \aj, 162, 158

\bibitem[{{Montalto} {et~al.}(2012){Montalto}, {Gregorio}, {Boué}, {Mortier}, {Boisse}, {Oshagh}, {Maturi}, {Figueira}, {Sousa}, \& {Santos}}]{wasp3}
{Montalto}, M., {Gregorio}, J., {Boué}, G., {et~al.} 2012, mnras, 427, 2757

\bibitem[{{Morris} {et~al.}(2021){Morris}, {Heng}, {Brandeker}, {Swan}, \& {Lendl}}]{2021A&A...651L..12M}
{Morris}, B.~M., {Heng}, K., {Brandeker}, A., {Swan}, A., \& {Lendl}, M. 2021, \aap, 651, L12

\bibitem[{{Nymeyer} {et~al.}(2011){Nymeyer}, {Harrington}, {Hardy}, {Stevenson}, {Campo}, {Madhusudhan}, {Collier-Cameron}, {Loredo}, {Blecic}, {Bowman}, {Britt}, {Cubillos}, {Hellier}, {Gillon}, {Maxted}, {Hebb}, {Wheatley}, {Pollacco}, \& {Anderson}}]{NymeyerEtal2011apjWASP18b}
{Nymeyer}, S., {Harrington}, J., {Hardy}, R.~A., {et~al.} 2011, \apj, 742, 35

\bibitem[{{Oddo} {et~al.}(2023){Oddo}, {Dragomir}, {Brandeker}, {Osborn}, {Collins}, {Stassun}, {Astudillo-Defru}, {Bieryla}, {Howell}, {Ciardi}, {Quinn}, {Almenara}, {Brice{\~n}o}, {Collins}, {Col{\'o}n}, {Conti}, {Crouzet}, {Furlan}, {Gan}, {Gnilka}, {Goeke}, {Gonzales}, {Harris}, {Jenkins}, {Jensen}, {Latham}, {Law}, {Lund}, {Mann}, {Massey}, {Murgas}, {Ricker}, {Relles}, {Rowden}, {Schwarz}, {Schlieder}, {Shporer}, {Seager}, {Srdoc}, {Torres}, {Twicken}, {Vanderspek}, {Winn}, \& {Ziegler}}]{2023AJ....165..134O}
{Oddo}, D., {Dragomir}, D., {Brandeker}, A., {et~al.} 2023, \aj, 165, 134

\bibitem[{{Pagano} {et~al.}(2024){Pagano}, {Scandariato}, {Singh}, {Lendl}, {Queloz}, {Simon}, {Sousa}, {Brandeker}, {Cameron}, {Sulis}, {Van Grootel}, {Wilson}, {Alibert}, {Alonso}, {Anglada}, {B{\'a}rczy}, {Navascues}, {Barros}, {Baumjohann}, {Beck}, {Beck}, {Benz}, {Billot}, {Bonfils}, {Borsato}, {Broeg}, {Bruno}, {Carone}, {Charnoz}, {Corral van Damme}, {Csizmadia}, {Cubillos}, {Davies}, {Deleuil}, {Deline}, {Delrez}, {Demangeon}, {Demory}, {Ehrenreich}, {Erikson}, {Fortier}, {Fossati}, {Fridlund}, {Gandolfi}, {Gillon}, {G{\"u}del}, {G{\"u}nther}, {Helling}, {Hoyer}, {Isaak}, {Kiss}, {Kopp}, {Lam}, {Laskar}, {Lecavelier des Etangs}, {Magrin}, {Maxted}, {Mordasini}, {Munari}, {Nascimbeni}, {Olofsson}, {Ottensamer}, {Pall{\'e}}, {Peter}, {Piotto}, {Pollacco}, {Ragazzoni}, {Rando}, {Rauer}, {Reimers}, {Ribas}, {Rieder}, {Santos}, {S{\'e}gransan}, {Smith}, {Stalport}, {Steller}, {Szab{\'o}}, {Thomas}, {Udry}, {Venturini}, \& {Walton}}]{Pagano2024}
{Pagano}, I., {Scandariato}, G., {Singh}, V., {et~al.} 2024, \aap, 682, A102

\bibitem[{{Parmentier} \& {Crossfield}(2018)}]{Parmentier2018}
{Parmentier}, V. \& {Crossfield}, I. J.~M. 2018, in Handbook of Exoplanets, ed. H.~J. {Deeg} \& J.~A. {Belmonte}, 116

\bibitem[{{Parmentier} {et~al.}(2021){Parmentier}, {Showman}, \& {Fortney}}]{Parmentier2021}
{Parmentier}, V., {Showman}, A.~P., \& {Fortney}, J.~J. 2021, \mnras, 501, 78

\bibitem[{{Parmentier} {et~al.}(2013){Parmentier}, {Showman}, \& {Lian}}]{Parmentier2013}
{Parmentier}, V., {Showman}, A.~P., \& {Lian}, Y. 2013, \aap, 558, A91

\bibitem[{{Parviainen} {et~al.}(2022){Parviainen}, {Wilson}, {Lendl}, {Kitzmann}, {Pall{\'e}}, {Serrano}, {Meier Valdes}, {Benz}, {Deline}, {Ehrenreich}, {Guterman}, {Heng}, {Demangeon}, {Bonfanti}, {Salmon}, {Singh}, {Santos}, {Sousa}, {Alibert}, {Alonso}, {Anglada}, {B{\'a}rczy}, {Barrado y Navascues}, {Barros}, {Baumjohann}, {Beck}, {Beck}, {Billot}, {Bonfils}, {Brandeker}, {Broeg}, {Cabrera}, {Charnoz}, {Collier Cameron}, {Corral Van Damme}, {Csizmadia}, {Davies}, {Deleuil}, {Delrez}, {Demory}, {Erikson}, {Farinato}, {Fortier}, {Fossati}, {Fridlund}, {Gandolfi}, {Gillon}, {G{\"u}del}, {Hoyer}, {Isaak}, {Kiss}, {Kopp}, {Laskar}, {Lecavelier des Etangs}, {Lovis}, {Magrin}, {Maxted}, {Mecina}, {Nascimbeni}, {Olofsson}, {Ottensamer}, {Pagano}, {Peter}, {Piazza}, {Piotto}, {Pollacco}, {Queloz}, {Ragazzoni}, {Rando}, {Rauer}, {Ribas}, {Scandariato}, {S{\'e}gransan}, {Simon}, {Smith}, {Steller}, {Szab{\'o}}, {Thomas}, {Udry}, {Van Grootel}, \& {Walton}}]{Parviainen2022}
{Parviainen}, H., {Wilson}, T.~G., {Lendl}, M., {et~al.} 2022, \aap, 668, A93

\bibitem[{Perez-Becker \& Showman(2013)}]{Perez2013}
Perez-Becker, D. \& Showman, A.~P. 2013, The Astrophysical Journal, 776, 134

\bibitem[{Perryman(2011)}]{Perryman}
Perryman, M. 2011, The Exoplanet Handbook (Cambridge University Press, New York)

\bibitem[{{Pollacco} {et~al.}(2008{\natexlab{a}}){Pollacco}, {Skillen}, {Collier Cameron}, {Loeillet}, {Stempels}, {Bouchy}, {Gibson}, {Hebb}, {Hébrard}, {Joshi}, {McDonald}, {Smalley}, {Smith}, {Street}, {Udry}, {West}, {Wilson}, {Wheatley}, {Aigrain}, {Alsubai}, {Benn}, {Bruce}, {Christian}, {Clarkson}, {Enoch}, {Evans}, {Fitzsimmons}, {Haswell}, {Hellier}, {Hickey}, {Hodgkin}, {Horne}, {Hrudková}, {Irwin}, {Kane}, {Keenan}, {Lister}, {Maxted}, {Mayor}, {Moutou}, {Norton}, {Osborne}, {Parley}, {Pont}, {Queloz}, {Ryans}, \& {Simpson}}]{Pollacco2008}
{Pollacco}, D., {Skillen}, I., {Collier Cameron}, A., {et~al.} 2008{\natexlab{a}}, MNRAS, 385, 1576

\bibitem[{{Pollacco} {et~al.}(2008{\natexlab{b}})}]{2008MNRAS.385.1576P}
{Pollacco}, D. {et~al.} 2008{\natexlab{b}}, \mnras, 385, 1576

\bibitem[{{Pollack} {et~al.}(1986){Pollack}, {Podolak}, {Bodenheimer}, \& {Christofferson}}]{Pollack1986}
{Pollack}, J.~B., {Podolak}, M., {Bodenheimer}, P., \& {Christofferson}, B. 1986, \icarus, 67, 409

\bibitem[{{Ricker} {et~al.}(2015){Ricker}, {Winn}, {Vanderspek}, {Latham}, {Bakos}, {Bean}, {Berta-Thompson}, {Brown}, {Buchhave}, {Butler}, {Butler}, {Chaplin}, {Charbonneau}, {Christensen-Dalsgaard}, {Clampin}, {Deming}, {Doty}, {De Lee}, {Dressing}, {Dunham}, {Endl}, {Fressin}, {Ge}, {Henning}, {Holman}, {Howard}, {Ida}, {Jenkins}, {Jernigan}, {Johnson}, {Kaltenegger}, {Kawai}, {Kjeldsen}, {Laughlin}, {Levine}, {Lin}, {Lissauer}, {MacQueen}, {Marcy}, {McCullough}, {Morton}, {Narita}, {Paegert}, {Palle}, {Pepe}, {Pepper}, {Quirrenbach}, {Rinehart}, {Sasselov}, {Sato}, {Seager}, {Sozzetti}, {Stassun}, {Sullivan}, {Szentgyorgyi}, {Torres}, {Udry}, \& {Villasenor}}]{2015JATIS...1a4003R}
{Ricker}, G.~R., {Winn}, J.~N., {Vanderspek}, R., {et~al.} 2015, Journal of Astronomical Telescopes, Instruments, and Systems, 1, 014003

\bibitem[{{Roman} {et~al.}(2021){Roman}, {Kempton}, {Rauscher}, {Harada}, {Bean}, \& {Stevenson}}]{Roman2021}
{Roman}, M.~T., {Kempton}, E. M.~R., {Rauscher}, E., {et~al.} 2021, \apj, 908, 101

\bibitem[{{Rostron} {et~al.}(2014){Rostron}, {Wheatley}, {Anderson}, {Collier Cameron}, {Fortney}, {Harrington}, {Knutson}, \& {Pollacco}}]{Rostron2014}
{Rostron}, J.~W., {Wheatley}, P.~J., {Anderson}, D.~R., {et~al.} 2014, \mnras, 441, 3666

\bibitem[{{Roth} {et~al.}(2024){Roth}, {Parmentier}, \& {Hammond}}]{Roth2024}
{Roth}, A., {Parmentier}, V., \& {Hammond}, M. 2024, \mnras, 531, 1056

\bibitem[{{Scandariato} {et~al.}(2022){Scandariato}, {Singh}, {Kitzmann}, {Lendl}, {Brandeker}, {Bruno}, {Bekkelien}, {Benz}, {Gutermann}, {Maxted}, {Bonfanti}, {Charnoz}, {Fridlund}, {Heng}, {Hoyer}, {Pagano}, {Persson}, {Salmon}, {Van Grootel}, {Wilson}, {Asquier}, {Bergomi}, {Gambicorti}, {Hasiba}, {Alibert}, {Alonso}, {Anglada}, {B{\'a}rczy}, {Barrado y Navascues}, {Barros}, {Baumjohann}, {Beck}, {Beck}, {Billot}, {Bonfils}, {Broeg}, {Cabrera}, {Collier Cameron}, {Csizmadia}, {Davies}, {Deleuil}, {Deline}, {Delrez}, {Demangeon}, {Demory}, {Erikson}, {Fortier}, {Fossati}, {Gandolfi}, {Gillon}, {G{\"u}del}, {Isaak}, {Kiss}, {Laskar}, {Lecavelier des Etangs}, {Lovis}, {Magrin}, {Nascimbeni}, {Olofsson}, {Ottensamer}, {Pall{\'e}}, {Parviainen}, {Peter}, {Piotto}, {Pollacco}, {Queloz}, {Ragazzoni}, {Rando}, {Rauer}, {Ribas}, {Santos}, {S{\'e}gransan}, {Serrano}, {Simon}, {Smith}, {Sousa}, {Steller}, {Szab{\'o}}, {Thomas}, {Udry}, {Ulmer}, \& {Walton}}]{Scandariato2022}
{Scandariato}, G., {Singh}, V., {Kitzmann}, D., {et~al.} 2022, \aap, 668, A17

\bibitem[{{Schanche} {et~al.}(2020){Schanche}, {H{\'e}brard}, {Collier Cameron}, {Dalal}, {Smalley}, {Wilson}, {Boisse}, {Bouchy}, {Brown}, {Demangeon}, {Haswell}, {Hellier}, {Kolb}, {Lopez}, {Maxted}, {Pollacco}, {West}, \& {Wheatley}}]{Schanche2020}
{Schanche}, N., {H{\'e}brard}, G., {Collier Cameron}, A., {et~al.} 2020, \mnras, 499, 428

\bibitem[{{Schneider} {et~al.}(2022){Schneider}, {Carone}, {Decin}, {J{\o}rgensen}, {Molli{\`e}re}, {Baeyens}, {Kiefer}, \& {Helling}}]{Schneider2022}
{Schneider}, A.~D., {Carone}, L., {Decin}, L., {et~al.} 2022, \aap, 664, A56

\bibitem[{{Schwartz} \& {Cowan}(2015)}]{Schwartz2015}
{Schwartz}, J.~C. \& {Cowan}, N.~B. 2015, \mnras, 449, 4192

\bibitem[{{Schwartz} {et~al.}(2017){Schwartz}, {Kashner}, {Jovmir}, \& {Cowan}}]{Schwartz2017}
{Schwartz}, J.~C., {Kashner}, Z., {Jovmir}, D., \& {Cowan}, N.~B. 2017, \apj, 850, 154

\bibitem[{{Schwarz}(1978)}]{Schwarz1978anstaBIC}
{Schwarz}, G. 1978, Annals of Statistics, 6, 461

\bibitem[{{Scuflaire} {et~al.}(2008){Scuflaire}, {Th{\'e}ado}, {Montalb{\'a}n}, {Miglio}, {Bourge}, {Godart}, {Thoul}, \& {Noels}}]{scuflaire2008}
{Scuflaire}, R., {Th{\'e}ado}, S., {Montalb{\'a}n}, J., {et~al.} 2008, \apss, 316, 83

\bibitem[{{Seager}(2010)}]{Seager2010}
{Seager}, S. 2010, {Exoplanet Atmospheres: Physical Processes}

\bibitem[{Seager \& Dotson(2010)}]{seager2010exoplanets}
Seager, S. \& Dotson, R. 2010, Exoplanets, Space Science Series (University of Arizona Press)

\bibitem[{{Siegel}(1982)}]{siegelslopes}
{Siegel}, A. 1982, Biometrika, 69, 242

\bibitem[{{Singh} {et~al.}(2022){Singh}, {Bonomo}, {Scandariato}, {Cibrario}, {Barbato}, {Fossati}, {Pagano}, \& {Sozzetti}}]{Singh2022}
{Singh}, V., {Bonomo}, A.~S., {Scandariato}, G., {et~al.} 2022, \aap, 658, A132

\bibitem[{{Skrutskie} {et~al.}(2006){Skrutskie}, {Cutri}, {Stiening}, {Weinberg}, {Schneider}, {Carpenter}, {Beichman}, {Capps}, {Chester}, {Elias}, {Huchra}, {Liebert}, {Lonsdale}, {Monet}, {Price}, {Seitzer}, {Jarrett}, {Kirkpatrick}, {Gizis}, {Howard}, {Evans}, {Fowler}, {Fullmer}, {Hurt}, {Light}, {Kopan}, {Marsh}, {McCallon}, {Tam}, {Van Dyk}, \& {Wheelock}}]{Skrutskie2006}
{Skrutskie}, M.~F., {Cutri}, R.~M., {Stiening}, R., {et~al.} 2006, \aj, 131, 1163

\bibitem[{{Stevenson} {et~al.}(2012{\natexlab{a}}){Stevenson}, {Harrington}, {Fortney}, {Loredo}, {Hardy}, {Nymeyer}, {Bowman}, {Cubillos}, {Bowman}, \& {Hardin}}]{StevensonEtal2012apjSpitzerHD149026b}
{Stevenson}, K.~B., {Harrington}, J., {Fortney}, J.~J., {et~al.} 2012{\natexlab{a}}, \apj, 754, 136

\bibitem[{{Stevenson} {et~al.}(2012{\natexlab{b}}){Stevenson}, {Harrington}, {Lust}, {Lewis}, {Montagnier}, {Moses}, {Visscher}, {Blecic}, {Hardy}, {Cubillos}, \& {Campo}}]{StevensonEtal2012apjGJ346nonPlanets}
{Stevenson}, K.~B., {Harrington}, J., {Lust}, N.~B., {et~al.} 2012{\natexlab{b}}, \apj, 755, 9

\bibitem[{{Stevenson} {et~al.}(2010){Stevenson}, {Harrington}, {Nymeyer}, {Madhusudhan}, {Seager}, {Bowman}, {Hardy}, {Deming}, {Rauscher}, \& {Lust}}]{StevensonEtal2010natGJ436b}
{Stevenson}, K.~B., {Harrington}, J., {Nymeyer}, S., {et~al.} 2010, \nat, 464, 1161

\bibitem[{{Stock} {et~al.}(2022){Stock}, {Kitzmann}, \& {Patzer}}]{Stock2022}
{Stock}, J.~W., {Kitzmann}, D., \& {Patzer}, A. B.~C. 2022, \mnras, 517, 4070

\bibitem[{{Street} {et~al.}(2007){Street}, {Christian}, {Clarkson}, {Collier Cameron}, {Enoch}, {Kane}, {Lister}, {West}, {Wilson}, {Evans}, {Fitzsimmons}, {Haswell}, {Hellier}, {Hodgkin}, {Horne}, {Irwin}, {Keenan}, {Norton}, {Osborne}, {Pollacco}, {Ryans}, {Skillen}, {Wheatley}, \& {Barnes}}]{Street2007}
{Street}, R.~A., {Christian}, D.~J., {Clarkson}, W.~I., {et~al.} 2007, \mnras, 379, 816

\bibitem[{{Sudarsky} {et~al.}(2000){Sudarsky}, {Burrows}, \& {Pinto}}]{Sudarsky2000}
{Sudarsky}, D., {Burrows}, A., \& {Pinto}, P. 2000, \apj, 538, 885

\bibitem[{{Sulis} {et~al.}(2023){Sulis}, {Lendl}, {Cegla}, {Rodr{\'\i}guez D{\'\i}az}, {Bigot}, {Van Grootel}, {Bekkelien}, {Cameron}, {Maxted}, {Simon}, {Lovis}, {Scandariato}, {Bruno}, {Nardiello}, {Bonfanti}, {Fridlund}, {Persson}, {Salmon}, {Sousa}, {Wilson}, {Krenn}, {Hoyer}, {Santerne}, {Ehrenreich}, {Alibert}, {Alonso}, {Anglada}, {B{\'a}rczy}, {Barrado y Navascues}, {Barros}, {Baumjohann}, {Beck}, {Beck}, {Benz}, {Billot}, {Bonfils}, {Borsato}, {Brandeker}, {Broeg}, {Cabrera}, {Charnoz}, {Corral van Damme}, {Csizmadia}, {Davies}, {Deleuil}, {Deline}, {Delrez}, {Demangeon}, {Demory}, {Erikson}, {Fortier}, {Fossati}, {Gandolfi}, {Gillon}, {G{\"u}del}, {Heng}, {Isaak}, {Kiss}, {Laskar}, {Lecavelier des Etangs}, {Magrin}, {Munari}, {Nascimbeni}, {Olofsson}, {Ottensamer}, {Pagano}, {Pall{\'e}}, {Peter}, {Piotto}, {Pollacco}, {Queloz}, {Ragazzoni}, {Rando}, {Rauer}, {Ribas}, {Rieder}, {Santos}, {S{\'e}gransan}, {Smith}, {Steinberger}, {Steller}, {Szab{\'o}}, {Thomas}, {Udry}, {Walton}, \&
  {Wolter}}]{2023A&A...670A..24S}
{Sulis}, S., {Lendl}, M., {Cegla}, H.~M., {et~al.} 2023, \aap, 670, A24

\bibitem[{{Szab{\'o}} {et~al.}(2021){Szab{\'o}}, {Gandolfi}, {Brandeker}, {Csizmadia}, {Garai}, {Billot}, {Broeg}, {Ehrenreich}, {Fortier}, {Fossati}, {Hoyer}, {Kiss}, {Lecavelier des Etangs}, {Maxted}, {Ribas}, {Alibert}, {Alonso}, {Anglada Escud{\'e}}, {B{\'a}rczy}, {Barros}, {Barrado}, {Baumjohann}, {Beck}, {Beck}, {Bekkelien}, {Bonfils}, {Benz}, {Borsato}, {Busch}, {Cabrera}, {Charnoz}, {Collier Cameron}, {Van Damme}, {Davies}, {Delrez}, {Deleuil}, {Demangeon}, {Demory}, {Erikson}, {Fridlund}, {Futyan}, {Garc{\'\i}a Mu{\~n}oz}, {Gillon}, {Guedel}, {Guterman}, {Heng}, {Isaak}, {Lacedelli}, {Laskar}, {Lendl}, {Lovis}, {Luntzer}, {Magrin}, {Nascimbeni}, {Olofsson}, {Osborn}, {Ottensamer}, {Pagano}, {Pall{\'e}}, {Peter}, {Piazza}, {Piotto}, {Pollacco}, {Queloz}, {Ragazzoni}, {Rando}, {Rauer}, {Santos}, {Scandariato}, {S{\'e}gransan}, {Serrano}, {Sicilia}, {Simon}, {Smith}, {Sousa}, {Steller}, {Thomas}, {Udry}, {Van Grootel}, {Walton}, \& {Wilson}}]{2021A&A...654A.159S}
{Szab{\'o}}, G.~M., {Gandolfi}, D., {Brandeker}, A., {et~al.} 2021, \aap, 654, A159

\bibitem[{{Tan} \& {Komacek}(2019)}]{Tan2019}
{Tan}, X. \& {Komacek}, T.~D. 2019, \apj, 886, 26

\bibitem[{{Tennyson} {et~al.}(2013){Tennyson}, {Hill}, \& {Yurchenko}}]{Tennyson2013}
{Tennyson}, J., {Hill}, C., \& {Yurchenko}, S.~N. 2013, in American Institute of Physics Conference Series, Vol. 1545, Eighth International Conference on Atomic and Molecular Data and Their Applications: ICAMDATA-2012, ed. J.~D. {Gillaspy}, W.~L. {Wiese}, \& Y.~A. {Podpaly}, 186--195

\bibitem[{{Tennyson} {et~al.}(2020){Tennyson}, {Yurchenko}, {Al-Refaie}, {Clark}, {Chubb}, {Conway}, {Dewan}, {Gorman}, {Hill}, {Lynas-Gray}, {Mellor}, {McKemmish}, {Owens}, {Polyansky}, {Semenov}, {Somogyi}, {Tinetti}, {Upadhyay}, {Waldmann}, {Wang}, {Wright}, \& {Yurchenko}}]{Tennyson2020}
{Tennyson}, J., {Yurchenko}, S.~N., {Al-Refaie}, A.~F., {et~al.} 2020, \jqsrt, 255, 107228

\bibitem[{{ter Braak} \& {Vrugt}(2008)}]{terBraak2008SnookerDEMC}
{ter Braak}, C. J.~F. \& {Vrugt}, J.~A. 2008, Statistics and Computing, 18, 435

\bibitem[{{Waldmann} {et~al.}(2015{\natexlab{a}}){Waldmann}, {Rocchetto}, {Tinetti}, {Barton}, {Yurchenko}, \& {Tennyson}}]{Waldmann2015b}
{Waldmann}, I.~P., {Rocchetto}, M., {Tinetti}, G., {et~al.} 2015{\natexlab{a}}, \apj, 813, 13

\bibitem[{{Waldmann} {et~al.}(2015{\natexlab{b}}){Waldmann}, {Tinetti}, {Rocchetto}, {Barton}, {Yurchenko}, \& {Tennyson}}]{Waldmann2015a}
{Waldmann}, I.~P., {Tinetti}, G., {Rocchetto}, M., {et~al.} 2015{\natexlab{b}}, \apj, 802, 107

\bibitem[{{Wilson} {et~al.}(2022){Wilson}, {Goffo}, {Alibert}, {Gandolfi}, {Bonfanti}, {Persson}, {Collier Cameron}, {Fridlund}, {Fossati}, {Korth}, {Benz}, {Deline}, {Flor{\'e}n}, {Guterman}, {Adibekyan}, {Hooton}, {Hoyer}, {Leleu}, {Mustill}, {Salmon}, {Sousa}, {Suarez}, {Abe}, {Agabi}, {Alonso}, {Anglada}, {Asquier}, {B{\'a}rczy}, {Barrado Navascues}, {Barros}, {Baumjohann}, {Beck}, {Beck}, {Billot}, {Bonfils}, {Brandeker}, {Broeg}, {Bryant}, {Burleigh}, {Buttu}, {Cabrera}, {Charnoz}, {Ciardi}, {Cloutier}, {Cochran}, {Collins}, {Col{\'o}n}, {Crouzet}, {Csizmadia}, {Davies}, {Deleuil}, {Delrez}, {Demangeon}, {Demory}, {Dragomir}, {Dransfield}, {Ehrenreich}, {Erikson}, {Fortier}, {Gan}, {Gill}, {Gillon}, {Gnilka}, {Grieves}, {Grziwa}, {G{\"u}del}, {Guillot}, {Haldemann}, {Heng}, {Horne}, {Howell}, {Isaak}, {Jenkins}, {Jensen}, {Kiss}, {Lacedelli}, {Lam}, {Laskar}, {Latham}, {Lecavelier des Etangs}, {Lendl}, {Lester}, {Levine}, {Livingston}, {Lovis}, {Luque}, {Magrin}, {Marie-Sainte}, {Maxted}, {Mayo},
  {McLean}, {Mecina}, {M{\'e}karnia}, {Nascimbeni}, {Nielsen}, {Olofsson}, {Osborn}, {Osborne}, {Ottensamer}, {Pagano}, {Pall{\'e}}, {Peter}, {Piotto}, {Pollacco}, {Queloz}, {Ragazzoni}, {Rando}, {Rauer}, {Redfield}, {Ribas}, {Ricker}, {Rieder}, {Santos}, {Scandariato}, {Schmider}, {Schwarz}, {Scott}, {Seager}, {S{\'e}gransan}, {Serrano}, {Simon}, {Smith}, {Steller}, {Stockdale}, {Szab{\'o}}, {Thomas}, {Ting}, {Triaud}, {Udry}, {Van Eylen}, {Van Grootel}, {Vanderspek}, {Viotto}, {Walton}, \& {Winn}}]{2022MNRAS.511.1043W}
{Wilson}, T.~G., {Goffo}, E., {Alibert}, Y., {et~al.} 2022, \mnras, 511, 1043

\bibitem[{{Wong} {et~al.}(2021){Wong}, {Kitzmann}, {Shporer}, {Heng}, {Fetherolf}, {Benneke}, {Daylan}, {Kane}, {Vanderspek}, {Seager}, {Winn}, {Jenkins}, \& {Ting}}]{Wong2021}
{Wong}, I., {Kitzmann}, D., {Shporer}, A., {et~al.} 2021, \aj, 162, 127

\bibitem[{{Wright} {et~al.}(2010){Wright}, {Eisenhardt}, {Mainzer}, {Ressler}, {Cutri}, {Jarrett}, {Kirkpatrick}, {Padgett}, {McMillan}, {Skrutskie}, {Stanford}, {Cohen}, {Walker}, {Mather}, {Leisawitz}, {Gautier}, {McLean}, {Benford}, {Lonsdale}, {Blain}, {Mendez}, {Irace}, {Duval}, {Liu}, {Royer}, {Heinrichsen}, {Howard}, {Shannon}, {Kendall}, {Walsh}, {Larsen}, {Cardon}, {Schick}, {Schwalm}, {Abid}, {Fabinsky}, {Naes}, \& {Tsai}}]{Wright2010}
{Wright}, E.~L., {Eisenhardt}, P. R.~M., {Mainzer}, A.~K., {et~al.} 2010, \aj, 140, 1868

\bibitem[{{Zhang} {et~al.}(2018){Zhang}, {Knutson}, {Kataria}, {Schwartz}, {Cowan}, {Showman}, {Burrows}, {Fortney}, {Todorov}, {Desert}, {Agol}, \& {Deming}}]{Zhang2018}
{Zhang}, M., {Knutson}, H.~A., {Kataria}, T., {et~al.} 2018, \aj, 155, 83

\bibitem[{{Zhao} {et~al.}(2012){Zhao}, {Monnier}, {Swain}, {Barman}, \& {Hinkley}}]{Zhao2012}
{Zhao}, M., {Monnier}, J.~D., {Swain}, M.~R., {Barman}, T., \& {Hinkley}, S. 2012, ApJ, 744, 122

\end{thebibliography}

\label{LastPage}

\end{document}